%% file: MultiNuisance.tex
\documentclass[a4paper,10pt]{scrartcl}
\usepackage{amsmath}
\usepackage{amssymb}
\usepackage{amsxtra}
\usepackage{amsthm}
\usepackage{mathrsfs}
\usepackage{a4wide}
\usepackage{subfigure}
\usepackage{tikz}
\usepackage{pgfplots}
\usepackage{natbib}
\usepackage{algorithm}
\usepackage{algpseudocode}
\usepackage{url}
\pgfplotsset{compat=newest}
\newlength{\fwidth}
\newlength{\fheight}

\newcommand{\R}{\mathbb{R}}
\newcommand{\N}{\mathbb{N}}

\renewcommand{\P}[2]{\mathbb{P}_{#1}\left[#2\right]}
\newcommand{\E}[2]{\mathbb{E}_{#1}\left[#2\right]}

\newcommand{\V}[2]{\mathbb{V}_{#1}\left[#2\right]}

\newcommand{\1}{\mathbf{1}}
\newcommand{\ms}{\mu \mathrm{s}}
\newcommand{\nm}{\mathrm{nm}}
\newcommand{\LINK}{\url{https://go.uniwue.de/math-ams}}

\newtheorem{theorem}{Theorem}[section]

\newtheorem{lem}[theorem]{Lemma}
\newtheorem{cor}[theorem]{Corollary}
\newtheorem{remark}[theorem]{Remark}

\newtheorem{example}[theorem]{Example}
\newtheorem{assumption}{Assumption}

\begin{document}

	\begin{center}
	\begin{minipage}{.8\textwidth}
		\centering 
		\LARGE Multiscale scanning with nuisance parameters\\[0.5cm]
		
		\normalsize
		\textsc{Claudia K\"onig}\\[0.1cm]
		\verb+claudia-juliane.koenig@mathematik.uni-goettingen.de+\\
		Institute for Mathematical Stochastics, University of G\"ottingen\\[0.3cm]
		
		\textsc{Axel Munk}\\[0.1cm]
		\verb+munk@math.uni-goettingen.de+\\
		Institute for Mathematical Stochastics, University of G\"ottingen\\
		and\\
		Felix Bernstein Institute for Mathematical Statistics in the Bioscience, University of G\"ottingen\\
		and\\
		DFG Cluster of Excellence "Multiscale Bioimaging - From Molecular Machines to Networks of Excitable Cells", University of G\"ottingen Medical Center\\[0.3cm]
		
		\textsc{Frank Werner}\footnotemark[1]\\[0.1cm]
		\verb+frank.werner@uni-wuerzburg.de+\\
		Julius-Maximilians-Universität Würzburg (JMU), Institute of Mathematics, Würzburg, Germany
	\end{minipage}
\end{center}

\footnotetext[1]{Corresponding author}	

\begin{abstract}
	We develop a multiscale scanning method to find anomalies in a $d$-dimensional random field in the presence of nuisance parameters. This covers the common situation that either the baseline-level or additional parameters such as the variance are unknown and have to be estimated from the data. We argue that state of the art approaches to determine asymptotically correct critical values for multiscale scanning statistics will in general fail when such parameters are naively replaced by plug-in estimators. Instead, we suggest to estimate the nuisance parameters on the largest scale and to use (only) smaller scales for multiscale scanning. We prove a uniform invariance principle for the resulting adjusted multiscale statistic (AMS), which is widely applicable and provides a computationally feasible way to simulate asymptotically correct critical values. We illustrate the implications of our theoretical results in a simulation study and in a real data example from super-resolution STED microscopy. This allows us to identify interesting regions inside a specimen in a pre-scan with controlled family-wise error rate.
\end{abstract}


\textit{AMS classification numbers:} 60F17, 62H10, 60G50, 62F03.  \\[0.3cm]
\date{\today}
	\section{Introduction}\label{sec:intro}

Scanning a $d$-dimensional random field $Y = \left(Y_i\right)_{i\in \left\{1,...,n\right\}^d}$ of random variables (r.v.'s) $Y_i$ for anomalies in their distribution is a prominent topic in statistical research which is of great practical use. Applications range from the detection of anomalous clusters in social and biological networks and graphs \citep{accd11,cn14,srs16,bamh20} to detection of structural changes in imaging \citep{cmv02,adh06,s18,kkm23}, genomics \citep{rdj14,fhms14}, brain research \citep{lpstv21} and cell microscopy \citep{pwm18}, to mention a few.

\subsection{Setup}

In this paper, we consider a parametric model of the form
\begin{equation}\label{eq:model}
	Y_i \sim F_{\theta_i, \xi},\qquad i \in I_n^d := \left\{1,...,n\right\}^d,
\end{equation}
where the $Y_i$ are independent, real valued r.v.'s distributed to some unknown $F_{\theta_i, \xi}$ from a given, known class of distributions $\mathcal F$ indexed in $\theta$ and $\xi$. {Here, $i$ denotes a multi-index in a $d$-dimensional grid $I_n^d = \{1,...,n\}^d$. In what follows}, $\theta_i \in \Theta$ denotes the parameter of interest, $\xi \in \Xi$ a collection of nuisance parameters, and $\Theta \subset \R^{d_1}$ and $\Xi\subset \R^{d_2}$ are the corresponding parameter spaces. {Note, that $\theta_i$ may vary between observations $Y_i$, whereas the nuisance parameter $\xi$ is assumed to be the same for all $i \in I_n^d$}, but we stress that our analysis can also be transferred to varying $\xi$. An exemplary case for model \eqref{eq:model} would be a Gaussian field $Y_i\sim \mathcal N \left(\theta_i, \Sigma\right)$, where anomalies $\theta_i$ occur in the mean as deviation from a {(known or unknown)} ground truth $\theta_0 \in\Theta$ {and} the unknown{, positive semi-definite covariance matrix} $\Sigma$ is the nuisance parameter. For the moment, we do not pose structural assumptions on $F_{\theta, \xi}$, but require some smoothness and tail properties later on (see Assumption \ref{ass:model} below).

\subsection{Anomaly detection}

Given a baseline or reference intensity $\theta_0 \in \Theta$ and a region $R \subset I_n^d$, the problem to detect an anomaly inside $R$ can be formulated as the testing problem
\begin{equation}\label{eq:single_testing}
	H_{R}:~\forall~i \in R: \theta_i = \theta_0 \qquad\text{vs.}\qquad 
	K_{R}:~\exists~i \in R \text{ s.t. }\theta_i \neq \theta_0.
\end{equation}
{Since $R$ depends on $n$, the hypotheses $H_R$ and alternatives $K_R$ also depend (implicitly) on $n$, which we suppress to ease notation}. Note, that in practice often $d_1 = 1$, {i.e. $\theta_i \in \R$,} and it might be more relevant to consider alternatives of the form $\theta_i > \theta_0$ or $\theta_i < \theta_0$. For the sake of brevity, we consider only two-sided {alternatives as in \eqref{eq:single_testing}} here, but we emphasize that the one-sided situation can be developed analogous{ly} (see Remark \ref{rem:one_sided} below).

Depending on the structure of $R$, the likelihood ratio test (LRT) is known to have {favourable} properties, see e.g. {Sections 3 and 12 in} \citet{lr05}. Therefore, throughout this paper, we consider the corresponding log LRT statistic
\begin{equation}\label{eq:TR}
	T_R \left(Y, \theta_0, \xi\right) := \sqrt{2 \log \left(\frac{\sup_{\theta \in\Theta} \prod_{i \in R} f_{\theta, \xi} \left(Y_i\right)}{\prod_{i \in R} f_{\theta_0, \xi} \left(Y_i\right)}\right)}
\end{equation}
for \eqref{eq:single_testing}, where $f_{\theta, \xi}$ denotes the density or probability distribution function of $F_{\theta, \xi}$. Whenever this statistic exceeds a suitably chosen threshold (depending on $R$ and the prescribed level $\alpha \in \left(0,1\right)$), the corresponding LRT will reject {$H_{R}$ in \eqref{eq:single_testing}}.

\subsection{Multiscale scanning {for a known reference distribution}}\label{subsec:scanning}

However, as position and size of anomalies will in general be unknown, the region of interest $R$ is not known in advance. Therefore, we aim to \textit{scan} over $\mathcal{R}_n \subset 2^{I_n^d}$ (with $2^X $ being the power set of $X$), a set of regions of interest (e.g. the system of all discrete cubes inside $I_n^d$). {If the system $\mathcal R_n$ covers a range of {(}spatial{)} scales, then this approach is denoted as} \textit{multiscale scanning}{, which} has a long history in statistics, see e.g. \citet{k97} for an early reference and \citet{adh06,w10,accd11,n12,mbs13,cw13,sac16,pwm18,kmw20} for more recent works. A particular instance is multiscale change point detection, see e.g. \citet{cws22} {or \citet{kblm22}}, and the references given there. 

{Following on from} from \eqref{eq:single_testing}, {we will treat} multiscale scanning as a multiple testing problem
\begin{equation}\label{eq:multiple_testing}
	H_{R} \text{ vs. }K_{R} \qquad\mathbf{simultaneously}\text{ over } \mathcal R_n.
\end{equation}
The resulting scanning test then has to combine the local test statistics $T_R$, $R \in \mathcal R_n$, in \eqref{eq:TR} {in a suitable way}, such that the number of false rejections over $\mathcal R_n$ can be controlled. To this end, we aim to {design the scanning test, s.t. we} bound (asymptotically) the family wise error rate (FWER) by {a preassigned error level} $\alpha \in \left(0,1\right)$, i.e. \citep[see e.g.][]{m81,d14} 
\begin{equation}\label{eq:FWER}
	\sup\limits_{R \in \mathcal R_n} \P{H_{R}}{\text{``any false rejection in }R\text{''}} \leq \alpha + o(1) \qquad\text{as}\qquad n\to \infty.
\end{equation}
To combine the local test statistics in {such} a multiscale fashion, it is pertinent to employ a scale calibration $\widetilde{\omega}{_n}, \omega{_n} : {(0,\infty)} \to \left(0,\infty\right)$ to avoid small regions $R \in \mathcal R_n$ from dominating the overall statistic. Here, and in what follows, the cardinality $\left|R\right| \in \mathbb N$ of a region $R \in \mathcal R_n$ is denoted as \textit{scale}. This leads to
\begin{align}\label{eq:Tn}
	T_n \equiv T_n (Y, \mathcal R_n, \theta_0, \xi):=\max_{R\in\mathcal R_n} \widetilde{\omega}{_n}\left(\left|R\right|\right)\left[ T_R\left(Y, \theta_0, \xi\right)-\omega{_n}\left(\left|R\right|\right)\right],
\end{align}
which we then call a (calibrated) \textit{multiscale test statistic} whenever a range of scales is covered by $\mathcal R_n$. An appropriate choice of the calibration $\widetilde{\omega}{_n}$ and $\omega{_n}$ is of high importance for the overall {performance of the} method and will determine its detection properties. Different choices of $\widetilde{\omega}{_n}$ and $\omega{_n}$ have been proposed in the literature, mostly to achieve a non-degenerate limiting behaviour of $T_n$ or to optimize the detection power of the multiple test on all or specific scales. Besides uncalibrated scan statistics used e.g. in \citet{k97,a97,a98,km08,n12,pywr22}, we mention \citet{ds01}, who proposed a calibration ensuring non-degenerateness of the limiting distribution, and a more recent multivariate analogue by \citet{ds18}. Building on this, \citet{dw08} proposed a calibration, which was later shown to be superior over the uncalibrated scan statistic in \citet{cw13}, and which is since then widely used in the community, see e.g. \citet{rw13,smd13,fms14,kmw20}. More recently, other calibrations have been suggested, e.g. leading to Gumbel limits \citep{sac16,pwm18}, see also \citet{w20} for a recent review on this topic. In this study, we mostly leave the choice of $\widetilde{\omega}{_n}$ and $\omega{_n}$ to the user up to some mild condition (cf. Assumption \ref{ass:Omega}) on its growth posed below. This choice may e.g. reflect a preference for detecting signals of certain size, which is driven by the application at hand.

The multiscale scanning methodology is now as follows. Given $Y$, the values $T_R \left(Y, \theta_0, \xi\right)$, $R \in \mathcal R_n$ are computed, and all hypotheses $H_{R}$ such that
\begin{equation}\label{eq:reject}
	T_R \left(Y, \theta_0, \xi\right) \geq c_{|R|}(\eta) := \frac{\eta}{\widetilde{\omega}{_n}(|R|)} + \omega{_n}(|R|)
\end{equation}
are rejected. {In general}, the scale calibration $\widetilde{\omega}{_n}, \omega{_n}$ leads to \textit{scale dependent} local critical values, see also \citet{w10,psm17}. {However, t}he global threshold ${\eta\in\R}$ is universal, and to be chosen such that 
\begin{equation}\label{eq:threshold}
	\P{{H_{I_n^d}}}{T_n \left(Y, \mathcal R_n, \theta_0, \xi\right)>\eta} \leq \alpha + o(1)\qquad\text{as}\qquad n \to \infty,
\end{equation}
for some $\alpha \in \left(0,1\right)$. Since controlling the false positives over $H_{I_n^d}$ implies control over any system of sub-hypotheses $H_{R,n}$, $R \in \mathcal R_n$, {i.e.}
\[
\sup\limits_{R \in \mathcal R_n} \P{H_{R}}{\text{``any false rejection in }R\text{''}}  \leq \P{{H_{I_n^d}}}{\text{``any false rejection in }I_n^d\text{''}},
\]
{it follows that} \eqref{eq:threshold} ensures \eqref{eq:FWER} to hold true, and hence any of the local rejections is (asymptotically) correct with probability $\geq 1-\alpha$. 

{This principle to control the FWER underlies (often implicitly) many modern multiscale scanning methods}. One central example is to scan an independent Gaussian field $Y$ for anomalies (elevated means), where $\theta_ 0 =\mu_0 = 0$ and $\xi = \Sigma_0$ is the known covariance. 
In this case, optimality properties of this method, depending on the chosen calibration $\widetilde \omega{_n}$ and $\omega{_n}$, are known, see e.g. \citet{cw13} for a statement about minimax optimality when using just a single scale, or \citet{sac16} and \citet{kmw20} for an oracle property revealing that scanning over several scales is asymptotically as efficient as scanning over the {signal's} scale only. 

\paragraph{Choosing the global threshold $\eta$.} The remaining and potentially difficult question is how to choose the threshold $\eta$ such that \eqref{eq:threshold} is valid. {To this end} knowledge of the distribution of $T_n$ under $\mathbb P_{{H_{I_n^d}}}$ {is beneficial,} which is not easy to obtain, in general. If the distributions $F_{\theta,\xi} = F_{\mu, \sigma^2}$ are Gaussians with known nuisance parameter $\xi = \sigma^2$ (variance), and if also the reference intensity $\mu = \theta_0$ (mean) is known, then it readily turns out that all local test statistics $T_R$ are constituted by Gaussians again, as $T_R\left(Y, \mu,\sigma\right) = \left|R\right|^{-1/2} \left| \sum_{i \in R} (Y_i - \mu)/\sigma \right|$ in this case. Consequently, $T_n$ can be simulated as the maximum over finitely many (dependent) Gaussian random variables. This way, it is possible to obtain empirical quantiles for $T_n$ under $H_0$ and to choose $\eta$ accordingly.

For many other distributional settings such as Binomial or Poisson distributions, the situation is more delicate. First steps have been taken by \citet{a97,a98,asn16}, who proposed distributional approximations of $T_n$ in a {Poisson} setting {under knowledge of a baseline intensity $\theta_0$}. However, those results are limited to one fixed scale. Another approach is based on tail bounds for $T_R$ as proven in \citet{w22}, which can be combined with a Bonferroni correction to ensure \eqref{eq:threshold} also in a multiscale setting. Complete multiscale approximations of $T_n$ in \eqref{eq:Tn} have been proposed in case of Bernoulli and Binomial fields by \citet{w10} or in a {more} general exponential family setting by \citet{kmw20} under knowledge of $\theta_0$ and $\xi$. The latter result is based on an invariance principle, which allows (given the baseline parameter $\theta_0$ and the nuisance parameters $\xi$) to approximate the distribution of $T_n$ by {the distribution in the Gaussian case as above, i.e.}
\begin{equation}\label{eq:Mn}
	M_n \equiv M_n\left(\mathcal R_n\right) := \max_{R\in\mathcal R_n} \widetilde{\omega}{_n}(|R|)\left[ \left|R\right|^{-1/2} \left| \sum\limits_{i \in R} X_i \right| - \omega{_n}(|R|) \right]
\end{equation}
with i.i.d. standard normal r.v.'s $X_i$, $i \in I_n^d$. Consequently, it is possible to simulate an appropriate threshold $\eta$ in \eqref{eq:threshold} in practice as soon as $\theta_0$ and $\xi$ are known. The rationale behind approximating $T_n$ in \eqref{eq:Tn} by $M_n$ in \eqref{eq:Mn} is, that for sufficiently large scales, a CLT implies that $T_R \approx \left|R\right|^{-1/2} \left| \sum\limits_{i \in R} X_i \right|$ in distribution. However, to ensure that $M_n$ in fact approximates $T_n$, this CLT has to hold true uniformly over all sets in $\mathcal R_n$, i.e. it is required to couple the process {(or at least its maximum)} $\left(T_R\right)_{R \in \mathcal R_n}$ with $\left(\left|R\right|^{-1/2} \left| \sum\limits_{i \in R} X_i\right|\right)_{R \in \mathcal R_n}$ {(or the maximum, respectively)} and to account for the calibration. Based on classical {strong approximation} results \citep{kmt75}, this approach has been introduced in \citet{fms14}, and more recently extended in \citet{pwm18,kmw20} by using explicit couplings for the suprema of empirical processes developed by \citet{cck14}.

\subsection{Our contribution and outline of the paper}

In contrast to existing work, the focus of this study is on the case that $\theta_0$ and{/or} the nuisance parameter $\xi$ in \eqref{eq:model} are unknown, {a typical situation in practice}. Note, that this changes the overall situation completely, as the local LRT statistics $T_R \left(Y, \theta_0, \xi\right)$ are no longer accessible. A common remedy which immediately comes to mind is to pre-estimate both $\theta_0$ and $\xi$ from $Y$ globally and then use the corresponding estimators $\hat \theta_0$, $\hat \xi$ as a plug-in in $T_R$ and consequently in the multiscale extension $T_n$ in \eqref{eq:Tn}. When one aims to \textit{estimate} $\theta_0$ only, then such an approach is known as pseudo maximum likelihood, and has e.g. been proposed in \citet{gs81}. Under certain regularity conditions, they prove asymptotic normality of the resulting estimator, but this result is limited to a single scale and cannot be transferred to multiscale testing. In fact, the {situation becomes more delicate: The} distribution of $\hat T_n := T_n \left(Y,\mathcal R_n, \hat \theta_0, \hat \xi\right)$ differs {even} asymptotically from that of $T_n \left(Y, \mathcal R_n, \theta_0, \xi\right)$ (and hence also from that of $M_n\left(\mathcal R_n\right)$ in \eqref{eq:Mn}) in general, and this is the case even in the most simple Gaussian example.

\begin{figure}[!htb]
	\scriptsize
	\centering
	\setlength{\fwidth}{3cm}
	\setlength{\fheight}{3cm}
	\begin{tabular}{cc}
		\input{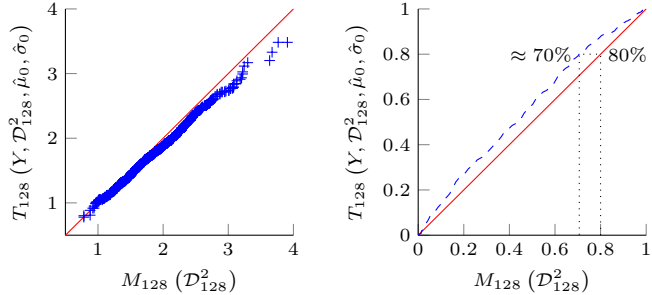}
		&
		\input{full_PP.tikz}
	\end{tabular}
	\caption{{QQ (left) and PP-Plots (right)} of the empirical distributions of $M_n\left(\mathcal R\right) = T_n \left(Y,\mathcal R,0,1\right)$ and the naive approximation $\hat T_n$ in the Gaussian case using all scales $1-128$, i.e. $\mathcal R_n = \mathcal D_n^d$ being the set of all rectangular regions, drawn from 2000 Monte Carlo runs.}
	\label{fig:gauss_dist}
\end{figure}

\paragraph{Failure of naive plug-in in the Gaussian model.} 
Consider the {univariate} homogeneous Gaussian model for $d_1 = 1$ as the best understood example of \eqref{eq:model}. Here, $F_{\theta_i, \xi} = \mathcal N \left(\mu_i, \sigma^2\right)$ is parametrized by its mean $\mu_i \in\R$ and variance $\sigma^2 > 0$. In this case, it can readily be seen that replacing {the baseline signal} $\mu_0$ and / or $\sigma$ by estimators does actually influence the distribution of $T_n$ and requires corrected critical values (thresholds) $c_{|R|}(\eta)$ for the local LRT test statistics. To illustrate this, we have depicted the empirical distributions of $T_n \left(Y,\mathcal R, \mu_0, \sigma_0\right) = M_n(\mathcal R)$ as in \eqref{eq:Mn}, and a corresponding variant with standard estimators $\hat \mu_0 = \frac{1}{n^d} \sum_{i \in I_n^d} Y_i$ and $\hat \sigma_0^2 = \frac{1}{n^d-1} \sum_{i \in I_n^d} \left(X_i - \hat \mu_0\right)^2$ for $n = 128$, $d = 2$, and {$\mathcal R = \mathcal D_n^d$ being the set of all rectangles in $2^{I_n^d}$} in Figure \ref{fig:gauss_dist}. As an exemplary scale calibration, we use the common choice $\widetilde \omega{_n} \equiv 1$ and $\omega{_n} \left(|R|\right) = \sqrt{2 \log \left(n^d / |R|\right) + 1}$, see \citet{dw08,fms14}.  

It is clearly visible that the distribution of the true scan statistic $T_n$ differs especially for large quantiles substantially from the one involving any of the estimators. If, e.g., the $80\%$-quantile of $M_n$ is used to calibrate $\hat T_n$, then the corresponding FWER in \eqref{eq:FWER} will be $\approx 30\%$ {instead of $20\%$}.

{The reason for this can already be explained in the easier setting when the} true variance $\sigma_0^2$ is known, i.e. for $T_n \left(Y, \mathcal R, \hat \mu_0, \sigma_0\right)$ with estimated global baseline $\hat \mu_0$. {In this case,} the log LRT statistic $T_R \left(Y,\mu_0,\sigma_0\right)$ compares the local average of $Y$ on $R$ with $\mu_0$, i.e. in case of unknown $\mu_0$ with the global average $\hat \mu_0$ on $I_n^d$. Consequently, for large regions, the value of $T_R \left(Y, \hat \mu_0, \sigma_0\right)$ will be negligible (in particular for $R = I_n^d$ even constantly be $0$), but $T_R \left(Y, \mu_0, \sigma_0\right)$ is not (for $R = I_n^d$, this is proportional to a $\chi^2_{n^d-1}$ distributed random variable). So, as long as the penalization $\widetilde \omega{_n}, \omega{_n}$ does not hinder these large scales from contributing to $T_n \left(Y, \mathcal R, \hat \mu_0, \sigma_0\right)$ (and $T_n \left(Y, \mathcal R, \mu_0, \sigma_0\right)$), the corresponding distributions must be different. On the other hand, a penalization hindering large scales from contributing to $T_n$ is not desired, as this will decrease the overall detection power of the corresponding LRT tests on such scales. It is well known, see e.g. \citet{ds01}, that the penalization chosen here indeed ensures the largest scales to contribute to $T_n \left(Y, \mathcal R, \mu_0, \sigma_0\right)$. Hence, we find that in this case, also the asymptotic distribution (which for this scale penalization is a supremum of a Gaussian process, see \citet{ds01}) should be different. 

Note, that the above observation is not restricted to {a} Gaussian model only, {it effectively} occurs in any model {where the nuisance parameter is estimated and naively plugged into the scan statistic}. The largest scales will in general contribute to the distribution-free statistic $M_n$ in \eqref{eq:Mn}, but will under mild assumptions on $T_R$ (local means) not contribute to $T_n$ in \eqref{eq:Tn} as argued above.

\paragraph{Our contribution.} This then raises the question how to adjust the threshold $\eta$ in \eqref{eq:threshold} properly, an issue which has often been overlooked and has not been discussed in the literature. A notable exception is \citet{wp22}, where the tail bounds for $T_R \left(Y, \hat \theta_0, \hat \xi\right)$ in \cite{w22} are combined with a Bonferroni correction to construct corrected values for $\eta$ in \eqref{eq:threshold}. This approach is, however, limited to relatively small systems $\mathcal R_n$ of regions. In this paper, we will close the aforementioned gap {and develop an adjusted multiscale scanning (AMS) procedure. This is based on a novel uniform invariance principle which can then be used to determine correct thresholds $\eta$ in \eqref{eq:threshold} for estimated values of $\theta_0$ and $\xi$. More precisely, we show (see Section \ref{sec:theory}), as long as the scales in $\mathcal R_n$ are not too small or too large compared to the sample size $n^{d}$,} that $\hat T_n$ can be approximated in distribution by $M_n \left(\mathcal R_n\right)$ as in \eqref{eq:Mn}, which can be pre-simulated given $\mathcal R_n$ in advance. From this it becomes {immediate} apparent that working with thresholds, which do not incorporate the error in estimating $\theta_0$ and $\xi$, does not provide the correct FWER, even asymptotically. 

\paragraph{Outline of the paper.} {The remainder of this study is organized as follows. In Section \ref{sec:pm} we will give a detailed description of the AMS procedure, address computational issues and provide a link to its implementation in MATLAB$^\copyright$. We will present the employed uniform invariance principle in Section \ref{sec:theory} and also discuss the posed assumptions. Section \ref{sec:sim} is devoted to simulations in a Poisson model, illustrating the finite sample performance of AMS. Following up on this, we present an analysis of innovative real-world data from smart super-resolution microscopy in Section \ref{sec:data_analysis}. We end this paper by a brief discussion in Section \ref{sec:discussion}. All proofs and further simulations as well as a detailed description of an efficient implementation of AMS are provided in a supplement.}

\section{Proposed method: adjusted multiscale scanning}\label{sec:pm}

{The uniform invariance principle in Theorem \ref{thm:approximation_estimated_intensity} below can briefly be described as follows. Given estimators $(\hat{\theta}_n)_{n \in \N}$ and $(\hat{\xi})_{n \in\N}$ of $\theta_0$ and $\xi$, the distribution of $T_n \left(Y, \mathcal R_n\left(r_n,m_n\right) \hat \theta_n, \hat \xi_n\right)$ can be approximated uniformly (over $\xi$) by that of $M_n \left(\mathcal R_n\left(r_n,m_n\right)\right)$ as $n \to \infty$, where $\mathcal R_n (c,d) := \left\{R \in \mathcal R_n ~\big|~ c \leq \left|R\right| \leq d\right\}$ denotes a scale-restricted set of regions and $r_n \nearrow \infty$, $m_n \nearrow \infty$ as $n \to \infty$ are suitable sequences depending on the consistency rate of $\hat \theta_n$ and $\hat \xi_n$.} In other words, we obtain a uniform invariance principle on all not too small and all not too large scales. In view of Figure \ref{fig:gauss_dist}, it is obvious that {such restrictions} are necessary to distinguish the baseline signal from local deviations, hence unavoidable in our context.

The AMS method is now carried out as depicted in Algorithm \ref{algo:AMS}, which by construction ensures asymptotic control of the FWER, i.e. \eqref{eq:FWER} holds true.
\begin{algorithm}[H]
	\centering
	\caption{Adjusted multiscale scanning - AMS}
	\label{algo:AMS}
	\begin{algorithmic}[1]
		\Require Sample size $n \in \N$, dimension $d \in \N$, scale system $\mathcal R_n$, penalization $\widetilde \omega_n, \omega_n$, scale bounds $r_n < m_n$, data $Y$, FWER level $\alpha \in (0,1)$
		\State Simulate the distribution of $M_n \left(\mathcal R_n\left(r_n,m_n\right)\right)$
		\State Compute the (empirical) $(1-\alpha)$-quantile $q_{1-\alpha}$ of $M_n$
		\Statex\Comment{Steps 1 and 2 can be precomputed and the results can be stored independent of $Y$}
		\State Set $\mathcal S \leftarrow \emptyset$
		\State Compute the estimators $\hat \theta_0$, $\hat \xi$ based on $Y$ for the baseline intensity $\theta_0$ and the nuisance parameters $\xi$
		\For{$R \in \mathcal R_n \left(r_n, m_n\right)$}
		\State Compute $T \leftarrow T_R\left(Y, \hat \theta_0, \hat \xi\right)$
		\State Compute $c\leftarrow c_{|R|}(q_{1-\alpha})$ as in \eqref{eq:reject}
		\If{$T \geq c$}
		\State $\mathcal S \leftarrow \mathcal S\cup \{R\}$
		\EndIf
		\EndFor
	\end{algorithmic}
\end{algorithm}

{Hence, given the data, AMS provides a System of sets $\mathcal S \subset \mathcal R_n$ such that with (asymptotic) probability $\geq 1-\alpha$ all $R \in \mathcal S$ contain at least one anomaly. For an illustration and details for Poisson regression, we refer to Section \ref{sec:sim} and for a data application to Section \ref{sec:data_analysis}.}

{In {the supplement (see Section \ref{sec:implementation})} we will provide details on a computationally efficient implementation of this procedure, which can be carried out by means of fast Fourier transforms as soon as $T_R$ in \eqref{eq:TR} is given by a function of the local mean $\bar Y_R := \sum_{i \in R} Y_i$. This leads to an algorithm for the evaluation of $M_n$ in \eqref{eq:Mn} and $\left(T_R\right)_{R \in \mathcal R_n}$ in $\mathcal O \left(d N n^d \log \left(n\right) \right)$ operations, where $N$ is the number of scales contained in $\mathcal R_n$. The corresponding MATLAB$^\copyright$-code is available under \LINK.}

\section{Assumptions and detailed theory}\label{sec:theory}

In this section we will now discuss our assumptions and {the uniform invariance principle} in detail. Furthermore, we will examine several examples to demonstrate the validity of our assumptions in realistic scenarios. 

\subsection{Notation}

{In the following, we will describe an invariance principle which is uniform over the nuisance parameters $\xi \in V \subseteq \Xi$ and holds true for all $\theta_0 \in U \subseteq \Theta$. The necessity to restrict to subsets $U$ and $V$ (see Assumption \ref{ass:model} below for details) is obvious even in the Gaussian case: if the variance $\sigma^2 > 0$ is a nuisance parameter, degenerative cases occurring when $\sigma^2 \searrow 0$ or $\sigma^2 \nearrow \infty$ have to be excluded. The same is true when the variance $\sigma^2 > 0$ is the parameter of interest.}

{To allow for a consistent notation expressing uniformity, we will denote the law of the field $Y$ in \eqref{eq:model} under $H_R$ by $\mathbb P_{\theta_0,\xi,R}$ and let $\mathbb P^*_{\theta_0,n} = \sup\limits_{R \in \mathcal R_n}\sup_{\xi \in V}\mathbb P_{\theta_0,\xi,R}$. This describes exactly the probability which needs to be controlled to obtain a FWER. Furthermore, let $\mathbb P^*_{0,n} = \sup_{\theta_0 \in U} \mathbb P^*_{\theta_0,n}$, which will be needed both in our assumptions and in the formulation of the invariance principle. }

To formulate our invariance principle, we will make use of stochastic $\mathcal O$-notation, see Sect. 2.2 in \citet{vdV98}.

%

\subsection{Assumptions and examples}

We start with our assumptions on the distributions $F_{\theta, \xi}$ in model \eqref{eq:model}. {Let}
\begin{equation}\label{eq:mean_variance}
	m(\theta, \xi) :=  \E{\theta,\xi}{Y}\qquad\text{and}\qquad  v(\theta, \xi):= \V{\theta,\xi}{Y}.
\end{equation}
Here and in what follows, $Y$ denotes a generic random variable with distribution $Y \sim F_{\theta,\xi}$.
\begin{assumption}[Model assumptions]\label{ass:model}
	Let $Y \sim F_{\theta, \xi}$, $\left(\theta,\xi\right) \in \Theta \times \Xi \subseteq \R^{d_1}\times \R^{d_2}$ with $d_1, d_2  \in \mathbb N$. For {the true parameters} $\theta_0 \in \Theta, \xi_0 \in \Xi$ assume {that there exists a non-empty open neighbourhood} $U \times V \subset \Theta \times \Xi$ of $\left(\theta_0, \xi_0\right)$ {with the following properties:}
	\begin{enumerate}
		\item[(a)] \textbf{Uniformly sub-exponential tails:} There exist constants $c_1 > 1, c_2 > 0$ such that
		\begin{align*}
			\sup_{\left(\theta, \xi\right)\in U \times V}\P{\theta,\xi}{|Y|>t} \leq c_1 \exp\left(- c_2 t\right) \qquad\text{for all}\qquad t > 0.
		\end{align*}
		\item[(b)] \textbf{{Positive variance}:} There exists a constant $c_v > 0$ such that {the variance function $v$ from \eqref{eq:mean_variance}} satisfies $v > c_v$ on $U \times V$.
		\item[(c)] \textbf{Bounded derivatives of {the expectation} $m$:} There exists a constant $C_m < \infty$ such that $\|\nabla_\theta m \|_2$, $\| \nabla_\xi m \|_2\leq C_m$ on $U \times V$.
		\item[(d)] \textbf{Bounded and non-vanishing derivatives of $v$:} There exist constants $0 < c_v < C_v < \infty$ such that $c_v \leq \|\nabla_\theta v\|_2, \|\nabla_\xi v\|_2 \leq C_v$ on $U \times V$.
		\item[(e)] \textbf{Taylor approximation:} There exists a constant $C_T < \infty$ such that the likelihood ratio statistic $T_R$ in \eqref{eq:TR} can be approximated by local means $\bar{Y}_R = |R|^{-1}\sum\limits_{i \in R}Y_i$ in the sense that
		\begin{equation}\label{eq:lrt_approx}
			\left| T_R^2 \left(Y,\theta, \xi\right) - |R|\left(\frac{\bar{Y}_R - m\left(\theta, \xi\right)}{\sqrt{v\left(\theta, \xi\right)}}\right)^2 \right|  \leq C_T |R|\left( \frac{\left|\bar{Y}_R - m\left(\theta, \xi\right)\right|}{\sqrt{v\left(\theta, \xi\right)}}\right)^3 \left(1+ o_{\mathbb P^*_{0,n}} \left(1\right)\right)
		\end{equation}
		holds true{, i.e. uniformly} for all $R \in \mathcal R$ and $\left(\theta, \xi\right) \in U \times V$.
	\end{enumerate}
\end{assumption}
We briefly comment on these assumptions.
\begin{remark}
	\begin{enumerate}
		\item[a)] Assumption \ref{ass:model}(a) {implies that the functions $m$ and $v$ in \eqref{eq:mean_variance} exist and are finite on $U \times V$. Thus, it} allows us to control moments of $Y_i$ and hence of local means $\bar Y_R = \sum_{i \in R} Y_i$ simultaneously over regions $R \in \mathcal R_n$. 
		\item[b)] Assumption \ref{ass:model}(b) seems natural, as otherwise no CLT on the small scales can hold true and hence no distributional limit is to be expected.
		\item[c)] Assumptions \ref{ass:model}(c) and (d) will be used to control the impact caused by replacing $\theta_0$ and $\xi$ in \eqref{eq:Tn} by the estimators $\hat \theta_n$ and $\hat \xi_n$, respectively.
		\item[d)] Assumption \ref{ass:model}(e) is required to link $T_R$ to the local means $\bar Y_R$ sufficiently well.
	\end{enumerate}
\end{remark}

For a better understanding of Assumption \ref{ass:model} let us discuss some examples:
\begin{example}[Complete exponential family model]
	Suppose that  $\left\{F_{\theta, \xi}\right\}_{\left(\theta,\xi\right) \in \Theta \times \Xi}$ is a natural exponential family with natural parameter space $\Theta \times \Xi$. In this case, each $Y_i$ has sub-exponential tails (which implies Assumption \ref{ass:model}(a)) and the differentiability properties from Assumption \ref{ass:model}(c)--(e) hold true \citep[see e.g.][]{b86}. 
	
	This includes both the homogeneous and the heterogeneous Gaussian model as well as the Poisson model, which we employ for our data example in Section \ref{sec:data_analysis} on fluorescence microscopy.
\end{example}

\begin{example}[Partial exponential family model]\label{ex:distributions}
	Suppose that for each fixed value of $\xi \in \Xi$, the collection $\left\{F_{\theta, \xi}\right\}_{\theta \in \Theta}$ is a natural exponential family. In this case, Assumption \ref{ass:model}(a) is again satisfied due to standard theory on exponential families, but the differentiability properties in Assumption \ref{ass:model}(c)--(e) have to be verified case by case. As one particular example we consider the 
	Gamma distribution $\Gamma(\alpha, \beta)$ with shape parameter $\alpha>0$, rate $\beta>0${, and density} $f(x)= 1/\Gamma(\alpha)  \beta^\alpha x^{\alpha-1} \exp \left( -\beta x \right)$. If the shape parameter $\alpha$ is a nuisance parameter, we obtain $m(\beta, \alpha)= \alpha/\beta$, $v(\beta, \alpha)= \alpha/\beta^2$ and
	\begin{align*}
		\frac{\partial}{\partial \beta} m(\beta, \alpha) = - \frac{\alpha}{\beta^2},  \qquad \frac{\partial}{\partial \beta} v(\beta, \alpha) = - \frac{2\alpha}{\beta^3}, \qquad \frac{\partial}{\partial \alpha} m(\beta, \alpha) = \frac{1}{\beta}, \qquad \frac{\partial}{\partial \alpha} v(\beta, \alpha) = \frac{1}{\beta^2},
	\end{align*}
	which proves the boundedness conditions in Assumption \ref{ass:model}(c) and (d). Furthermore, \eqref{eq:lrt_approx} follows directly from the exponential family structure \citep[see e.g.][]{kmw20}. 
\end{example}

We also give a counterexample to show the limitations of our Assumptions:
\begin{example}[Weibull distribution]
	Consider the Weibull distribution with shape parameter $k>0$ and rate parameter $\lambda>0$. If $k\geq 1$, then all random variables $Y_i$ will have sub-exponential tails \citep[see e.g.][]{s85,fb97}, but 
	\begin{align*}
		\log \left(\frac{\sup\limits_{\theta \in \Theta} \prod_{i \in R} f_{(\theta, \xi)}(Y_i)}{\prod_{i \in R} f_{(\theta_0, \xi)} (Y_i)} \right)&= k \log(\lambda_0) - \log \left( \sum\limits_{i \in R} Y_i^k\right) + \frac{\sum\limits_{i \in R} Y_i^k}{\lambda_0^k} -1\\
		|R|\left(\frac{\bar{Y}_R - m(\theta_0, \xi)}{\sqrt{v(\theta_0, \xi)}}\right)^2 &= |R| \left( \frac{\bar{Y}_R^2 - 2 \lambda \Gamma\left(1+ \frac{1}{k} \right)\bar{Y}_R +\lambda^2 \left(\Gamma\left(1+ \frac{1}{k} \right)\right)^2}{\lambda^2 \left[\Gamma\left(1+ \frac{2}{k} \right) - \left(\Gamma\left(1+ \frac{1}{k} \right)\right)^2 \right]}\right),
	\end{align*}
	which shows that \eqref{eq:lrt_approx} cannot hold true.
\end{example}

Next we continue with our assumptions on the multiscale scanning procedure described in terms of $T_n$ in \eqref{eq:Tn}. To approximate $T_n$ by $M_n$ in \eqref{eq:Mn}, we first have to restrict {$\# \mathcal R_n$,} the cardinality of $\mathcal R_n$.
\begin{assumption}[Polynomial growth of $\mathcal R_n$]\label{ass:R}~
	There exists constants $c_3, c_4>0$ such that
	\begin{equation}\label{eq:finite_cardinality}
		\#\mathcal R_n\leq c_3 n^{c_4}.
	\end{equation}
\end{assumption}
As discussed in \citet{kmw20}, Assumption \ref{ass:R} is rather mild and allows e.g. for $\mathcal R_n$ being the set of all hyper-rectangles, hyper-cubes or half-spaces in $\left[0,1\right]^d$.

Our next assumption is on the scale calibration $\widetilde{\omega}{_n}, \omega{_n}$ in \eqref{eq:Tn}:
\begin{assumption}\label{ass:Omega}
	Suppose $\omega{_n},\widetilde{\omega}{_n}: \left(0,\infty\right) \to \left(0,\infty\right)$ have the following properties:
	\begin{enumerate}
		\item For each $n \in \mathbb N$, $\omega{_n}$ and $\widetilde \omega{_n}$ are decreasing {and differentiable}.
		\item There exist $C_{\omega} > 0$, $\alpha, \widetilde{\alpha}>0$ and $\beta, \widetilde{\beta} \in \R$ such that
		\begin{align*}
			\omega{_n}(r) & \leq C_\omega\left(\log \frac{n^d}{r} \right)^\alpha, \qquad \left|\omega{_n}^\prime(r) \right|  \leq C_\omega\left(\log \frac{n^d}{r} \right)^\beta \frac{1}{r},
		\end{align*}
		for all $r > 0$ and $n \in \N${, and the same holds for $\widetilde{\omega}{_n}$ and $\widetilde{\omega}{_n}^\prime$ with $\widetilde{\alpha}$ and $\widetilde{\beta}$ instead}.
	\end{enumerate}
\end{assumption}
Let us briefly discuss this assumption for common choices considered in the literature:
\begin{example}\label{ex:omega}
	\begin{enumerate}
		\item[a)] \citet{dw08,cw13,rw13} consider
		\begin{align}
			\widetilde{\omega}{_n}\equiv 1, \qquad 
			\omega{_n}(|R|)= \sqrt{2 \nu (\log(n^d/|R|))+1},\label{pen_ds01}
		\end{align}
		where $\nu  \geq 1$ depends on the complexity of the candidate region. These terms fulfil Assumption \ref{ass:Omega} with $\alpha=\frac{1}{2}, \beta=-\frac{1}{2}$ and $\widetilde{\alpha}= \widetilde{\beta}=0.$
		\item[b)] \citet{sac16} consider
		\begin{align}
			\widetilde{\omega}{_n}(|R|)& = \sqrt{2(\log(n^d/|R|))},\nonumber\\
			\omega{_n}(|R|)&= \sqrt{2(\log(n^d/|R|))} + \frac{\left(4d-1\right) \log \left(\sqrt{2(\log(n^d/|R|))}\right) - \log\left(\sqrt{2\pi}\right)}{\sqrt{2(\log(n^d/|R|))}} ,\label{pen_sac}
		\end{align}
		and therefore Assumption \ref{ass:Omega} is fulfilled with $\alpha = \widetilde{\alpha} = \frac12$ and $\beta = \widetilde{\beta} = - \frac12$.
		\item[c)] \citet{pwm18} consider 
		\begin{equation}\label{pen_pwm}
			\omega{_n}(|R|) := \widetilde{\omega}{_n}(|R|):= \sqrt{2 \log \left( \frac{Cn^d}{|R|}\right)} + C_d \frac{\log \left(\sqrt{2 \log \left( \frac{Cn^d}{|R|}\right)} \right)}{\sqrt{2 \log \left( \frac{Cn^d}{|R|}\right)}},
		\end{equation}
		with $C>1$ and $C_d$ depending on the dimension and the system of considered scales.
		These fulfil Assumption \ref{ass:Omega} with $\alpha= \widetilde{\alpha}= \frac{1}{2}$ and $\beta= \widetilde{\beta}= -\frac{1}{2}.$
	\end{enumerate}
\end{example}

\subsection{Main result and implications}

Now we are in position to state our uniform invariance principle in detail. Therefore let $(\hat{\theta}_n)_{n \in \N}$ and $(\hat{\xi})_{n \in\N}$ be sequences of estimators for $\theta_0$ and $\xi$ based on $Y$ such that they are $s_n$ (e.g. $s_n = n^{{d/2}}$) consistent under the null hypothesis, i.e. 
\begin{equation}\label{eq:estimators}
	\|\hat{\theta}_n - \theta_0\|_2 = \mathcal O_{\mathbb P^*_{0,n}} \left(s_n^{-1}\right)\qquad\text{and}\qquad \|\hat{\xi}_n - \xi\|_2 = \mathcal O_{\mathbb P^*_{0,n}} \left(s_n^{-1}\right),
\end{equation}
where {$s_n \nearrow \infty$, $n \to \infty$}. Note, that the MLEs for $\theta_0$ and $\xi$ satisfy \eqref{eq:estimators} under Assumption \ref{ass:model} with $s_n = n^{d/2}$. Furthermore let 
\[
\gamma= 12+ 6 \widetilde{\alpha}+ 2 \max\left\{\frac12, \alpha, \widetilde{\alpha}\right\} + 2 \max\left\{\beta, \widetilde{\beta},0\right\}
\]
with the parameters $\alpha, \widetilde{\alpha}, \beta, \widetilde{\beta}$ from Assumption \ref{ass:Omega} and suppose that
\begin{align}
	\log^{\gamma}(n) = o\left(r_n\right) \text{ as }n \to \infty\label{eq:r_n},\qquad m_n =  o \left(s_n\right)  \text{ as }n \to \infty.
\end{align}

\begin{theorem}[Gaussian approximation]\label{thm:approximation_estimated_intensity}
	Let $\mathcal R_n$ be a set of candidate regions and let $\omega{_n}, \widetilde \omega{_n}$ be scale calibrations. Grant Assumptions \ref{ass:model}, \ref{ass:R} and \ref{ass:Omega}. Furthermore let $(r_n)_n\subset(0,\infty)$ and $\left(m_n\right)_n \subset \left(0,\infty\right)$ be sequences such that \eqref{eq:r_n} holds true and let $(\hat{\xi}_n)_{n\in\N}$ and $(\hat{\theta}_n)_{n \in\N}$ are sequences of estimators such that \eqref{eq:estimators} is valid. 
	
	Then, for a {sequence of fields} $Y = \left(Y_i\right)_{i \in I_n^d}$ with i.i.d. random variables $Y_i \sim F_{\theta_0, \xi}$, it holds
	\begin{align}\label{eq:approx_in_prob}
		\left|T_n \left(Y,  \mathcal R_n\left(r_n,m_n\right), \hat{\theta}, \hat{\xi} \right) - M_n\left(\mathcal R_n\left(r_n,m_n\right)\right) \right| = \mathcal O_{\mathbb P^*_{0,n}}\left( \frac{\log^{\widetilde{\alpha}} \left(n\right)}{r_n} \sqrt{\frac{m_n}{s_n}}+\left(\frac{\log^\gamma\left(n\right)}{r_n}\right)^{\frac18}\right),
	\end{align}
	as $n \to \infty$.
\end{theorem}

{The proof of Theorem \ref{thm:approximation_estimated_intensity} will be given in the supplement. Note, that \eqref{eq:approx_in_prob} is in fact uniform over $U \times V$, and hence properly extends the classical scanning approximation of $T_n$ by $M_n$ to the case of nuisance parameters.}

\begin{cor}
	Let the assumptions of Theorem \ref{thm:approximation_estimated_intensity} hold true, $Y = \left(Y_i\right)_{i \in I_n^d}$ as in \eqref{eq:model} (independent, but not necessarily identically distributed), and $\alpha \in \left[0,1\right]$. If 
	\[
	\overline{\mathcal R}_n := \left\{R \in \mathcal R_n ~\big|~ r_m \leq |R| \leq m_n, T_R \left(Y, \hat \theta_n, \hat \xi_n\right) \geq \frac{q_{1-\alpha}}{\widetilde{\omega}{_n} \left(\left|R\right|\right)} + \omega{_n} \left(\left|R\right|\right)\right\}
	\]
	with the $\left(1-\alpha\right)$-quantile $q_{1-\alpha}$ of $M_n$ as in \eqref{eq:Mn} 	{denotes all rejected regions given the data}, then \eqref{eq:FWER} holds true {uniformly over $\xi \in V$ for any $\theta_0 \in U$}, i.e.
	\[
	\mathbb P_{\theta_0,n}^* \left[\forall~R \in \overline{\mathcal R}_n ~\exists~i \in R \text{ s.t. } \theta_i \neq \theta_0\right] \geq 1-\alpha + o(1) \qquad\text{as}\qquad n \to \infty.
	\]
\end{cor}

\begin{remark}\label{rem:one_sided}
	In the situation that all local alternatives {(and tests)} are one-sided, i.e. $\theta_i > \theta_0$, {Theorem \ref{thm:approximation_estimated_intensity}} can readily be generalized by setting
	\[
	\tilde T_R \left(Y,\hat \theta_n, \hat \xi_n\right) = \begin{cases} T_R \left(Y, \hat \theta_n,\hat\xi_n\right) & \text{if }\bar Y_R > \hat \theta_n, \\ 0 & \text{else},\end{cases}
	\]
	as well as
	\[
	\tilde T_n \left(Y,\mathcal R_n,\hat \theta_n, \hat \xi_n\right)  := \max_{R\in\mathcal R_n} \widetilde{\omega}{_n}\left(\left|R\right|\right)\left[ \tilde T_R\left(Y, \theta_0, \xi\right)-\omega{_n}\left(\left|R\right|\right)\right]
	\]
	and
	\[
	\tilde M_n\left(\mathcal R_n\right) := \max_{R\in\mathcal R_n, \bar X_R > 0} \widetilde{\omega}{_n}(|R|)\left[ \left|R\right|^{-1/2} \left| \sum\limits_{i \in R} X_i \right| - \omega{_n}(|R|) \right].
	\]
\end{remark}

\begin{remark}
	Let us briefly compare the invariance principle in the paper at hand with previous results, especially those from \citet{kmw20}. There, an approximation of $T_n$ in \eqref{eq:Tn} by $M_n$ in \eqref{eq:Mn} has been proven in the case that $F_{\theta,\xi}$ is a natural exponential family, both $\theta_0$ and $\xi$ are known, and the calibration functions $\omega{_n}, \widetilde{\omega}{_n}$ are explicitly chosen as in \cite{dw08}, see also Example \ref{ex:omega} below. {For the situation in this paper, we have extended} this result by dropping all three restrictions: First of all, we allow for distributions $F_{\theta,\xi}$ beyond the exponential family setting. 
	{The major theoretical contribution of the paper at hand is the uniformity of the invariance principle over a set of baseline intensities $\theta_0 \in U$ and nuisance parameters $\xi \in V$. Finally, the invariance principle offers a greater flexibility as we allow for a wide choice of scale calibrations.}
\end{remark}


\section{Simulations}\label{sec:sim}

In this section, we will investigate the {behaviour} of {AMS} in a simulation study. {Details on {the efficient} implementation of {AMS} are given in the supplement (see Section \ref{sec:implementation}).}

\subsection{Validity of the Gaussian approximation of AMS}

\begin{figure}[!htb]
	\scriptsize
	\centering
	\setlength{\fwidth}{3cm}
	\setlength{\fheight}{3cm}
	\begin{tabular}{cc}
		\input{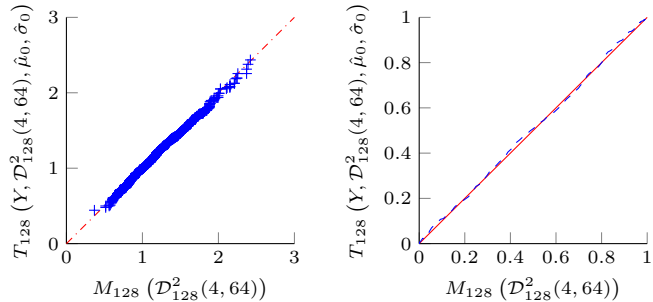} & 
		\input{restricted_PP.tikz}
	\end{tabular}
	\caption{{QQ (left) and PP-Plots (right) of the empirical distributions of $M_n\left(\mathcal R\right) = T_n \left(Y,\mathcal R,0,1\right)$ and the naive approximation $\hat T_n$ in the Gaussian case under scale restrictions with $r_{128} = 4$ and $m_{128} = 64$}, drawn from 2000 Monte Carlo runs. We emphasize that the only difference to Figure \ref{fig:gauss_dist} is the restriction {to} scales {of} sizes $4-64$.}
	\label{fig:gauss_dist_limited}
\end{figure}

We illustrate Theorem \ref{thm:approximation_estimated_intensity} in {the} homogeneous Gaussian setting. Figure \ref{fig:gauss_dist_limited} shows the empirical distribution of $T_n \left(Y, \mathcal R_n\left(r_n,m_n\right), \hat \theta_n, \hat \xi_n\right)$ and $M_n \left(\mathcal R_n\left(r_n,m_n\right)\right)$ with the same $n$ and $\widetilde{\omega}{_n}, \omega{_n}$ as there for  $r_{128} = 4$ and $m_{128} = 64$.

We find the distributional fit to be remarkably accurate in all situations, which supports the applicability of the asymptotic result from Theorem \ref{thm:approximation_estimated_intensity}. The distribution of $M_n$ itself will be shown in Figure \ref{fig:invariance_dist} for the situation considered in our simulations.

\subsection{Simulations of level and power in a Poisson model}\label{sec:poisson_sim}

In this section, we will study the finite sample properties of the adjusted multiscale test. {We consider a Poisson field, where $F_{\theta, \xi}$ is a Poisson distribution $\text{Poi} \left(\lambda\right)$ with unknown background intensity $\lambda_0$, which will be estimated by the sample mean, i.e $\hat\lambda_n := \bar Y_{I_n^d}$. Then \eqref{eq:estimators} is clearly satisfied with $s_n = n^{d/2}$.} The empirical level and power of the AMS procedure {will be simulated} by means of Monte Carlo simulations with $M = 1000$ runs. As a default setup, we use $d = 2$, $n = 128$ and $\mathcal R_n$ to be the set of all rectangles in $I_n^d$ with even side-lengths between $4$ and $14$ pixels. As penalization functions $\widetilde \omega{_n}$ and $\omega{_n}$, we choose the previously mentioned penalization \eqref{pen_ds01} of \cite{dw08}. {For the sake of completeness, } empirical quantiles of $M_n$ in \eqref{eq:Mn} for this situation are shown in Table \ref{tab:quantiles_128} {in the supplement}, {and} the empirical density and CDF {both for the calibrated and the uncalibrated statistic} are displayed in Figure \ref{fig:invariance_dist} {also in the supplement}. {Additional simulations of level and power in a (simpler) homogeneous Gaussian model are also presented there.}

{In the simulations of level and power below, we will compare three possible methods: the AMS method, an oracle (benchmark) procedure, where the local LRT statistics $T_R$ from \eqref{eq:TR} are evaluated using the true parameter $\lambda_0$, and a bootstrap method which does not rely on the Gaussian approximation from Theorem \ref{thm:approximation_estimated_intensity} to obtain quantiles, but rather estimates all parameters from the available data and then simulates the empirical distribution of $T_n\left(Y, \mathcal R_n, \hat \lambda_0\right)$ directly. In these simulations we use only $\tilde M= 100$ Monte Carlo runs to simulate these quantiles as the bootstrap method is considerably slower (roughly $\tilde M$-times), since the simulation of quantiles have to be carried out online (i.e. depending on the data $Y$), and not offline (i.e. beforehand as for the AMS and the oracle procedure).}

For power simulations, we consider a $a^2 = 8\times8$ sized anomaly positioned in the {centre} of the $128\times 128$ image with amplitude $\lambda = 1.6$ in the centre, where the background intensity $\lambda_0 = 1$.

{To investigate the influence of $d$ and $n$, we additionally perform simulations with $d= 1,2$ and $n = 32,64,96,128$ {for the AMS procedure only}. Therein we determine the necessary signal strength (mean) of an anomaly of size $a^d$ ($a \in \{4,5,...,13,14\}$ a fixed number), again positioned in the centre of the domain, such that the empirical power is $\approx 90\%$. The corresponding functions $\lambda = \lambda(a)$, are in fact an upper bound for the (finite sample) separation rate between $H_0$ and the prescribed alternative. The (asymptotic) separation rate, i.e. the asymptotic detection boundary, provides a sharp notion of the minimal signal strength necessary for detection, which cannot be overcome by any test. For the identification of cuboid signals with known size $a$ and mean $\mu$ in Gaussian white noise as considered here, this has been derived in the literature, see e.g. \cite{cw13,emw18,k21}, as, translated to our notation,
	\begin{equation}\label{eq:separation_rate}
		\mu \sim \sqrt{\frac{2}{a^d} \log \left(\frac{n^d}{a^d}\right)}.
	\end{equation}
	We emphasize that AMS treats a more difficult situation (i.e. $\theta_0, \xi$ and $a$ unknown), and will hence not be able to reach this boundary, but as we will show below {and in the supplement}, a certain comparison is still possible.}

{The results of the simulations as described before are depicted in Figure \ref{fig:poisson}. Sub-panels (a) and (b) show the signal and the corresponding data for the default setup. Sub-panel (c) shows the empirical level for different background intensities $\lambda_0$. We again find that the level is kept quite stable, even for considerably small values of $\lambda_0$. The little bump around $\lambda_0 = 0.3$ emphasizes that the quantiles are chosen according to the Gaussian approximation in Theorem \ref{thm:approximation_estimated_intensity}. {Furhermore, also the bootstrap-based procedure consistently keeps its nominal level.} Sub-panel (d) shows the empirical power for $\lambda_0 = 1$ against different amplitudes $\lambda\geq \lambda_0$. Similarly to the Gaussian case, there is nearly no power loss caused by adaptation for an unknown $\lambda_0$. {Similarly, the power of the bootstrap-based procedure is not lower that those of the other two.} Sub-panel (e) depicts that value of $\lambda$, such that the AMS procedure in $d =1$ is able to detect a present anomaly of size $a$ marked on the $x$-axis with amplitude $\lambda$ with a power of $\approx 90\%$. We find that a larger anomaly clearly corresponds to a smaller amplitude required for detection, in agreement with the results on separation rates in the Gaussian case. Furthermore, we find that the influence of the sample size $n$ on the separation lines is small, despite for the smallest sample size $n = 32$ where the required value of $\lambda$ is moderately larger. This might be caused by the worse approximation of $T_n$ by $M_n$ in this case. Sub-panel (f) shows the same situation in dimension $d = 2$. Due to the larger anomaly, which now contains $a^2$ pixels, the corresponding values for $\mu$ are smaller. The conclusions drawn for the one-dimensional case apply here for all considered values of $n$. Compared with sub-panel (e) we furthermore conclude that the dimension $d$ does not have a big influence.}

\begin{figure}[!htb]
	\scriptsize
	\centering
	\setlength{\fwidth}{2.8cm}
	\setlength{\fheight}{2.8cm}
	\begin{tabular}{lll}
		\hspace*{1cm}\input{Poisson_means_sim.tikz}
		&
		\hspace*{.55cm}\input{Poisson_data_sim.tikz}
		&
		\setlength{\fwidth}{3.72cm}
		\hspace*{-.5cm}\input{poisson_level.tikz}	
		\\ 
		\hspace*{1cm}(a) Signal with $\lambda = 1.6$
		& 
		\hspace*{.55cm}(b) Corresponding data for (a)
		&
		\hspace*{.5cm}(c) Level vs. $\lambda_0$ 
		\\[0.2cm]
		\setlength{\fwidth}{3.72cm}
		\hspace*{-.1cm}\input{poisson_power.tikz}
		&
		\setlength{\fwidth}{3.72cm}
		\hspace*{-.3cm}\input{poisson_mean_for_power_d=1.tikz}
		&
		\setlength{\fwidth}{3.72cm}
		\hspace*{-.28cm}\input{poisson_mean_for_power_d=2.tikz}		
		\\
		\hspace*{1cm}(d) Power vs. $\lambda$ for $\lambda_0 = 1$
		&
		\hspace*{.55cm}(e) separation rate, $d = 1$, $\lambda_0 = 1$
		&
		\hspace*{.5cm}(f) separation rate, $d = 2$, $\lambda_0 = 1$		
	\end{tabular}
	\caption{{Simulation results in a Poisson model with $\alpha = 0.1$. Panels (a) and (b): Signal under the alternative ($\lambda_0 = 1$, $\lambda = 1.6$) used for this simulation and corresponding data. Panel (c): Empirical level for the oracle multiscale scan ($\lambda_0$ known), AMS ($\lambda_0$ unknown, i.e. considered as a nuisance parameter){, and the bootstrap-based procedure} for different values of $\lambda_0$, smoothed with a moving average filter of width $10$. Panel (d): Empirical power versus $\lambda$ for a $8 \times 8$ anomaly as in (a), smoothed with a moving average filter of width $10$ {for visualization purposes}. Panels (e) and (f): Value of $\lambda$ required to obtain an empirical power of $\beta = 0.9$ versus size of a (squared) anomaly with amplitude $\lambda$ {for AMS}.}}
	\label{fig:poisson}
\end{figure}

\section{Data Analysis}\label{sec:data_analysis}

{A major motivation for our work comes from super-resolution fluorescence microscopy imaging. Fluorescence microscopy is employed to visualize structures of interest (e.g. proteins or protein complexes) in a sample. These are {labelled} biochemically by fluorescent markers, and then scanned spatially along a grid with a diffraction-limited laser spot {centred} at the current grid point. Whenever a marker is hit by the incoming light it excites with a certain probability, and if so, afterwards light of a different wavelength is emitted and recorded by photon-detectors \citep[cf.][for details]{aem15}. This excitation-emission-recording procedure is repeated several (say $t \in\N$) times, called the measurement time, before the focus is moved to the next grid point.}

{To obtain a resolution below the Abbe diffraction limit (see \citet{kmw21} for an explanation in a statistical context), nowadays different techniques are known. Here we will focus on stimulated emission depletion (STED) microscopy, developed by Stefan Hell (see e.g. \citet{hw94,h07} for details), which has been awarded the Nobel {prize} for Chemistry in 2014. It is based on an additional donut-shaped depletion beam, which suppresses fluorescence in the corresponding region and hence reduces effectively the area in which light is emitted. This way, the effective resolution can be decreased dramatically from roughly $200-300\nm$ for a standard (confocal) laser microscope to $\leq 50 \nm$, say.}

\subsection{Mathematical setup}

Mathematically, the nature of photon counting in super-resolution microscopy leads to a Poisson field of independent observations \citep[see e.g.][]{msw20}, 
\begin{align}\label{eq:model_pois}
	Y_i \sim \text{Poi}(t\lambda_i), \quad i \in I_n^d:= \{1, \ldots,n\}^d.
\end{align}

{However, $t$ cannot be chosen arbitrary, as each marker is only able to pass through the cycle of excitation and emission a limited number of times before it bleaches. Hence, to avoid masking effects by bleaching, the total illumination time per pixel (called pixel dwell time) has to be set to a (not too large) number $T > 0$. However, even if large areas in the image are mostly empty, these regions have to be scanned and by doing so all markers in the image are hit by light and hence either excited or at least stressed. Therefore, methods which are able to localize interesting structures within a short pre-scan (i.e. with short measurement time $t \ll T$) have achieved great interest recently in many applications and can -- in the present context -- be seen as an instance of \emph{smart} microscopy, see \citet{s20}. Some heuristic methods have been suggested \citep[such as RESCue and others, see][]{vge20}, which stop measuring at a pixel as soon as a predefined number of photons has been detected (known as spatially-controlled illumination microscopy (SCIM), see e.g. the review by \citet{kvnmh16}). In principle this can be interpreted as scanning each grid point with a substantially smaller pixel dwell time $t\ll T$ and deciding upon the corresponding measurement $Y_i$ if the full pixel dwell time $T$ is applied or if the measurement at the current grid point is stopped. Another approach exploits machine learning to pre-select regions of interest based on shortly measured (i.e. very noisy) data, see e.g. \citet{pwsg19,mszgwm22}. All the previously described methods suffer from the fact that they are not able to provide statistical guarantees for correctness of the selected (or more intensively scanned) regions. As we will demonstrate now, AMS does provide a methodology that allows to identify interesting regions with prescribed statistical error.}

\subsection{Used data sets}

{The data set to be shown below {and those in the supplement} have been measured in the Institut für Nanophotonik (IFNANO) by Ren\'e Siegmund especially to apply AMS with different initial scan times $t$. {It is available together with the codes under \LINK.} The measurement device is a STED super-resolution microscope with an effective resolution of $\sim 60 \nm$. For each $t \in \{1 \ms,2 \ms,...,100\ms \}$ a data set $Y_t$ with corresponding dwell time $t$ (in microseconds $\ms$) is recorded. The corresponding value $t = 100 \ms$ can be seen as the maximal value $T$, which was chosen to avoid bleaching of the fluorescent dyes. In that sense, we will consider $t = T$ as \textit{ground truth}, since no better measurement is available or can realistically be measured. We aim to achieve a localization of all or most regions of interest by means of AMS based on data with $t \ll T$, since this then yields a statistically rigorous method to obtain the spatially-controlled illumination described above with (asymptotically) controlled FWER.}

{The raw data image is corrupted by background noise resulting from other contributions such as out-of-focus markers or external photon sources, which also can be modelled as {Poisson} random variables, see e.g. \citet{aem15,msw20}. Since the background intensity $\lambda_0 > 0$ varies between experiments, it has to be estimated from the data by the global mean $\hat \lambda_n = \bar Y_{I_n^d}$.}

\subsection{Applying AMS}

{In the data sets described above, available also under \LINK, we have} $n = 400$, $d = 2$, and we choose $\mathcal R_n$ to be the subset of all rectangles in $I_n^d$ with scale between $r_n = 4$ and $m_n \in \{10,20\}$, depending on the investigated structure of interest. The details will be explained below. As we are only interested in active regions, we consider the one-sided testing problem to find regions $R$ where $\lambda_i > \lambda_0$, which can be achieved by restricting the corresponding local rejections $R$ with $\bar Y_R = \frac{1}{|R|} \sum_{i \in R} Y_i > \hat \lambda_n$ (recall Remark \ref{rem:one_sided}). Furthermore we simulate the threshold $\eta$ from an asymptotically distribution-free Gaussian approximation $M_n$ of the test statistic (see \eqref{eq:Mn}, with the corresponding adjustment that $\bar X_R > 0$, cf. also Remark \ref{rem:one_sided}). We then choose $\eta$ as the $0.9$-quantile $q_{0.9}^{M_n}$ of this adjusted $M_n$. Our main Theorem \ref{thm:approximation_estimated_intensity} implies that this asymptotically controls the FWER at level $\alpha$. The corresponding results will be shown as \textit{significance (fluorescence) maps}, depicting all significant rectangles with colour indicating their size in square-nanometers $\nm^2$. If two rectangles overlap, the colour is chosen according to the smaller one, which leads to a map indicating the smallest scale of significance. The major result of this paper implies that the AMS method ensures the FWER control in \eqref{eq:FWER}, and hence each coloured region contains an anomaly with overall probability $\geq 0.9$, i.e. the probability that any of these detections is false is less then $0.1$.

\subsection{Results: tubulin in neonatal fibroblast}\label{sec:data}

{In the supplement, we present the performance of AMS on two data sets consisting of crimson beads of different sizes. As therein the structures of interest are spheres of known diameter, this can be seen as a validation on experimental data which supports the validity of AMS. Here we discuss a more complicated data set, the formation of the cell skeleton network by} tubulin in neonatal fibroblast. The corresponding structures are filaments and hence, compared to the spheres {for crimson beads}, no longer fully contained in single rectangles. To better cover this filamentary structure, which is expected to have width of $60-80\nm$, we choose $m_n = 10$ here, i.e. we restrict to boxes with maximal length $200 \nm$. Again, we aim to detect active regions inside the field of view based on noisy $t = 25 \ms$ data. It can readily be seen from sub-panel (a) of Figure \ref{fig:fibro}, that the corresponding data is quite noisy, but filamentary structures are clearly visible. Sub-panel (d) of the same figure shows the corresponding $T = 100 \ms$ measurement, revealing a folded filamentary structure in the upper right part of the field of view, which was hardly visible by eye in sub-panel (a). With the previously described setup, AMS is able to detect even the folded filamentary structure in the upper left part with a high accuracy, cf. sub-panels (b) and (c). We furthermore find that the method {does} not yield any false positives despite the relatively high noise-level in the $t = 25 \ms$ measurement, as the comparison with the full $T = 100 \ms$ ground truth data proves that all significant regions belong to fibroblasts.

\begin{figure}[!htb]
	\scriptsize
	\centering
	\setlength{\fwidth}{3.1cm}
	\setlength{\fheight}{3.1cm}
	\begin{tabular}{p{4.2cm}p{4.9cm}p{4.5cm}}
		
		\input{fibro_limited_data.tikz}
		&
		\input{fibro_result_limited.tikz}
		&
		\input{fibro_full_data.tikz}
		\\
		(a) $t = 25 \ms$, raw data
		&
		(b) significance map, $t = 25 \ms$
		&
		(c) $T = 100 \ms$, raw data, with detected regions from (b) in red
		\\
		\input{fibro_full_data2.tikz}
		&
		\input{fibro_result_full.tikz}
		&
		\input{fibro_full_segments.tikz}
		\\
		(d) $T = 100 \ms$, raw data
		&
		(e) significance map, $T = 100 \ms$
		&
		(f) segmentation derived from (b) using only the smallest scale
	\end{tabular}
	\caption{Raw tubulin data (photon counts, panels (a) and (d)) and corresponding significance map at level $\alpha = 0.1$ (smallest significance in $\nm^2$, panels (b) and (e)). All significant boxes are plotted with colour indicating their size in $\nm^2$. Due to the FWER control, with probability $\geq 1-\alpha = 0.9$, all boxes marked in (b) contain anomalies. Using only the smallest scale significance, this leads to the segmentation in (f) showing active regions in red.}
	\label{fig:fibro}
\end{figure}

{At this point, let us also compare the AMS methodology with direct pixel-wise thresholding. This is e.g. comparable to the techniques used in the previously mentioned SCIM methods (cf. \cite{vge20,kvnmh16}. Therefore, we consider simply each pixel, for which the corresponding value $Y_i$ in \eqref{eq:model_pois} exceeds the threshold $\eta \in \{2,3\}$, as an anomaly. Note, that these methods do not lead to any sort of statistical error control. In order to do so, we employ the \cite{bh95} procedure to obtain a pixel-wise thresholding with controlled false discovery rate (FDR). The results are shown in Figure \ref{fig:fibro_thresholding}.}

\begin{figure}[!htb]
	\scriptsize
	\centering
	\setlength{\fwidth}{3.1cm}
	\setlength{\fheight}{3.1cm}
	\begin{tabular}{p{4.9cm}p{4.9cm}p{4.9cm}}
		
		\input{fibro_pixelwise_Hard1.tikz}
		&
		\input{fibro_pixelwise_Hard2.tikz}
		&
		\input{fibro_pixelwise_FDR.tikz}
		\\
		(a) Pixels with value $\geq \eta = 2$
		&
		(b) Pixels with value $\geq \eta = 3$
		&
		(c) Detected pixels at FDR $\leq 0.9$
	\end{tabular}
	\caption{Pixel-wise thresholding applied to the raw tubulin data from Figure \ref{fig:fibro}(a). All marked pixels exceeded the corresponding thresholds. Panels (a) and (b) show hard thresholding with no statistical guarantee for $\eta = 2$ and $\eta = 3$ respectively, and panel (c) shows the result of the Benjamin-Hochberg procedure with FDR $\leq 0.9$.}
	\label{fig:fibro_thresholding}
\end{figure}

{Apparently, none of the {thresholding methods} is able to effectively detect the filamentary structures within the specimen {(as they do not borrow signal strength from neighbouring pixels)}. Even though the hard thresholding with $\eta = 3$ seems to work best, it still contains noise pixels in all parts of the image. The FDR-controlled method seems to suffer severely from the inhomogeneity of the image, namely that the right side appears brighter than the left, and hence the corresponding $p$-values of the pixel-wise tests are smaller there. Overall, we can conclude that due to its multiscale structure the AMS procedure clearly outperforms simple pixel-wise thresholding and hence yields a substantial advantage over stat-of-the-art SCIM methods.}

{In the supplement, we also compare the resulting significance maps for different values of $\alpha$. This shows that the AMS methodology is {remarkably} stable w.r.t. the significance level $\alpha$ and its choice in practical application appears not as a critical issue.}

\section{Discussion}\label{sec:discussion}

In this paper, we have developed AMS, a multiscale scanning method allowing for unknown baseline and nuisance parameters. The performance of AMS has been investigated both in simulations and on real data examples from super-resolution microscopy, where it can be seen as an instance of smart microscopy equipped with statistical guarantees. In particular, it outperforms pixel-wise thresholding procedures, including FDR-controlled ones. Under a minimal condition on the scales, AMS selects all regions deviating from the global null at guaranteed FWER. To this end we have proven a uniform invariance principle which allows to approximate the multiscale statistic by a distribution-free Gaussian version. This can be used for the computation of (asymptotically) correct critical values.

{We give an FFT based algorithm to perform AMS which roughly scales as $dNn^d$ where $N$ is the number of scales contained in $\mathcal R_n$. This effectively delimits the practical performance of AMS to small dimensions $d \leq 3$, say. However, a combination with ideas from \citet{w10} or using random regions $R$ might allow to extend AMS to a larger $d$ in almost linear time. {I}t is expected that this comes at the expense of losing detection power compared to the use of the full grid $I_n^d$.}

{The bootstrap version of AMS appeared to keep the nominal level in our simulation study and we speculate that a proof of its formal validity is possible extending our uniform invariance principle to a version conditioned on the data. This might offer an interesting venue for further theoretical research.}

{An alternative option to the AMS approach might be based on sample splitting. Therefore, one could separate the spatial observations $Y_i$, $i \in I := I_n^d$ into two subsets $I= I_1 \cup I_2$, $I_1 \cap I_2 = \emptyset$, say e.g. $I_1 = \{1,3,...,n-1\}^d$ and $I_2 = \{2,4,....n\}^d$ for even $n$, and use $\left(Y_i\right)_{i \in I_1}$ for estimation of $\theta_0$ and $\xi$ only. This eases the asymptotic analysis and the statistic $T_n \left(\left(Y_i\right)_{i \in I_2}, \mathcal R_n, \hat \theta, \hat \xi\right)$ will then asymptotically behave exactly like $T_{n/2} \left(Y, \mathcal R_{n/2}, \theta_0,\xi\right)$, and Theorem \ref{thm:approximation_estimated_intensity} is applicable. However, note that this leads to a severe reduction of the actual resolution of the method (precisely by a factor of $2$ due to omitting every second observation) and hence to a reduced detection power.}

{A further interesting possible extension of the AMS methodology would be to random fields with spatial dependency. For $d = 1$ we suggest in the case of simple linear time series error models to follow the approach in \citet{fms14}, Section 6.1.1 to ease computational effort. Instead of the full likelihood ratio statistic, simply the local LRT statistics for the independent case have to be reweighed by estimates of the total variance. For general dimensions this is more involved. We believe that it might be possible to extend the weak invariance principle for $m$-dependent fields in Thm. 2.3 in \citet{kkm23} to the multiscale setup in this paper, however, at the expense of a large technical effort.}

{Finally, as mentioned by a referee, it would be of interest to extend AMS to the situation when no parametric family $F_{\theta,\xi}$ is assumed. In this case one aims to scan against an (unknown) baseline $F_0$, say. Consequently, not only certain nuisance parameters are unknown, rather the entire distribution $F_0$ and all possible alternatives $F_i, i \in I_n^d$. Already for $d = 1$ this leads to a challenging situation, as the local LRT statistics have to be replaced by some \textit{non-parametric analogue}, e.g. local median statistics, see \citet{vbm22}. How to extend this to general $d$ is not obvious at all and an interesting task for future research.}

\section*{Data availability statement}

{The data underlying this article, including all software and scripts to reproduce the simulations and data analysis shown in this work are available under \LINK.}

\section*{Acknowledgment}

Financial support by the German Research Foundation DFG through subprojects A04 and A07 of CRC 755 and the DFG Cluster of Excellence "MBExC Multiscale Bioimaging - from Molecular Machines to Networks of Excitable Cells" is gratefully acknowledged. We also thank Ren\'e Siegmund {and Alexander Egner} from IFNANO for providing data and helpful discussions. Furthermore, helpful comments and questions raised by the editor, an associate editor and three referees, which led to an improved presentation of the manuscript, are gratefully acknowledged. {We are also grateful to the associate editor for suggesting the bootstrap version of AMS}.

\bibliography{MultiNuisance}
\bibliographystyle{apalike} 

\appendix
	
\section{Implementation}\label{sec:implementation}

As already mentioned in the introduction (Section \ref{sec:intro}), for a generic set $\mathcal R_n$ all local test statistics $T_R$ in \eqref{eq:TR} have to computed individually. To this end, for example, the approximating rectangular system from \citet{w10} can be used to compute local averages on these. Depending on the structure of $\mathcal R_n$, other efficient computational schemes can be employed. In the following we will focus on the situation that there is a global shape $B \subset I_n^d$ such that every $R \in \mathcal R_n$ is a rescaled and shifted version of $B$. More precisely, for each $R \in \mathcal R_n$ there exist $t, h \in I_n^d$ with $t_i + h_i \leq n$ for all $1 \leq i\leq d$ such that
\begin{equation}\label{eq:R_struct}
	\1_R \left(x\right) = \1_B \left(\frac{x-t}{h}\right), \qquad x \in I_n^d,
\end{equation}
where division is meant component-wise and $\1$ denotes the indicator function. For the set $\mathcal R_n$ of all hyper-rectangles, this is for example the case with $B = \1_{I_n^d}$. 

If $R$ obeys \eqref{eq:R_struct}, we obtain
\[
\bar Y_R = \frac{1}{\left|R\right|} \sum_{i \in R} Y_i = \frac{1}{\left|R\right|} \sum_{i \in I_n^d} Y_i \1_B\left(\frac{i-t}{h}\right) = \frac{1}{\left|R\right|} \left(Y \ast \1_B \left(\frac{\cdot}{h}\right)\right) \left(t\right),
\]
where $\ast$ denotes a discrete convolution. Employing the fast Fourier transform (FFT), {see \cite{ct65},} this allows to compute all $T_R$ with a fixed scale $h$ by means of {three} FFTs via
\[
\left(\bar Y_R\right)_{h} = \frac{1}{h^d \left|B\right|} \text{FFT}^{-1} \left(\text{FFT} \left(Y\right) \cdot \text{FFT}\left( \1\left(\frac{\cdot}{h}\right)\right)\right).
\]
This operation has a computational complexity almost linear in the data, i.e. $\mathcal O \left(d n^d \log\left(n\right)\right)$, and hence if each $R \in \mathcal R_n$ obeys \eqref{eq:R_struct}, then $T_n$ in \eqref{eq:Tn} can be computed within $\mathcal O \left(d N n^d\log\left(n\right)\right)$ where $N$ is the number of different scales. The same holds true for the evaluation of $M_n$ in \eqref{eq:Mn}. {A corresponding implementation in MATLAB$^\copyright$ is available under \LINK.}

\section{Simulations of level and power in a homogeneous Gaussian model}\label{sec:gaussian_sim}

At first, we present empirical quantiles of $M_n$ in \eqref{eq:Mn} for the situation considered in \ref{sec:poisson_sim} {and} the empirical density and CDF {both for the calibrated and the uncalibrated statistic}, see Table \ref{tab:quantiles_128} and Figure \ref{fig:invariance_dist} below.

\begin{table}[!htb]
	\caption{\label{tab:quantiles_128}	Empirical quantiles $q_{1-\alpha}$ with different values of $\alpha$ for the distribution of $M_n$ in \eqref{eq:Mn} in the case $n = 128, d = 2$ and $\mathcal R_n$ being the set of all rectangles in $I_n^d$ with side-lengths between $4$ and $14$.
	}
	\centering
	\fbox{%
		\begin{tabular}{*{6}{c}}
			$\alpha$ & $0.2$ & $0.1$ & $0.05$ & $0.025$ & $0.01$ \\
			$q_{1-\alpha}$& $1.2906$ & $1.4677$ & $1.6278$ & $1.7841$ & $1.9768$\\
		\end{tabular}
	}
\end{table}

\begin{figure}[!htb]
	\scriptsize
	\centering
	\setlength{\fwidth}{3cm}
	\setlength{\fheight}{3cm}
	\begin{tabular}{ll}
		\input{density.tikz}
		&
		\input{cdf.tikz}
	\end{tabular}
	\caption{Empirical density (left) and CDF (right) of $M_{n}$ in \eqref{eq:Mn} for $n = 128$ in $d = 2$ dimensions with $\mathcal R_n$ being the set of all rectangles in $I_n^d$ with even side-lengths between $4$ and $14$ pixels, displayed by a standard kernel density estimator, drawn from 2000 Monte Carlo runs, either calibrated (\ref{cal}) with penalization functions $\widetilde \omega{_n}$ and $\omega{_n}$ as in \eqref{pen_ds01}, or uncalibrated (\ref{uncal}).}
	\label{fig:invariance_dist}
\end{figure}

Secondly, we consider a homogeneous Gaussian model as already discussed in the introduction (see Section \ref{sec:intro}), i.e. $F_{\theta, \xi}$ is a normal distribution $\mathcal N \left(\mu, \sigma^2\right)$, and the variance $\sigma^2$ is considered as the nuisance parameter and estimated by the sample variance
\[
\hat\sigma^2_n := \frac{1}{n^d-1} \sum_{i \in I_n^d} \left(Y_i - \bar Y_{I_n^d}\right)^2.
\]
Then \eqref{eq:estimators} is clearly satisfied under $\mathbb P_0$ with $s_n = n^{d/2}$. 

{The results of the same simulations as described in Section \ref{sec:poisson_sim} are depicted in Figure \ref{fig:gauss}. Sub-panels (a) and (b) display the signal and the corresponding data for the  default setup ($d = 2$, $n = 128$) with an anomaly of size $8 \times 8$ pixels and amplitude $\mu = 0.5$ in the centre, where the variance $\sigma^2 = 1$. Sub-panel (c) shows the empirical level for different variances $\sigma^2$. We find that the level is kept quite stable over a large range of variances. Sub-panel (d) shows the empirical power for $\sigma^2 = 1$ against different means $\mu$. Notably, there is nearly no power loss of the AMS method compared to the oracle caused by adaptation for an unknown $\sigma^2$. Sub-panel (e) is devoted to simulations in dimension $d = 1$ with $\sigma^2 = 1$ and different values of $n \in \{32,64,96,128\}$. The lines depict that value of $\mu$, such that the AMS procedure is able to detect a present anomaly of size $a$ marked on the $x$-axis with amplitude $\mu$ with a power of $\approx 90\%$. We find that a larger anomaly clearly corresponds to a smaller amplitude required for detection. This is in agreement with the asymptotic separation rate in \eqref{eq:separation_rate}, and remarkably, despite the presence of nuisance parameters and the additional difficulty that the length $a$ is not known, the finite sample shapes of all separation lines in sub-panel (e) are of this form. Furthermore, we find that the influence of the sample size $n$ on the separation lines is small. Sub-panel (f) shows the same situation in dimension $d = 2$. Due to the larger anomaly, which now contains $a^2$ pixels, the corresponding values for $\mu$ are chosen smaller. However, the conclusions drawn for the one-dimensional case apply here similarly. Compared with sub-panel (e) we furthermore conclude that the dimension $d$ does not have a big influence, and again we see a good agreement with the asymptotic separation rate in \eqref{eq:separation_rate}.}

\begin{figure}[!htb]
	\scriptsize
	\centering
	\setlength{\fwidth}{2.8cm}
	\setlength{\fheight}{2.8cm}
	\begin{tabular}{lll}
		\hspace*{1cm}\input{Gaussian_means_sim.tikz}
		&
		\hspace*{.5cm}\input{Gaussian_data_sim.tikz}
		&
		\setlength{\fwidth}{3.72cm}
		\hspace*{-.5cm}\input{gauss_level.tikz}
		\\ 
		\hspace*{1cm}(a) Signal with $\mu =0.5$
		& 
		\hspace*{.5cm}(b) Corresponding data for (a)
		&
		\hspace*{.55cm}(c) Level vs. $\sigma$
		\\[0.2cm]
		\setlength{\fwidth}{3.72cm}
		\hspace*{-.1cm}\input{gauss_power.tikz}
		&
		\setlength{\fwidth}{3.72cm}
		\hspace*{-.32cm}\input{gauss_mean_for_power_d=1.tikz}
		&
		\setlength{\fwidth}{3.72cm}
		\hspace*{-.5cm}\input{gauss_mean_for_power_d=2.tikz}		
		\\
		\hspace*{1cm}(d) Power vs. $\mu$ for $\sigma^2 = 1$
		&
		\hspace*{.5cm}(e) separation rate, $d = 1$, $\sigma^2 = 1$
		&
		\hspace*{.55cm}(f) separation rate, $d = 2$, $\sigma^2 = 1$		
	\end{tabular}
	
	\caption{{Simulation results in a homogeneous Gaussian model with $\alpha = 0.1$. Panels (a) and (b): Signal under the alternative $(\mu = 0.5)$ used for this simulation and corresponding data for $\sigma^2= 1$. As the signal is standardized, the standard deviation $\sigma$ is proportional to the signal-to-noise ratio (SNR). Panel (c): Empirical level for the oracle multiscale scan ($\sigma^2$ known) and AMS ($\sigma^2$ unknown, i.e. considered as a nuisance parameter) for different values of $\sigma$, smoothed with a moving average filter of width $10$. Panel (d): Empirical power versus $\mu$ for a $8 \times 8$ anomaly as in (a), smoothed with a moving average filter of width $10$. Panels (e) and (f): Value of $\mu$ required to obtain an empirical power of $\beta = 0.9$ versus size of a (squared) anomaly with amplitude $\mu$.}}
	\label{fig:gauss}
\end{figure}	

\section{Additional data examples}

Similar to the setting in Section \ref{sec:data} we investigate two further data sets.

\subsection{Data example 1: Single sized crimson beads}

As a first data example, we investigate $48 \nm$ crimson beads, i.e. carboxylate modified microspheres of $48\nm$ diameter filled with fluorescent markers. {Taking into account the microscopes effective resolution of about $60$nm, we expect to see circular anomalies of size $\sim 110$nm in the data. Therefore we choose $r_n$ and $m_n$ such that the smallest box-size is $80$nm, and the largest box-size is $400$nm. Since each pixel has size $20 \nm \times 20 \nm$, this corresponds to $r_n = 4$ and $m_n = 20$.} The measured data for an illumination time of $t =5 \ms$ and $T = 100 \ms$ are shown in Figure \ref{fig:beads}(a) and (d). The corresponding results for AMS are shown in sub-panels (b) and (e). Surprisingly, {the test detects most beads based on the $5 \mu\mathrm{s}$ data, even though some of them are barely} visible by eye there. Regarding the $T = 100 \ms$ data as ground truth, we depict this together with the regions found in (b) in sub-panel (e), {revealing} that there no false positive detections. Finally, the $T = 100 \ms$ data can be used to derive a segmentation into active and inactive regions depicted in (f).

\begin{figure}[!htb]
	\scriptsize
	\centering
	\setlength{\fwidth}{3.1cm}
	\setlength{\fheight}{3.1cm}
	\begin{tabular}{p{4.2cm}p{4.9cm}p{4.5cm}}
		
		\input{beads_limited_data.tikz}
		&
		\input{beads_result_limited.tikz}
		&
		\input{beads_full_data.tikz}
		\\
		(a) $t = 5 \ms$, raw data 
		&
		(b) significance map, $t = 5 \ms$
		&
		(c) $T = 100 \ms$, raw data, with detected regions from (b) in red
		\\		
		\input{beads_full_data2.tikz}
		&
		\input{beads_result_full.tikz}
		&
		\input{beads_full_segments.tikz}
		\\
		(d) $T = 100 \ms$, raw data
		&
		(e) significance map, $T = 100 \ms$
		&
		(f) segmentation derived from (e) using only the smallest scale
	\end{tabular}
	\caption{Raw crimson bead data (photon counts, panels (a) and (d)) and corresponding significance maps with level $\alpha = 0.1$ (smallest significance in $\nm^2$, panels (b) and (d)). All significant boxes are plotted with colour indicating their size in $\nm^2$. Due to the FWER control, with probability $\geq 1-\alpha = 0.9$, all boxes marked contain anomalies.  Using only the smallest scale significance, this leads to the segmentation in (f) showing active regions in red.}
	\label{fig:beads}
\end{figure}

\subsection{Data example 2: mixture of differently sized crimson beads}

As a second data example, we now consider a mixture of $48 \nm$ and $200 \nm$ crimson beads. Consequently, the data shown in Figure \ref{fig:mixed_beads}(a) and (d) consists of two structures of different sizes, namely around $48 \nm$ and $200 \nm$. Due to the number of markers inside the spheres, the larger structures are significantly brighter than the small ones. We investigate $t = 15 \ms$ and $T = 100 \ms$. The AMS procedure is applied with the same parameters as before (i.e. again $m_n = 20$, corresponding to a largest box size of $400 \nm$), and the result is shown in sub-panels (b) and (e). It is immediately visible that it detects again nearly all of the small structures already from the measurements taken with $15 \ms$ dwell time, which is surprising as some of them are not visible by eye in the data. Once again, regarding the $T = 100 \ms$ data as ground truth, we depict this together with the regions found in (b) in sub-panel (e), {revealing} that there no false positive detections. Finally, the $T = 100 \ms$ data can be used to derive a segmentation into active and inactive regions depicted in (f).

\begin{figure}[!htb]
	\scriptsize
	\centering
	\setlength{\fwidth}{3.1cm}
	\setlength{\fheight}{3.1cm}
	\begin{tabular}{p{4.2cm}p{4.9cm}p{4.5cm}}
		
		\input{mixed_beads_limited_data.tikz}
		&
		\input{mixed_beads_result_limited.tikz}
		&
		\input{mixed_beads_full_data.tikz}
		\\
		(a) $t = 15 \ms$, raw data (logarithmic)
		&
		(b) significance map, $t = 15 \ms$
		&
		(c) $T = 100 \ms$, raw data (logarithmic), with detected regions from (b) in red
		\\		
		\input{mixed_beads_full_data2.tikz}
		&
		\input{mixed_beads_result_full.tikz}
		&
		\input{mixed_beads_full_segments.tikz}
		\\
		(d) $T = 100 \ms$, raw data (logarithmic)
		&
		(e) significance map, $T = 100 \ms$
		&
		(f) segmentation derived from (e) using only the smallest scale
	\end{tabular}
	\caption{Raw crimson bead data (photon counts, logarithmic scale, panels (a) and (d)) and corresponding significance maps with level $\alpha = 0.1$ (smallest significance in $\nm^2$, panels (b) and (e)). All significant boxes are plotted with colour indicating their size in $\nm^2$. Due to the FWER control, with probability $\geq 1-\alpha = 0.9$, all boxes marked contain anomalies.  Using only the smallest scale significance, this leads to the segmentation in (f) showing active regions in red.}
	\label{fig:mixed_beads}
\end{figure}

\section{Results for different values of $\alpha$}

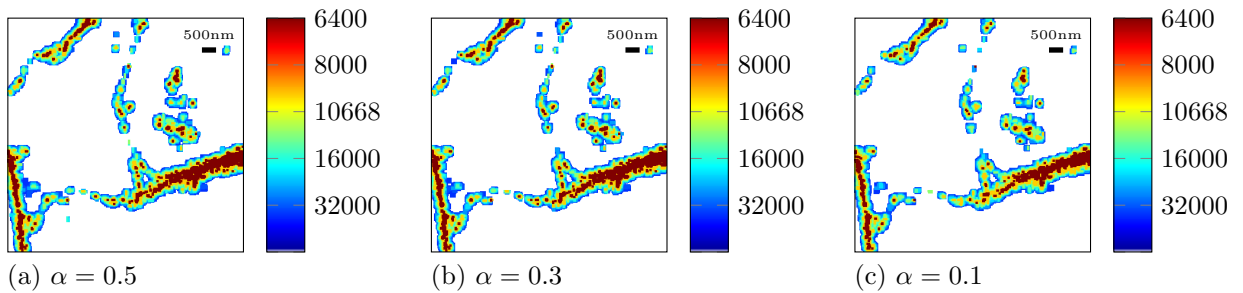
\begin{figure}[!htb]
	\scriptsize
	\centering
	\setlength{\fwidth}{3.1cm}
	\setlength{\fheight}{3.1cm}
	\begin{tabular}{lll}
		\input{fibro_alpha_05.tikz}
		& 
		\input{fibro_alpha_07.tikz}
		&
		\input{fibro_result_limited.tikz}
		\\
		(a) $\alpha = 0.5$ 
		& 
		(b) $\alpha = 0.3$	
		&
		(c) $\alpha = 0.1$
	\end{tabular}
	\caption{Significance maps from the raw tubulin data in Figure \ref{fig:fibro}(a) for different values of $\alpha$.}
	\label{fig:fibro3}
\end{figure}

We also compare the resulting significance maps for different values of $\alpha$ for the data set from Section \ref{sec:data}, see Figure \ref{fig:fibro3}. It turns out that the AMS methodology is {remarkably} stable w.r.t. the significance level $\alpha$.

\section{Proofs}

We start with some helpful abbreviations. For the whole section, let $X = \left(X_i\right)_{i \in I_n^d}$ be a field of i.i.d. standard Gaussians, $X_i \sim \mathcal N\left(0,1\right)$. As we rely heavily on {centred} and standardized partial sums, we introduce for $R \in \mathcal R_n$ the quantity
\[
Y_R\left(\theta, \xi\right):= \left|R\right|^{-1}\sum\limits_{i \in R} \left( \frac{Y_i-m\left(\theta, \xi\right)}{\sqrt{v\left(\theta, \xi\right)}}\right) = \frac{\bar Y_R - m\left(\theta, \xi\right)}{\sqrt{v\left(\theta,\xi\right)}}, \qquad \left(\theta, \xi\right) \in \Theta \times \Xi
\]
with the mean and variance functions $m$ and $v$ from Assumption \ref{ass:model}. Also recall our abbreviation
\[
\bar X_R := \frac{1}{\left|R\right|} \sum_{i \in R} X_i.
\]

\subsection{Preparations}

Next we recall a a series of helpful results taken from \citet{kmw20}.

\begin{lem}[Coupling]\label{lem:coupling}
	Let Assumptions \ref{ass:model}(a) and \ref{ass:R} hold true, and let $Y = \left(Y_i\right)_{i \in I_n^d}$ be an array of i.i.d. random variables $Y_i \sim F_{\theta, \xi}$ with $\theta \in U, \xi \in V$ and let $\left(r_n\right)_{n \in \N} \subset \mathbb N$ be a sequence. Then, on the same probability space, there exists an array $X = \left(X_i\right)_{i \in I_n^d}$ of i.i.d. standard Gaussians $X_i \sim \mathcal N\left(0,1\right)$ such that
	\[
	\mathbb P^*_{0,n}\left[\left|\max\limits_{ R \in \mathcal{R}_n\left(r_n, n^d\right)} \left|R\right|^{\frac12}\left|Y_R\left(\theta, \xi\right)\right|  - \max\limits_{ R \in \mathcal{R}_n\left(r_n, n^d\right)}\left|R\right|^{\frac12} \left|\bar X_R\right| \right| > \delta \right] \leq C \delta^{-3} \left(\frac{\log^{10}n}{r_n} \right)^{1/2}
	\]
	for all $\delta > 0$ with some universal constant $C> 0$.
\end{lem}
\begin{proof}
	As the distributions $F_{\theta, \xi}$ have uniformly sub-exponential tails, this follows directly from Theorem 4.3 in \citet{kmw20} in combination with a symmetrization argument as employed in the proof of Theorem 2.5 there.
\end{proof}
Note, that in terms of $\mathcal O$ notation, Lemma \ref{lem:coupling} yields
\begin{equation}\label{eq:coupling}
	\max\limits_{ R \in \mathcal{R}_n\left(r_n, n^d\right)} \left|R\right|^{\frac12}\left|Y_R\left(\theta, \xi\right)\right|  - \max\limits_{ R \in \mathcal{R}_n\left(r_n, n^d\right)}\left|R\right|^{\frac12}\left|\bar X_R\right|  = \mathcal O_{\mathbb P^*_{0,n}} \left(\left(\frac{\log^{10} \left(n\right)}{r_n}\right)^{\frac16}\right),
\end{equation}
uniformly in $\left(\theta, \xi\right)\in U \times V$.

\begin{lem}[Taylor expansion, see \protect{Lemma 5.1 in \citet{kmw20}}]\label{lem:taylor_exact}
	Let $Y = \left(Y_i\right)_{i \in I_n^d}$ be an array of i.i.d. random variables $Y_i \sim F_{\theta, \xi}$ and let $\left(r_n\right)_{n \in \N} \subset \mathbb N$ be a sequence satisfying \eqref{eq:r_n}. If Assumptions \ref{ass:model}(a) and \ref{ass:R} hold true, then
	
	\[
	\max\limits_{R \in  \mathcal{R}_n \left(r_n, n^d\right)} \left| T_R\left(Y, \theta, \xi\right) -  \left|R\right|^{\frac12} Y_R \left(\theta, \xi\right)\right| = \mathcal O_{\mathbb P^*_{0,n}} \left(\left( \frac{\log^3(n)}{r_n}\right)^{1/4}\right)
	\]
	uniformly for $\theta \in U, \xi \in V$.
\end{lem}
Note, that the previous results remain true if we replace the upper bound $n^d$ for the scale size by $m_n$ with any sequence $\left(m_n\right)_{n \in \N} \subset \N$ satisfying $r_n \leq m_n$ for all $n \in \N$.

\begin{lem}[Uniform remainder]\label{lem:bound}
	Let Assumptions \ref{ass:model}(a) and \ref{ass:R} hold true, and let $Y = \left(Y_i\right)_{i \in I_n^d}$ be an array of i.i.d. random variables $Y_i \sim F_{\theta, \xi}$ with fixed $\theta \in U, \xi \in V$. If  $\left(r_n\right)_{n \in \N} \subset \mathbb N$ is a sequence satisfying \eqref{eq:r_n}, then we obtain	
	\[
	\max\limits_{R \in \mathcal{R}_n(r_n, m_n)} \left|R\right|\left| Y_R \left(\theta_0, \xi\right) \right|^3 = \mathcal O_{\mathbb P^*_{0,n}}\left(\left(\frac{\log^3\left(n\right)}{r_n}\right)^{\frac12} \right)
	\]
	uniformly for $\theta \in U, \xi \in V$.
\end{lem}
\begin{proof}
	Note, that due to $\alpha, \widetilde{\alpha} >0$, \eqref{eq:r_n} implies especially that $\gamma \geq 4$ and hence $\log^4\left(n\right) = o \left(r_n\right)$. 
	
	It is well known (see e.g. \citet{k11}), that for a standard Gaussian array $X = \left(X_i\right)_{i \in I_n^d}$, $X_i \stackrel{\text{i.i.d.}}{\sim} \mathcal N \left(0,1\right)$, one has
	\begin{equation}\label{eq:aux3}
		\E{}{\max_{R \in \mathcal R_n \left(r_n, m_n\right)} \left|R\right|^{\frac12} \left|\bar X_R\right|} \leq C \sqrt{\log\left(\# \mathcal R_n \left(r_n, m_n\right)\right)}
	\end{equation}
	with some constant $C > 0$. Due to Assumption \ref{ass:R} this implies that
	\[
	\frac{1}{\sqrt{\log\left(n\right)}} \max_{R \in \mathcal R_n \left(r_n, m_n\right)} \left|R\right|^{\frac12} \left|\bar X_R\right|  = \mathcal O_{\mathbb P}\left(1\right). 
	\]
	Together with \eqref{eq:coupling} this yields
	\[
	\frac{1}{\sqrt{\log\left(n\right)}} \max_{R \in \mathcal R_n \left(r_n, m_n\right)} \left|R\right|^{\frac12}\left|Y_R\left(\theta, \xi\right)\right| = \mathcal O_{\mathbb P^*_{0,n}}\left(1\right)
	\]
	uniformly for $\left(\theta, \xi\right)\in U \times V$, where we used that $\log^4\left(n\right)= o \left(r_n\right)$.  Consequently we find
	\begin{align*}
		\sqrt{\frac{r_n}{\log^3\left(n\right)}} \max_{R \in \mathcal R_n \left(r_n, m_n\right)}  \left|R\right|\left| Y_R \left(\theta_0, \xi\right) \right|^3 \leq  \sqrt{\frac{1}{\log^3\left(n\right)}} \max_{R \in \mathcal R_n \left(r_n, m_n\right)}\left|R\right|^{\frac32}\left| Y_R \left(\theta_0, \xi\right) \right|^3 =\mathcal O_{\mathbb P^*_{0,n}}
	\end{align*}
	uniformly for $\left(\theta, \xi\right)\in U \times V$, which proves the claim.
\end{proof}

\subsection{An adjusted Taylor expansion}

Now we are in position to derive some preparations for the proof of Theorem \ref{thm:approximation_estimated_intensity}. Therefore we first replace the likelihood ratio statistic $T_R$ by its Taylor expansion according to Assumption \ref{ass:model}(e):
\begin{theorem}\label{thm:taylor_estimated}
	Let $\left(r_n\right)_{n\in\N} \subset \N$ and $\left(m_n\right)_{n\in\N} \subset \N$ be sequences tending to $\infty$ and consider the local statistic $T_R$ from \eqref{eq:TR}. Suppose that $Y$ is as in \eqref{eq:model} and let Assumptions \ref{ass:model} and \ref{ass:R} be fulfilled. If the estimators $\hat \theta_n$ and $\hat \xi_n$ satisfy \eqref{eq:estimators}, then
	\begin{equation}\label{eq:taylor}
		\max\limits_{R \in  \mathcal{R}_n ( r_n, m_n)} \left| T_R(Y, \hat \theta_n, \hat \xi_n) -  \left|R\right|^{1/2}\left|Y_R \left(\theta_0, \xi\right)\right|\right| = \mathcal O_{\mathbb P^*_{0,n}}\left(\left(\frac{\log^7(n)}{r_n}\right)^{\frac14} + \sqrt{\frac{m_n}{s_n}}\right).
	\end{equation}
\end{theorem}

\begin{proof}
	Let $U \times V$ be the {neighbourhood} of $\left(\theta_0, \xi\right)$ as in Assumption \ref{ass:model}. By \eqref{eq:estimators}, the probability that $\hat \theta_n \in U$ and $\hat \xi_n \in V$ tends to $1$, as $n \to \infty$. Hence, it suffices to prove the result conditional on this event. Consequently we can assume $\hat \theta_n \in U$ and $\hat \xi_n \in V$ in the following.
	
	By the triangle inequality and Lemma \ref{lem:taylor_exact}, we find
	\begin{align*}
		&\max\limits_{R \in  \mathcal{R}_n ( r_n, m_n)} \left| T_R\left(Y, \hat \theta_n, \hat \xi_n\right) -   \left|R\right|^{\frac12} \left|Y_R\left(\theta_0, \xi\right)\right|\right| \\
		\leq &  \max\limits_{R \in  \mathcal{R}_n ( r_n, m_n)} \left| T_R(Y, \theta_0, \xi) -   |R|^{\frac12}\left|Y_R\left(\theta_0, \xi\right)\right|\right|\\
		&+ \max\limits_{R \in  \mathcal{R}_n ( r_n, m_n)} \left| T_R(Y, \hat \theta_n, \hat \xi_n) -   T_R\left(Y, \theta_0, \xi\right)\right| \\
		\leq &\sqrt{\max\limits_{R \in  \mathcal{R}_n ( r_n, m_n)} \left| T_R^2(Y, \hat \theta_n, \hat \xi_n) -   T_R^2\left(Y, \theta_0, \xi\right)\right|} + \mathcal O_{\mathbb P^*_{0,n}}\left(\left( \frac{\log^3(n)}{r_n}\right)^{\frac14}\right),
	\end{align*}
	where we exploited Lemma \ref{lem:taylor_exact} and $\left|a-b\right| \leq \sqrt{\left|a^2 - b^2\right|}$. By the triangle inequality and Assumption \ref{ass:model}(e) we find furthermore that
	\begin{align*}
		&\max\limits_{R \in \mathcal{R}_n(r_n, m_n)} \left|T_R^2(Y, \hat{\theta}, \hat{\xi}) - T_R^2(Y, \theta_0, \xi) \right|\\
		\leq& \max\limits_{R \in \mathcal{R}_n(r_n, m_n)} \left|T_R^2(Y, \hat{\theta}, \hat{\xi}) - \left|R\right| \left|Y_R\left(\hat \theta_n, \hat\xi_n\right)\right|^2 \right| + \max\limits_{R \in \mathcal{R}_n(r_n, m_n)} \left|T_R^2\left(Y, \theta_0, \xi\right)-\left|R\right| \left|Y_R\left(\theta_0, \xi\right)\right|^2\right|\\
		&+ \max\limits_{R \in \mathcal{R}_n(r_n, m_n)} \left|R\right| \left|Y_R\left(\hat \theta_n, \hat \xi_n\right)^2 - Y_R\left(\theta_0, \xi\right)^2 \right|\\
		\leq & C_T \max\limits_{R \in \mathcal{R}_n(r_n, m_n)} \left|R\right| \left|Y_R \left(\hat \theta_n, \hat \xi_n\right) \right|^3 + C_T \max\limits_{R \in \mathcal{R}_n(r_n, m_n)} \left|R\right|  \left|Y_R\left(\theta_0, \xi\right) \right|^3 \\
		&+ \max\limits_{R \in \mathcal{R}_n(r_n, m_n)} \left|R\right| \left|Y_R\left(\hat \theta_n, \hat \xi_n\right)^2-Y_R\left(\theta_0,\xi\right)^2\right| \\
		\leq & 2C_T \max\limits_{R \in \mathcal{R}_n(r_n, m_n)} \left|R\right| \left|Y_R\left(\theta_0, \xi\right) \right|^3  + \max\limits_{R \in \mathcal{R}_n(r_n, m_n)} \left|R\right|\left|Y_R\left(\hat \theta_n, \hat \xi_n\right)^2-Y_R\left(\theta_0,\xi\right)^2\right|\\ & + C_T\max\limits_{R \in \mathcal{R}_n(r_n, m_n)} \left|R\right| \left|Y_R\left(\hat \theta_n, \hat \xi_n\right)^3-Y_R\left(\theta_0,\xi\right)^3\right|,
	\end{align*}
	where we used $\left| \left|a\right|^3 - \left|b\right|^3\right| = \left| \left|a^3\right| - \left|b^3\right|\right| \leq \left| a^3 - b^3\right|$ for $a,b \in \R$. In view of Assumptions \ref{ass:model}(a) and \ref{ass:R}, the first term in the last display can be bounded using Lemma \ref{lem:bound}. The second and third term will be handled using the mean value theorem. Therefore, note that all derivatives of
	\[
	\left(\theta, \xi\right)\mapsto Y_R\left(\theta,\xi\right)^k, \qquad k \in \left\{2,3\right\}
	\]
	are a.s. bounded due to Assumption \ref{ass:model}(b)--(d) and the fact that $\bar Y_R$ is a.s. bounded in view of Lemma \ref{lem:bound}. This together with \eqref{eq:estimators} implies that
	\begin{align*}
		\max\limits_{R \in \mathcal{R}_n(r_n, m_n)} \left|R\right| \left| Y_R\left(\hat{\theta}, \hat{\xi}\right)^2 - Y_R\left(\theta_0, \xi\right)^2\right|\leq m_n C \left(\|\hat{\theta}_n-\theta_0 \|_2 + \|\hat{\xi}_n - \xi \|_2\right) = \mathcal O_{\mathbb P^*_{0,n}} \left(\frac{m_n}{s_n}\right),
	\end{align*}
	where $C$ is some generic constant $C > 0$. As $\max\limits_{R \in \mathcal{R}_n(r_n, m_n)} \left|R\right| \left|Y_R\left(\hat \theta_n, \hat \xi_n\right)^3-Y_R\left(\theta_0,\xi\right)^3\right|$ can be treated similarly, this yields the claim.
\end{proof}

\subsection{Proof of Theorem \ref{thm:approximation_estimated_intensity}}

Now we are ready to prove the Gaussian approximation result from Theorem \ref{thm:approximation_estimated_intensity}.
\begin{proof}[Proof of Theorem \ref{thm:approximation_estimated_intensity}]
	For notional simplicity, we will throughout this proof drop the index of $\widetilde{\omega}{_n}$ and $\omega{_n}$. Furthermore we abbreviate $\widetilde{\mathcal R}_n := \mathcal R_n \left(r_n, m_n\right)$. First of all, we estimate	
	\begin{align*}
		&\left|T_n \left(Y,  \widetilde{\mathcal R}_n, \hat{\theta}, \hat{\xi} \right) - M_n\left(\widetilde{\mathcal R}_n\right) \right|\\
		=& \left|\max_{R\in \widetilde{\mathcal R}_n} \widetilde{\omega} \left(\left|R\right|\right) \left(T_R \left(Y, \hat \theta_n, \hat \xi_n\right) - \omega \left(\left|R\right|\right) \right) - \max_{R\in \widetilde{\mathcal R}_n} \widetilde{\omega} \left(\left|R\right|\right) \left(\left|R\right|^{\frac12} \left|\bar X_R\right| - \omega \left(\left|R\right|\right) \right) \right| \\
		\leq& \left|\max_{R\in \widetilde{\mathcal R}_n} \widetilde{\omega} \left(\left|R\right|\right) \left(T_R \left(Y, \hat \theta_n, \hat \xi_n\right) -  \omega \left(\left|R\right|\right) \right) - \max_{R\in \widetilde{\mathcal R}_n} \widetilde{\omega} \left(\left|R\right|\right) \left(\left|R\right|^{\frac12} \left|Y_R \left(\theta_0, \xi\right)\right| - \omega \left(\left|R\right|\right) \right) \right| \\
		& + \left|\max_{R\in \widetilde{\mathcal R}_n} \widetilde{\omega} \left(\left|R\right|\right) \left(\left|R\right|^{\frac12} \left|Y_R \left(\theta_0, \xi\right)\right| -  \omega \left(\left|R\right|\right) \right) - \max_{R\in \widetilde{\mathcal R}_n} \widetilde{\omega} \left(\left|R\right|\right) \left(\left|R\right|^{\frac12} \left|\bar X_R\right|- \omega \left(\left|R\right|\right) \right) \right| \\
		\leq & \max_{R \in \widetilde{\mathcal R}_n} \widetilde{\omega} \left(\left|R\right|\right)  \max _{R \in \widetilde{\mathcal R}_n} \left| T_R \left(Y, \hat \theta_n, \hat \xi_n\right) - \left|R\right|^{\frac12} \left|Y_R \left(\theta_0, \xi\right)\right|\right| \\
		& + \left|\max_{R\in \widetilde{\mathcal R}_n} \widetilde{\omega} \left(\left|R\right|\right) \left(\left|R\right|^{\frac12} \left|Y_R \left(\theta_0, \xi\right)\right| -  \omega \left(\left|R\right|\right) \right) - \max_{R\in \widetilde{\mathcal R}_n} \widetilde{\omega} \left(\left|R\right|\right) \left(\left|R\right|^{\frac12} \left|\bar X_R\right| - \omega \left(\left|R\right|\right) \right) \right|,
	\end{align*}
	where we used $\left|\|x\|_\infty-\|y\|_\infty  \right|\leq \|x-y\|_\infty$. The first term can be controlled using Assumption \ref{ass:Omega} and Theorem \ref{thm:taylor_estimated} by
	\begin{multline*}
		\max_{R \in \widetilde{\mathcal R}_n} \widetilde{\omega} \left(\left|R\right|\right)  \max _{R \in \widetilde{\mathcal R}_n} \left| T_R \left(Y, \hat \theta_n, \hat \xi_n\right) - \left|R\right|^{\frac12} \left|Y_R \left(\theta_0, \xi\right)\right|\right| 
		\\= \mathcal O_{\mathbb P^*_{0,n}}\left( \frac{\log^{\widetilde{\alpha}} \left(n\right)}{r_n} \left(\left(\frac{\log^3\left(n\right)}{r_n} \right)^{\frac14} + \sqrt{\frac{m_n}{s_n}}\right)\right).
	\end{multline*}
	So it remains to estimate the second term, i.e. for a suitable sequence $\left(c_n\right)_{n\in\N}$ we want to show that
	\begin{multline}\label{eq:aux2}
		\lim_{M \to \infty} \limsup_{n  \to\infty} \mathbb P_0 \left[\left|\max_{R\in \widetilde{\mathcal R}_n} \widetilde{\omega} \left(\left|R\right|\right) \left(\left|R\right|^{\frac12} \left|Y_R \left(\theta_0, \xi\right)\right|  \right.\right.\right.\\\left.\left.\left.-  \omega \left(\left|R\right|\right) \right)- \max_{R\in \widetilde{\mathcal R}_n} \widetilde{\omega} \left(\left|R\right|\right) \left(\left|R\right|^{\frac12} \left|\bar X_R\right| - \omega \left(\left|R\right|\right) \right) \right|> c_n M\right]  = 0
	\end{multline}
	This is more subtle, as the coupling in Lemma \ref{lem:coupling} only yields a result for the difference of the maxima, and not for the maximum of the differences (as Theorem \ref{thm:taylor_estimated} does). We therefore borrow a slicing technique from \citet{pwm18}, dividing the set of scales is into families on which $\widetilde{\omega}$ and $\omega$ are almost constant, i.e. into
	\[
	\mathcal{R}_{n,j}:= \left\lbrace R \in \mathcal{R}_n \left| \right. \exp\left(\epsilon_j\right) \leq \left|R\right| <  \exp\left(\epsilon_{j+1}\right) \right\rbrace.
	\]
	Here and in what follows we choose
	\[
	\epsilon_1 := \log\left(r_n\right), \qquad \epsilon_j = \epsilon_1 + \frac{j-1}{J} \log \left(\frac{m_n}{r_n}\right), \quad j=2, ..., J.
	\]
	With this definition, for any choice $J \in \N_{\geq 2}$ we obtain
	\[
	\widetilde{\mathcal R}_n = \mathcal{R}_n(r_n, m_n) = \left\{ R \in \R_n ~\big|~ r_n \leq \left|R\right| \leq m_n \right\} = \bigcup\limits_{j\in J}\mathcal{R}_{n,j}.
	\]
	The parameter $J \in \N$ will be defined later. On $\mathcal R_{n,j}$ we can approximate $\widetilde{\omega}\left(\left|R\right|\right)$ and $\omega\left(\left|R\right|\right)$ by
	\begin{align*}
		\widetilde{\omega}_{j,n} := \widetilde{\omega}\left(\exp\left(\epsilon_j\right)\right) \qquad \text{and}\qquad \omega_{j,n} := \omega\left(\exp\left(\epsilon_j\right)\right),
	\end{align*}
	respectively. Then it holds
	\[
	\omega_{j+1,n} \leq \omega\left(\left|R\right|\right) \leq \omega_{j,n}\qquad\text{and}\qquad \widetilde{\omega}_{j+1,n} \leq \widetilde{\omega}\left(\left|R\right|\right) \leq \widetilde{\omega}_{j,n}
	\]
	for all $R \in \mathcal{R}_{n,j}$. Now we compute
	\begin{align*}
		&\max\limits_{R \in \mathcal{R}_{n,j}}\widetilde{\omega}(|R|) \left(\left|R\right|^{\frac12} \left|Y_R\left(\theta_0, \xi\right)\right|-\omega(|R|) \right)-\max\limits_{R \in \mathcal{R}_{n,j}}\widetilde{\omega}(|R|) \left(\left|R\right|^{\frac12} \left|\bar X_R\right|-\omega(|R|) \right)\\
		\leq& \left( \widetilde{\omega}_{j,n} \max\limits_{R \in \mathcal{R}_{n,j}} \left|R\right|^{\frac12}\left|Y_R \left(\theta_0, \xi\right) \right| - \widetilde{\omega}_{j+1,n} \max\limits_{R \in \mathcal{R}_{n,j}}\left|R\right|^{\frac12}\left|\bar X_R\right| \right) + \left(\widetilde{\omega}_{j,n} \omega_{j,n} - \widetilde{\omega}_{j+1,n}\omega_{j+1,n} \right)\\
		=& \widetilde{\omega}_{j,n} \left( \max\limits_{R \in \mathcal{R}_{n,j} } \left|R\right|^{\frac12} \left|Y_R \left(\theta_0, \xi\right) \right| - \max\limits_{R \in \mathcal{R}_{n,j}} \left|R\right|^{\frac12}\left| \bar X_R\right|\right) + \left(\widetilde{\omega}_{j,n} - \widetilde{\omega}_{j+1,n} \right) \max\limits_{ R \in \mathcal{R}_{n,j}}\left|R\right|^{\frac12}\left|\bar X_R\right|\\
		& + \left(\widetilde{\omega}_{j,n}\omega_{j,n} - \widetilde{\omega}_{j+1,n}\omega_{j+1,n} \right),
	\end{align*}
	and a similar relation holds true if the roles of $\bar X_R$ and $Y_R \left(\theta_0, \xi\right)$ are interchanged. To bound the differences involving the $\widetilde{\omega}_{j,n}$ terms, we exploit the rough bound
	\begin{equation}\label{eq:aux1}
		\widetilde{\omega}_{j,n} = \widetilde{\omega}\left(\exp(\epsilon_j) \right)\leq C_\omega \left(\log \left( \frac{n^d}{\exp\left(\epsilon_j\right)}\right) \right)^{\widetilde{\alpha}}\leq C_\omega\left(\log(n^d) \right)^{\widetilde{\alpha}}=:A_n
	\end{equation}
	for all $1 \leq j \leq J$, and a similar analogue for $\omega_{j,n}$. From the mean value theorem we get
	\begin{align*}
		\widetilde{\omega}_{j,n} - \widetilde{\omega}_{j+1,n} =&\ \widetilde{\omega}\left(\exp\left(\epsilon_{j+1}\right) \right)-\widetilde{\omega}\left(\exp\left(\epsilon_{j}\right) \right)\\
		\leq& \left|\exp\left(\epsilon_j\right)-\exp\left(\epsilon_{j+1}\right) \right| \max_{\zeta \in \left(\epsilon_j, \epsilon_{j+1}\right)} \widetilde{\omega}'\left(\exp\left(\zeta_j\right) \right)\\
		\leq& C_\omega\left( \log (n^d)\right)^{\max(\widetilde{\beta}, 0)} \left(\exp(\epsilon_{j+1}-\epsilon_j)-1\right)\\
		=& C_\omega \left( \log (n^d)\right)^{\max(\widetilde{\beta}, 0)}\left( \exp\left(\frac{1}{J} \log\left(\frac{m_n}{r_n}\right) \right)-1\right) \\
		= & C_\omega \left( \log (n^d)\right)^{\max(\widetilde{\beta}, 0)} \left(\left(\frac{m_n}{r_n}\right)^{\frac1J}-1\right)=:B_n
	\end{align*}
	Similar statements hold for $\omega_{j,n}$ and $\omega_{j,n}-\omega_{j+1,n}$ and hence using \eqref{eq:aux1} we obtain
	\begin{multline*}
		\left|\widetilde{\omega}_{j,n} \omega_{j,n} - \widetilde{\omega}_{j+1,n}\omega_{j+1,n}\right|
		= \left|\widetilde{\omega}_{j,n} (\omega_{j,n} - \omega_{j+1,n} ) - \omega_{j+1,n}\left(\widetilde{\omega}_{j,n}-\widetilde{\omega}_{j+1,n}\right)\right|\\
		\leq C_\omega^2  \left(\left(\frac{m_n}{r_n}\right)^{\frac1J}-1\right) \left( \left(\log(n^d) \right)^{\widetilde{\alpha} + \max\left\{\beta, 0\right\}} +  \left(\log(n^d) \right)^{\alpha + \max\left\{\widetilde{\beta}, 0\right\}} \right)=:C_n
	\end{multline*}
	This implies
	\begin{align*}
		&\left|\max\limits_{R \in \widetilde{\mathcal R}_n} \widetilde{\omega}\left(\left|R\right|\right) \left(\left|R\right|^{\frac12}\left|Y_R\left(\theta_0,\xi\right)\right|-\omega\left(\left|R\right|\right) \right)- \max\limits_{R \in \widetilde{\mathcal R}_n} \widetilde{\omega}\left(\left|R\right|\right) \left(\left|R\right|^{\frac12}\left|\bar X_R\right|-\omega\left(\left|R\right|\right) \right) \right|\\
		\leq & A_n \max_{1 \leq j \leq J} \left| \max\limits_{R \in \mathcal{R}_{n,j} } \left|R\right|^{\frac12} \left|Y_R \left(\theta_0, \xi\right) \right| - \max\limits_{R \in \widetilde{\mathcal R}_n} \left|R\right|^{\frac12}\left| \bar X_R\right|\right| + B_n \max\limits_{ R \in \mathcal{R}_{n,j}}\left|R\right|^{\frac12}\left|\bar X_R\right| + C_n,
	\end{align*}
	and hence
	\begin{align*}
		&\mathbb P^*_{0,n}\left[\left|\max_{R\in \widetilde{\mathcal R}_n} \widetilde{\omega} \left(\left|R\right|\right) \left(\left|R\right|^{\frac12} \left|Y_R \left(\theta_0, \xi\right)\right| -  \omega \left(\left|R\right|\right) \right)- \max_{R\in \widetilde{\mathcal R}_n} \widetilde{\omega} \left(\left|R\right|\right) \left(\left|R\right|^{\frac12} \left|\bar X_R\right| - \omega \left(\left|R\right|\right) \right) \right|> c_n M\right]\\
		\leq & \mathbb P^*_{0,n}\left[A_n \max_{1 \leq j \leq J} \left|  \max\limits_{R \in \mathcal{R}_{n,j} } \left|R\right|^{\frac12} \left|Y_R \left(\theta_0, \xi\right) \right| - \max\limits_{R \in \mathcal{R}_{n,j}} \left|R\right|^{\frac12}\left| \bar X_R\right|\right| > \frac{c_n M}{3}\right] \\
		& + \mathbb P^*_{0,n}\left[B_n \max\limits_{ R \in \mathcal{R}_{n}}\left|R\right|^{\frac12}\left|\bar X_R\right| > \frac{c_n M}{3}\right] + \mathbb P^*_{0,n}\left[C_n > \frac{c_n M}{3}\right] \\
		=:& \text{I} + \text{II} + \text{III}.
	\end{align*}
	To bound I, we use Lemma \ref{lem:coupling} and the union bound to obtain
	\[
	\text{I} \leq C\left|J\right| \frac{3^3A_n^3}{c_n^3M^3} \left(\frac{\log^{10} \left(n\right)}{r_n}\right)^{\frac12}.
	\]
	Here and in what follows, $C >0$ is some generic constant, the value of which can change from line to line. For II, we exploit Markov's inequality as well as \eqref{eq:aux3} and find in view of Assumption \ref{ass:R}, that
	\[
	\text{II} \leq \frac{\E{}{\max_{R \in \mathcal R_n} \left|R\right|^{\frac12} \left|\bar X_R\right|}}{c_n M 3^{-1} B_n^{-1}} \leq C \frac{B_n \sqrt{\log\left(n\right)}}{c_n M}
	\]
	with some constant $C > 0$. This shows that \eqref{eq:aux2} is satisfied as soon as
	\begin{enumerate}
		\item[(A)] $\left|J\right| A_n^3 c_n^{-3} \log^5 \left(n\right) r_n^{-\frac12} = \mathcal O \left(1\right)$, 
		\item[(B)] $B_n c_n^{-1} \sqrt{\log\left(n\right)} = \mathcal O \left(1\right)$, and
		\item[(B)] $C_n = o \left( c_n\right)$
	\end{enumerate}
	hold true. To obtain this, we set
	\[
	J := \lfloor \frac{\log^{\nu+1} \left(n\right)}{c_n}\rfloor
	\]
	with some parameter $\nu > 0$ to be determined. This implies
	\[
	\left|J\right| \frac{A_n^3}{c_n^3} \left(\frac{\log^{10}\left(n\right)}{r_n}\right)^{\frac12} \leq C \frac{\log^{6+3\widetilde{\alpha} + \nu}}{c_n^4 \sqrt{r_n}},
	\]
	which proves (B) for
	\begin{equation}\label{eq:cn}
		c_n := \left(\frac{\log^{12+6\widetilde{\alpha} + 2\nu}\left(n\right)}{r_n}\right)^{\frac18}.
	\end{equation}
	Next we use $m_n / r_n \leq n^d$ and use $\exp\left(x\right) - 1 \leq 2x$ for $x \in \left[0,1\right]$ to find
	\[
	\left(\frac{m_n}{r_n}\right)^{\frac1J} \leq n^{\frac{d}{J}} = \exp\left(\frac{d c_n}{\log^\nu\left(n\right)}\right) \leq 1+C \frac{c_n}{\log^\nu\left(n\right)}
	\]
	for any sufficiently large $n$ supposed that $c_n = o \left(\log^\nu\left(n\right)\right)$. Thus it holds
	\[
	B_n \leq C \frac{c_n}{\log^{\nu-\max\left\{\widetilde{\beta},0\right\}}\left(n\right)},
	\]
	and hence (A) is satisfied if $\nu \geq \max\left\{\widetilde{\beta},0\right\} + \frac12$. It can readily be seen that (C) is now satisfied as soon as we define
	\[
	\nu := \max\left\{\frac12, \alpha, \widetilde{\alpha}\right\} + \max\left\{\beta, \widetilde{\beta},0\right\}.
	\]
	Note that this also ensures the required condition $c_n = o \left(\log^\nu\left(n\right)\right)$ under \eqref{eq:r_n}. Hence \eqref{eq:aux2} is satisfied with $c_n$ as in \eqref{eq:cn}, and thus we have
	\begin{multline*}
		\left|T_n \left(Y,  \widetilde{\mathcal R}_n, \hat{\theta}, \hat{\xi} \right) - M_n\left(\widetilde{\mathcal R}_n\right) \right|\\= \mathcal O_{\mathbb P^*_{0,n}}\left( \frac{\log^{\widetilde{\alpha}} \left(n\right)}{r_n} \left(\left(\frac{\log^3\left(n\right)}{r_n} \right)^{\frac14} + \sqrt{\frac{m_n}{s_n}}\right)+\left(\frac{\log^{12+6\widetilde{\alpha} + 2\nu}\left(n\right)}{r_n}\right)^{\frac18}\right),
	\end{multline*}
	which together with $\frac{\log^{\tilde \alpha + 3/44}(n)}{r_n^{5/4}} = \mathcal O \left(\frac{\log^{\gamma/8}(n)}{r_n^{1/8}}\right)$ proves the claim.
\end{proof}

\end{document}

%% file: full_PP.tikz
\begin{tikzpicture}[baseline]

\begin{axis}[%
width=\fwidth,
height=\fheight,
scale only axis,
xmin=0,
xmax=1,
xlabel={$M_{128} \left(\mathcal D_{128}^2\right)$},
ymin=0,
ymax=1,
ylabel={$T_{128} \left(Y,\mathcal D_{128}^2,\hat \mu_0,\hat \sigma_0\right)$},
axis background/.style={fill=white},
axis x line*=bottom,
axis y line*=left
]

\addplot [color=red, solid]
table[row sep=crcr]{%
	0	0\\
	1	1\\
};

\addplot [color=black, dotted]
table[row sep=crcr]{%
	0.8	0\\
	0.8	0.8\\
	0.7065	0.8\\
	0.7065	0\\
};

\node[right] at (0.8,0.8) {$80\%$};

\node[left] at (0.7065,0.8) {$\approx 70\%$};

\addplot [color=blue, dashed]
  table[row sep=crcr]{%
0	0\\
0.001	0.0005\\
0.001	0.0015\\
0.001	0.0015\\
0.001	0.0015\\
0.001	0.0015\\
0.001	0.0015\\
0.002	0.002\\
0.002	0.002\\
0.0025	0.003\\
0.0025	0.003\\
0.0035	0.0035\\
0.0045	0.0035\\
0.0065	0.0035\\
0.0075	0.005\\
0.009	0.0075\\
0.0125	0.008\\
0.0135	0.01\\
0.0135	0.0145\\
0.015	0.019\\
0.018	0.0225\\
0.0215	0.025\\
0.022	0.03\\
0.0265	0.0345\\
0.028	0.0405\\
0.0295	0.0475\\
0.037	0.0565\\
0.0425	0.0645\\
0.0465	0.0725\\
0.0515	0.083\\
0.058	0.092\\
0.069	0.099\\
0.0765	0.109\\
0.088	0.125\\
0.101	0.139\\
0.1135	0.1565\\
0.128	0.1715\\
0.139	0.1845\\
0.1455	0.1985\\
0.1545	0.215\\
0.1655	0.236\\
0.1775	0.251\\
0.193	0.2635\\
0.2065	0.2755\\
0.2205	0.294\\
0.241	0.31\\
0.2585	0.3325\\
0.2805	0.345\\
0.2935	0.358\\
0.3115	0.3755\\
0.33	0.396\\
0.3455	0.4115\\
0.364	0.4345\\
0.378	0.456\\
0.393	0.466\\
0.412	0.485\\
0.4315	0.4955\\
0.452	0.517\\
0.465	0.54\\
0.486	0.5635\\
0.502	0.578\\
0.514	0.5985\\
0.5295	0.6165\\
0.5475	0.6325\\
0.564	0.6405\\
0.5825	0.6605\\
0.5955	0.672\\
0.606	0.69\\
0.616	0.7085\\
0.629	0.721\\
0.646	0.7375\\
0.6595	0.7455\\
0.667	0.7575\\
0.6775	0.775\\
0.6905	0.7835\\
0.7015	0.793\\
0.7115	0.809\\
0.7305	0.8155\\
0.741	0.8305\\
0.748	0.838\\
0.763	0.8475\\
0.7695	0.8535\\
0.779	0.861\\
0.7885	0.867\\
0.797	0.875\\
0.804	0.8825\\
0.813	0.8865\\
0.826	0.888\\
0.8385	0.892\\
0.849	0.8955\\
0.8545	0.8995\\
0.8615	0.903\\
0.8675	0.911\\
0.871	0.9145\\
0.876	0.9205\\
0.8825	0.925\\
0.889	0.9295\\
0.894	0.9315\\
0.8985	0.934\\
0.9025	0.9395\\
0.908	0.942\\
0.912	0.9455\\
0.9185	0.947\\
0.9235	0.9505\\
0.9265	0.9515\\
0.933	0.956\\
0.9375	0.9585\\
0.9415	0.9605\\
0.9445	0.9645\\
0.947	0.9655\\
0.95	0.9675\\
0.952	0.971\\
0.9545	0.973\\
0.957	0.976\\
0.9575	0.977\\
0.9615	0.979\\
0.965	0.9815\\
0.9655	0.982\\
0.966	0.982\\
0.9665	0.983\\
0.9695	0.983\\
0.9715	0.9845\\
0.972	0.985\\
0.974	0.985\\
0.9775	0.987\\
0.9785	0.9885\\
0.9805	0.9885\\
0.982	0.9905\\
0.982	0.993\\
0.9825	0.993\\
0.983	0.993\\
0.984	0.994\\
0.984	0.9945\\
0.984	0.9945\\
0.985	0.9945\\
0.9855	0.995\\
0.986	0.9955\\
0.986	0.9955\\
0.987	0.996\\
0.987	0.996\\
0.9875	0.996\\
0.9875	0.9965\\
0.9885	0.9965\\
0.9885	0.997\\
0.9895	0.997\\
0.9895	0.997\\
0.992	0.997\\
0.994	0.997\\
0.994	0.997\\
0.994	0.9975\\
0.995	0.9975\\
0.995	0.9975\\
0.995	0.9975\\
0.9955	0.998\\
0.9955	0.998\\
0.996	0.9985\\
0.997	0.9985\\
0.9975	0.9985\\
0.9975	0.9985\\
0.9975	0.9985\\
0.9975	0.9985\\
0.998	0.9985\\
0.998	0.9985\\
0.998	0.999\\
0.998	0.999\\
0.998	0.999\\
0.998	0.999\\
0.998	0.999\\
0.998	0.999\\
0.998	0.999\\
0.998	0.999\\
0.998	0.999\\
0.998	1\\
0.998	1\\
0.998	1\\
0.998	1\\
0.998	1\\
0.998	1\\
0.998	1\\
0.998	1\\
0.998	1\\
0.998	1\\
0.9985	1\\
0.9985	1\\
0.9985	1\\
0.999	1\\
0.999	1\\
0.999	1\\
0.999	1\\
0.999	1\\
0.9995	1\\
0.9995	1\\
0.9995	1\\
0.9995	1\\
0.9995	1\\
0.9995	1\\
0.9995	1\\
0.9995	1\\
0.9995	1\\
0.9995	1\\
};

\end{axis}
\end{tikzpicture}%

%% file: restricted_PP.tikz
\begin{tikzpicture}[baseline]

\begin{axis}[%
width=\fwidth,
height=\fheight,
scale only axis,
xmin=0,
xmax=1,
xlabel={$M_{128} \left(\mathcal D_{128}^2(4,64)\right)$},
ymin=0,
ymax=1,
ylabel={$T_{128} \left(Y,\mathcal D_{128}^2(4,64),\hat \mu_0,\hat \sigma_0\right)$},
axis background/.style={fill=white},
axis x line*=bottom,
axis y line*=left
]

\addplot [color=red, solid]
table[row sep=crcr]{%
	0	0\\
	1	1\\
};

\addplot [color=blue, dashed]
  table[row sep=crcr]{%
0	0\\
0.0005	0\\
0.0005	0\\
0.0005	0\\
0.0005	0\\
0.0005	0\\
0.0005	0\\
0.0005	0\\
0.0005	0.0005\\
0.0005	0.0005\\
0.0005	0.0005\\
0.0005	0.001\\
0.0005	0.001\\
0.0005	0.002\\
0.0005	0.003\\
0.0015	0.003\\
0.0015	0.003\\
0.0015	0.004\\
0.0015	0.0055\\
0.0025	0.0065\\
0.0025	0.0065\\
0.003	0.008\\
0.004	0.008\\
0.007	0.011\\
0.0095	0.015\\
0.012	0.0185\\
0.0135	0.02\\
0.0145	0.023\\
0.019	0.023\\
0.02	0.027\\
0.025	0.031\\
0.027	0.0325\\
0.0345	0.035\\
0.038	0.04\\
0.043	0.0475\\
0.0445	0.054\\
0.05	0.0625\\
0.054	0.0695\\
0.059	0.0745\\
0.064	0.0805\\
0.07	0.086\\
0.0785	0.093\\
0.0865	0.1005\\
0.0935	0.1075\\
0.1065	0.111\\
0.1125	0.1175\\
0.1245	0.1255\\
0.1345	0.1365\\
0.1465	0.1495\\
0.155	0.1615\\
0.1695	0.1745\\
0.1845	0.1845\\
0.1965	0.193\\
0.2165	0.2075\\
0.226	0.219\\
0.2365	0.234\\
0.249	0.253\\
0.2665	0.263\\
0.2785	0.271\\
0.286	0.286\\
0.2985	0.3025\\
0.3115	0.3225\\
0.3255	0.333\\
0.338	0.3455\\
0.352	0.361\\
0.3645	0.3715\\
0.3785	0.3875\\
0.3895	0.405\\
0.407	0.421\\
0.42	0.442\\
0.4395	0.4575\\
0.4545	0.4715\\
0.4715	0.484\\
0.487	0.4985\\
0.5035	0.5105\\
0.516	0.5245\\
0.53	0.5345\\
0.548	0.546\\
0.5645	0.564\\
0.585	0.575\\
0.596	0.585\\
0.6115	0.5995\\
0.6265	0.6155\\
0.6375	0.6285\\
0.646	0.6385\\
0.659	0.65\\
0.671	0.6605\\
0.686	0.6755\\
0.699	0.6875\\
0.707	0.699\\
0.7195	0.7135\\
0.73	0.7275\\
0.736	0.742\\
0.746	0.7525\\
0.76	0.7615\\
0.7715	0.768\\
0.7815	0.7775\\
0.7885	0.7845\\
0.796	0.7945\\
0.8055	0.8065\\
0.811	0.8185\\
0.814	0.8295\\
0.823	0.841\\
0.8285	0.8475\\
0.8355	0.851\\
0.845	0.857\\
0.8485	0.867\\
0.855	0.872\\
0.86	0.879\\
0.8675	0.887\\
0.8745	0.8915\\
0.8815	0.896\\
0.888	0.898\\
0.894	0.9025\\
0.897	0.9095\\
0.9025	0.9155\\
0.9055	0.919\\
0.9105	0.922\\
0.9125	0.929\\
0.917	0.9325\\
0.921	0.936\\
0.9235	0.937\\
0.93	0.938\\
0.937	0.943\\
0.939	0.944\\
0.942	0.9465\\
0.9435	0.949\\
0.944	0.951\\
0.9485	0.955\\
0.949	0.955\\
0.9515	0.96\\
0.953	0.964\\
0.9545	0.968\\
0.958	0.9705\\
0.96	0.9715\\
0.9605	0.974\\
0.9615	0.974\\
0.9635	0.9765\\
0.9695	0.9775\\
0.9695	0.9785\\
0.9705	0.9805\\
0.9715	0.9805\\
0.973	0.9815\\
0.975	0.9825\\
0.9775	0.984\\
0.9775	0.984\\
0.982	0.985\\
0.9845	0.985\\
0.9855	0.9865\\
0.986	0.987\\
0.986	0.9875\\
0.9865	0.9875\\
0.987	0.9895\\
0.9875	0.9895\\
0.9875	0.99\\
0.988	0.99\\
0.989	0.99\\
0.99	0.991\\
0.9905	0.991\\
0.991	0.991\\
0.992	0.991\\
0.992	0.991\\
0.992	0.992\\
0.992	0.9925\\
0.992	0.9935\\
0.992	0.9935\\
0.992	0.9945\\
0.992	0.9945\\
0.9925	0.995\\
0.993	0.995\\
0.993	0.996\\
0.993	0.996\\
0.995	0.996\\
0.995	0.996\\
0.995	0.996\\
0.995	0.9965\\
0.995	0.9975\\
0.995	0.9975\\
0.996	0.9975\\
0.996	0.9975\\
0.9965	0.9975\\
0.9975	0.9975\\
0.998	0.9985\\
0.998	0.9985\\
0.998	0.9985\\
0.998	0.9985\\
0.998	0.9985\\
0.998	0.9985\\
0.998	0.999\\
0.998	0.999\\
0.998	0.999\\
0.998	0.999\\
0.998	0.999\\
0.999	0.999\\
0.999	0.9995\\
0.9995	0.9995\\
0.9995	0.9995\\
0.9995	0.9995\\
1	0.9995\\
1	0.9995\\
};

\end{axis}
\end{tikzpicture}%

%% file: Poisson_means_sim.tikz
\begin{tikzpicture}[baseline]

\begin{axis}[%
width=\fwidth,
height=\fheight,
scale only axis,
point meta min=1,
point meta max=1.6,
axis on top,
xmin=0.5,
xmax=128.5,
y dir=reverse,
ymin=0.5,
ymax=128.5,
xtick = \empty,
ytick = \empty,
axis background/.style={fill=white},
legend style={legend cell align=left, align=left, draw=white!15!black},
colormap={mymap}{[1pt] rgb(0pt)=(0.2422,0.1504,0.6603); rgb(1pt)=(0.2444,0.1534,0.6728); rgb(2pt)=(0.2464,0.1569,0.6847); rgb(3pt)=(0.2484,0.1607,0.6961); rgb(4pt)=(0.2503,0.1648,0.7071); rgb(5pt)=(0.2522,0.1689,0.7179); rgb(6pt)=(0.254,0.1732,0.7286); rgb(7pt)=(0.2558,0.1773,0.7393); rgb(8pt)=(0.2576,0.1814,0.7501); rgb(9pt)=(0.2594,0.1854,0.761); rgb(11pt)=(0.2628,0.1932,0.7828); rgb(12pt)=(0.2645,0.1972,0.7937); rgb(13pt)=(0.2661,0.2011,0.8043); rgb(14pt)=(0.2676,0.2052,0.8148); rgb(15pt)=(0.2691,0.2094,0.8249); rgb(16pt)=(0.2704,0.2138,0.8346); rgb(17pt)=(0.2717,0.2184,0.8439); rgb(18pt)=(0.2729,0.2231,0.8528); rgb(19pt)=(0.274,0.228,0.8612); rgb(20pt)=(0.2749,0.233,0.8692); rgb(21pt)=(0.2758,0.2382,0.8767); rgb(22pt)=(0.2766,0.2435,0.884); rgb(23pt)=(0.2774,0.2489,0.8908); rgb(24pt)=(0.2781,0.2543,0.8973); rgb(25pt)=(0.2788,0.2598,0.9035); rgb(26pt)=(0.2794,0.2653,0.9094); rgb(27pt)=(0.2798,0.2708,0.915); rgb(28pt)=(0.2802,0.2764,0.9204); rgb(29pt)=(0.2806,0.2819,0.9255); rgb(30pt)=(0.2809,0.2875,0.9305); rgb(31pt)=(0.2811,0.293,0.9352); rgb(32pt)=(0.2813,0.2985,0.9397); rgb(33pt)=(0.2814,0.304,0.9441); rgb(34pt)=(0.2814,0.3095,0.9483); rgb(35pt)=(0.2813,0.315,0.9524); rgb(36pt)=(0.2811,0.3204,0.9563); rgb(37pt)=(0.2809,0.3259,0.96); rgb(38pt)=(0.2807,0.3313,0.9636); rgb(39pt)=(0.2803,0.3367,0.967); rgb(40pt)=(0.2798,0.3421,0.9702); rgb(41pt)=(0.2791,0.3475,0.9733); rgb(42pt)=(0.2784,0.3529,0.9763); rgb(43pt)=(0.2776,0.3583,0.9791); rgb(44pt)=(0.2766,0.3638,0.9817); rgb(45pt)=(0.2754,0.3693,0.984); rgb(46pt)=(0.2741,0.3748,0.9862); rgb(47pt)=(0.2726,0.3804,0.9881); rgb(48pt)=(0.271,0.386,0.9898); rgb(49pt)=(0.2691,0.3916,0.9912); rgb(50pt)=(0.267,0.3973,0.9924); rgb(51pt)=(0.2647,0.403,0.9935); rgb(52pt)=(0.2621,0.4088,0.9946); rgb(53pt)=(0.2591,0.4145,0.9955); rgb(54pt)=(0.2556,0.4203,0.9965); rgb(55pt)=(0.2517,0.4261,0.9974); rgb(56pt)=(0.2473,0.4319,0.9983); rgb(57pt)=(0.2424,0.4378,0.9991); rgb(58pt)=(0.2369,0.4437,0.9996); rgb(59pt)=(0.2311,0.4497,0.9995); rgb(60pt)=(0.225,0.4559,0.9985); rgb(61pt)=(0.2189,0.462,0.9968); rgb(62pt)=(0.2128,0.4682,0.9948); rgb(63pt)=(0.2066,0.4743,0.9926); rgb(64pt)=(0.2006,0.4803,0.9906); rgb(65pt)=(0.195,0.4861,0.9887); rgb(66pt)=(0.1903,0.4919,0.9867); rgb(67pt)=(0.1869,0.4975,0.9844); rgb(68pt)=(0.1847,0.503,0.9819); rgb(69pt)=(0.1831,0.5084,0.9793); rgb(70pt)=(0.1818,0.5138,0.9766); rgb(71pt)=(0.1806,0.5191,0.9738); rgb(72pt)=(0.1795,0.5244,0.9709); rgb(73pt)=(0.1785,0.5296,0.9677); rgb(74pt)=(0.1778,0.5349,0.9641); rgb(75pt)=(0.1773,0.5401,0.9602); rgb(76pt)=(0.1768,0.5452,0.956); rgb(77pt)=(0.1764,0.5504,0.9516); rgb(78pt)=(0.1755,0.5554,0.9473); rgb(79pt)=(0.174,0.5605,0.9432); rgb(80pt)=(0.1716,0.5655,0.9393); rgb(81pt)=(0.1686,0.5705,0.9357); rgb(82pt)=(0.1649,0.5755,0.9323); rgb(83pt)=(0.161,0.5805,0.9289); rgb(84pt)=(0.1573,0.5854,0.9254); rgb(85pt)=(0.154,0.5902,0.9218); rgb(86pt)=(0.1513,0.595,0.9182); rgb(87pt)=(0.1492,0.5997,0.9147); rgb(88pt)=(0.1475,0.6043,0.9113); rgb(89pt)=(0.1461,0.6089,0.908); rgb(90pt)=(0.1446,0.6135,0.905); rgb(91pt)=(0.1429,0.618,0.9022); rgb(92pt)=(0.1408,0.6226,0.8998); rgb(93pt)=(0.1383,0.6272,0.8975); rgb(94pt)=(0.1354,0.6317,0.8953); rgb(95pt)=(0.1321,0.6363,0.8932); rgb(96pt)=(0.1288,0.6408,0.891); rgb(97pt)=(0.1253,0.6453,0.8887); rgb(98pt)=(0.1219,0.6497,0.8862); rgb(99pt)=(0.1185,0.6541,0.8834); rgb(100pt)=(0.1152,0.6584,0.8804); rgb(101pt)=(0.1119,0.6627,0.877); rgb(102pt)=(0.1085,0.6669,0.8734); rgb(103pt)=(0.1048,0.671,0.8695); rgb(104pt)=(0.1009,0.675,0.8653); rgb(105pt)=(0.0964,0.6789,0.8609); rgb(106pt)=(0.0914,0.6828,0.8562); rgb(107pt)=(0.0855,0.6865,0.8513); rgb(108pt)=(0.0789,0.6902,0.8462); rgb(109pt)=(0.0713,0.6938,0.8409); rgb(110pt)=(0.0628,0.6972,0.8355); rgb(111pt)=(0.0535,0.7006,0.8299); rgb(112pt)=(0.0433,0.7039,0.8242); rgb(113pt)=(0.0328,0.7071,0.8183); rgb(114pt)=(0.0234,0.7103,0.8124); rgb(115pt)=(0.0155,0.7133,0.8064); rgb(116pt)=(0.0091,0.7163,0.8003); rgb(117pt)=(0.0046,0.7192,0.7941); rgb(118pt)=(0.0019,0.722,0.7878); rgb(119pt)=(0.0009,0.7248,0.7815); rgb(120pt)=(0.0018,0.7275,0.7752); rgb(121pt)=(0.0046,0.7301,0.7688); rgb(122pt)=(0.0094,0.7327,0.7623); rgb(123pt)=(0.0162,0.7352,0.7558); rgb(124pt)=(0.0253,0.7376,0.7492); rgb(125pt)=(0.0369,0.74,0.7426); rgb(126pt)=(0.0504,0.7423,0.7359); rgb(127pt)=(0.0638,0.7446,0.7292); rgb(128pt)=(0.077,0.7468,0.7224); rgb(129pt)=(0.0899,0.7489,0.7156); rgb(130pt)=(0.1023,0.751,0.7088); rgb(131pt)=(0.1141,0.7531,0.7019); rgb(132pt)=(0.1252,0.7552,0.695); rgb(133pt)=(0.1354,0.7572,0.6881); rgb(134pt)=(0.1448,0.7593,0.6812); rgb(135pt)=(0.1532,0.7614,0.6741); rgb(136pt)=(0.1609,0.7635,0.6671); rgb(137pt)=(0.1678,0.7656,0.6599); rgb(138pt)=(0.1741,0.7678,0.6527); rgb(139pt)=(0.1799,0.7699,0.6454); rgb(140pt)=(0.1853,0.7721,0.6379); rgb(141pt)=(0.1905,0.7743,0.6303); rgb(142pt)=(0.1954,0.7765,0.6225); rgb(143pt)=(0.2003,0.7787,0.6146); rgb(144pt)=(0.2061,0.7808,0.6065); rgb(145pt)=(0.2118,0.7828,0.5983); rgb(146pt)=(0.2178,0.7849,0.5899); rgb(147pt)=(0.2244,0.7869,0.5813); rgb(148pt)=(0.2318,0.7887,0.5725); rgb(149pt)=(0.2401,0.7905,0.5636); rgb(150pt)=(0.2491,0.7922,0.5546); rgb(151pt)=(0.2589,0.7937,0.5454); rgb(152pt)=(0.2695,0.7951,0.536); rgb(153pt)=(0.2809,0.7964,0.5266); rgb(154pt)=(0.2929,0.7975,0.517); rgb(155pt)=(0.3052,0.7985,0.5074); rgb(156pt)=(0.3176,0.7994,0.4975); rgb(157pt)=(0.3301,0.8002,0.4876); rgb(158pt)=(0.3424,0.8009,0.4774); rgb(159pt)=(0.3548,0.8016,0.4669); rgb(160pt)=(0.3671,0.8021,0.4563); rgb(161pt)=(0.3795,0.8026,0.4454); rgb(162pt)=(0.3921,0.8029,0.4344); rgb(163pt)=(0.405,0.8031,0.4233); rgb(164pt)=(0.4184,0.803,0.4122); rgb(165pt)=(0.4322,0.8028,0.4013); rgb(166pt)=(0.4463,0.8024,0.3904); rgb(167pt)=(0.4608,0.8018,0.3797); rgb(168pt)=(0.4753,0.8011,0.3691); rgb(169pt)=(0.4899,0.8002,0.3586); rgb(170pt)=(0.5044,0.7993,0.348); rgb(171pt)=(0.5187,0.7982,0.3374); rgb(172pt)=(0.5329,0.797,0.3267); rgb(173pt)=(0.547,0.7957,0.3159); rgb(175pt)=(0.5748,0.7929,0.2941); rgb(176pt)=(0.5886,0.7913,0.2833); rgb(177pt)=(0.6024,0.7896,0.2726); rgb(178pt)=(0.6161,0.7878,0.2622); rgb(179pt)=(0.6297,0.7859,0.2521); rgb(180pt)=(0.6433,0.7839,0.2423); rgb(181pt)=(0.6567,0.7818,0.2329); rgb(182pt)=(0.6701,0.7796,0.2239); rgb(183pt)=(0.6833,0.7773,0.2155); rgb(184pt)=(0.6963,0.775,0.2075); rgb(185pt)=(0.7091,0.7727,0.1998); rgb(186pt)=(0.7218,0.7703,0.1924); rgb(187pt)=(0.7344,0.7679,0.1852); rgb(188pt)=(0.7468,0.7654,0.1782); rgb(189pt)=(0.759,0.7629,0.1717); rgb(190pt)=(0.771,0.7604,0.1658); rgb(191pt)=(0.7829,0.7579,0.1608); rgb(192pt)=(0.7945,0.7554,0.157); rgb(193pt)=(0.806,0.7529,0.1546); rgb(194pt)=(0.8172,0.7505,0.1535); rgb(195pt)=(0.8281,0.7481,0.1536); rgb(196pt)=(0.8389,0.7457,0.1546); rgb(197pt)=(0.8495,0.7435,0.1564); rgb(198pt)=(0.86,0.7413,0.1587); rgb(199pt)=(0.8703,0.7392,0.1615); rgb(200pt)=(0.8804,0.7372,0.165); rgb(201pt)=(0.8903,0.7353,0.1695); rgb(202pt)=(0.9,0.7336,0.1749); rgb(203pt)=(0.9093,0.7321,0.1815); rgb(204pt)=(0.9184,0.7308,0.189); rgb(205pt)=(0.9272,0.7298,0.1973); rgb(206pt)=(0.9357,0.729,0.2061); rgb(207pt)=(0.944,0.7285,0.2151); rgb(208pt)=(0.9523,0.7284,0.2237); rgb(209pt)=(0.9606,0.7285,0.2312); rgb(210pt)=(0.9689,0.7292,0.2373); rgb(211pt)=(0.977,0.7304,0.2418); rgb(212pt)=(0.9842,0.733,0.2446); rgb(213pt)=(0.99,0.7365,0.2429); rgb(214pt)=(0.9946,0.7407,0.2394); rgb(215pt)=(0.9966,0.7458,0.2351); rgb(216pt)=(0.9971,0.7513,0.2309); rgb(217pt)=(0.9972,0.7569,0.2267); rgb(218pt)=(0.9971,0.7626,0.2224); rgb(219pt)=(0.9969,0.7683,0.2181); rgb(220pt)=(0.9966,0.774,0.2138); rgb(221pt)=(0.9962,0.7798,0.2095); rgb(222pt)=(0.9957,0.7856,0.2053); rgb(223pt)=(0.9949,0.7915,0.2012); rgb(224pt)=(0.9938,0.7974,0.1974); rgb(225pt)=(0.9923,0.8034,0.1939); rgb(226pt)=(0.9906,0.8095,0.1906); rgb(227pt)=(0.9885,0.8156,0.1875); rgb(228pt)=(0.9861,0.8218,0.1846); rgb(229pt)=(0.9835,0.828,0.1817); rgb(230pt)=(0.9807,0.8342,0.1787); rgb(231pt)=(0.9778,0.8404,0.1757); rgb(232pt)=(0.9748,0.8467,0.1726); rgb(233pt)=(0.972,0.8529,0.1695); rgb(234pt)=(0.9694,0.8591,0.1665); rgb(235pt)=(0.9671,0.8654,0.1636); rgb(236pt)=(0.9651,0.8716,0.1608); rgb(237pt)=(0.9634,0.8778,0.1582); rgb(238pt)=(0.9619,0.884,0.1557); rgb(239pt)=(0.9608,0.8902,0.1532); rgb(240pt)=(0.9601,0.8963,0.1507); rgb(241pt)=(0.9596,0.9023,0.148); rgb(242pt)=(0.9595,0.9084,0.145); rgb(243pt)=(0.9597,0.9143,0.1418); rgb(244pt)=(0.9601,0.9203,0.1382); rgb(245pt)=(0.9608,0.9262,0.1344); rgb(246pt)=(0.9618,0.932,0.1304); rgb(247pt)=(0.9629,0.9379,0.1261); rgb(248pt)=(0.9642,0.9437,0.1216); rgb(249pt)=(0.9657,0.9494,0.1168); rgb(250pt)=(0.9674,0.9552,0.1116); rgb(251pt)=(0.9692,0.9609,0.1061); rgb(252pt)=(0.9711,0.9667,0.1001); rgb(253pt)=(0.973,0.9724,0.0938); rgb(254pt)=(0.9749,0.9782,0.0872); rgb(255pt)=(0.9769,0.9839,0.0805)},
colorbar
]
\addplot [forget plot] graphics [xmin=0.5, xmax=128.5, ymin=0.5, ymax=128.5] {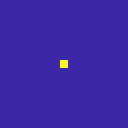};
\end{axis}
\end{tikzpicture}%

%% file: Poisson_data_sim.tikz
\begin{tikzpicture}[baseline]

\begin{axis}[%
width=\fwidth,
height=\fheight,
scale only axis,
point meta min=0,
point meta max=7,
axis on top,
xmin=0.5,
xmax=128.5,
y dir=reverse,
ymin=0.5,
ymax=128.5,
xtick = \empty,
ytick = \empty,
axis background/.style={fill=white},
legend style={legend cell align=left, align=left, draw=white!15!black},
colormap={mymap}{[1pt] rgb(0pt)=(0.2422,0.1504,0.6603); rgb(1pt)=(0.2444,0.1534,0.6728); rgb(2pt)=(0.2464,0.1569,0.6847); rgb(3pt)=(0.2484,0.1607,0.6961); rgb(4pt)=(0.2503,0.1648,0.7071); rgb(5pt)=(0.2522,0.1689,0.7179); rgb(6pt)=(0.254,0.1732,0.7286); rgb(7pt)=(0.2558,0.1773,0.7393); rgb(8pt)=(0.2576,0.1814,0.7501); rgb(9pt)=(0.2594,0.1854,0.761); rgb(11pt)=(0.2628,0.1932,0.7828); rgb(12pt)=(0.2645,0.1972,0.7937); rgb(13pt)=(0.2661,0.2011,0.8043); rgb(14pt)=(0.2676,0.2052,0.8148); rgb(15pt)=(0.2691,0.2094,0.8249); rgb(16pt)=(0.2704,0.2138,0.8346); rgb(17pt)=(0.2717,0.2184,0.8439); rgb(18pt)=(0.2729,0.2231,0.8528); rgb(19pt)=(0.274,0.228,0.8612); rgb(20pt)=(0.2749,0.233,0.8692); rgb(21pt)=(0.2758,0.2382,0.8767); rgb(22pt)=(0.2766,0.2435,0.884); rgb(23pt)=(0.2774,0.2489,0.8908); rgb(24pt)=(0.2781,0.2543,0.8973); rgb(25pt)=(0.2788,0.2598,0.9035); rgb(26pt)=(0.2794,0.2653,0.9094); rgb(27pt)=(0.2798,0.2708,0.915); rgb(28pt)=(0.2802,0.2764,0.9204); rgb(29pt)=(0.2806,0.2819,0.9255); rgb(30pt)=(0.2809,0.2875,0.9305); rgb(31pt)=(0.2811,0.293,0.9352); rgb(32pt)=(0.2813,0.2985,0.9397); rgb(33pt)=(0.2814,0.304,0.9441); rgb(34pt)=(0.2814,0.3095,0.9483); rgb(35pt)=(0.2813,0.315,0.9524); rgb(36pt)=(0.2811,0.3204,0.9563); rgb(37pt)=(0.2809,0.3259,0.96); rgb(38pt)=(0.2807,0.3313,0.9636); rgb(39pt)=(0.2803,0.3367,0.967); rgb(40pt)=(0.2798,0.3421,0.9702); rgb(41pt)=(0.2791,0.3475,0.9733); rgb(42pt)=(0.2784,0.3529,0.9763); rgb(43pt)=(0.2776,0.3583,0.9791); rgb(44pt)=(0.2766,0.3638,0.9817); rgb(45pt)=(0.2754,0.3693,0.984); rgb(46pt)=(0.2741,0.3748,0.9862); rgb(47pt)=(0.2726,0.3804,0.9881); rgb(48pt)=(0.271,0.386,0.9898); rgb(49pt)=(0.2691,0.3916,0.9912); rgb(50pt)=(0.267,0.3973,0.9924); rgb(51pt)=(0.2647,0.403,0.9935); rgb(52pt)=(0.2621,0.4088,0.9946); rgb(53pt)=(0.2591,0.4145,0.9955); rgb(54pt)=(0.2556,0.4203,0.9965); rgb(55pt)=(0.2517,0.4261,0.9974); rgb(56pt)=(0.2473,0.4319,0.9983); rgb(57pt)=(0.2424,0.4378,0.9991); rgb(58pt)=(0.2369,0.4437,0.9996); rgb(59pt)=(0.2311,0.4497,0.9995); rgb(60pt)=(0.225,0.4559,0.9985); rgb(61pt)=(0.2189,0.462,0.9968); rgb(62pt)=(0.2128,0.4682,0.9948); rgb(63pt)=(0.2066,0.4743,0.9926); rgb(64pt)=(0.2006,0.4803,0.9906); rgb(65pt)=(0.195,0.4861,0.9887); rgb(66pt)=(0.1903,0.4919,0.9867); rgb(67pt)=(0.1869,0.4975,0.9844); rgb(68pt)=(0.1847,0.503,0.9819); rgb(69pt)=(0.1831,0.5084,0.9793); rgb(70pt)=(0.1818,0.5138,0.9766); rgb(71pt)=(0.1806,0.5191,0.9738); rgb(72pt)=(0.1795,0.5244,0.9709); rgb(73pt)=(0.1785,0.5296,0.9677); rgb(74pt)=(0.1778,0.5349,0.9641); rgb(75pt)=(0.1773,0.5401,0.9602); rgb(76pt)=(0.1768,0.5452,0.956); rgb(77pt)=(0.1764,0.5504,0.9516); rgb(78pt)=(0.1755,0.5554,0.9473); rgb(79pt)=(0.174,0.5605,0.9432); rgb(80pt)=(0.1716,0.5655,0.9393); rgb(81pt)=(0.1686,0.5705,0.9357); rgb(82pt)=(0.1649,0.5755,0.9323); rgb(83pt)=(0.161,0.5805,0.9289); rgb(84pt)=(0.1573,0.5854,0.9254); rgb(85pt)=(0.154,0.5902,0.9218); rgb(86pt)=(0.1513,0.595,0.9182); rgb(87pt)=(0.1492,0.5997,0.9147); rgb(88pt)=(0.1475,0.6043,0.9113); rgb(89pt)=(0.1461,0.6089,0.908); rgb(90pt)=(0.1446,0.6135,0.905); rgb(91pt)=(0.1429,0.618,0.9022); rgb(92pt)=(0.1408,0.6226,0.8998); rgb(93pt)=(0.1383,0.6272,0.8975); rgb(94pt)=(0.1354,0.6317,0.8953); rgb(95pt)=(0.1321,0.6363,0.8932); rgb(96pt)=(0.1288,0.6408,0.891); rgb(97pt)=(0.1253,0.6453,0.8887); rgb(98pt)=(0.1219,0.6497,0.8862); rgb(99pt)=(0.1185,0.6541,0.8834); rgb(100pt)=(0.1152,0.6584,0.8804); rgb(101pt)=(0.1119,0.6627,0.877); rgb(102pt)=(0.1085,0.6669,0.8734); rgb(103pt)=(0.1048,0.671,0.8695); rgb(104pt)=(0.1009,0.675,0.8653); rgb(105pt)=(0.0964,0.6789,0.8609); rgb(106pt)=(0.0914,0.6828,0.8562); rgb(107pt)=(0.0855,0.6865,0.8513); rgb(108pt)=(0.0789,0.6902,0.8462); rgb(109pt)=(0.0713,0.6938,0.8409); rgb(110pt)=(0.0628,0.6972,0.8355); rgb(111pt)=(0.0535,0.7006,0.8299); rgb(112pt)=(0.0433,0.7039,0.8242); rgb(113pt)=(0.0328,0.7071,0.8183); rgb(114pt)=(0.0234,0.7103,0.8124); rgb(115pt)=(0.0155,0.7133,0.8064); rgb(116pt)=(0.0091,0.7163,0.8003); rgb(117pt)=(0.0046,0.7192,0.7941); rgb(118pt)=(0.0019,0.722,0.7878); rgb(119pt)=(0.0009,0.7248,0.7815); rgb(120pt)=(0.0018,0.7275,0.7752); rgb(121pt)=(0.0046,0.7301,0.7688); rgb(122pt)=(0.0094,0.7327,0.7623); rgb(123pt)=(0.0162,0.7352,0.7558); rgb(124pt)=(0.0253,0.7376,0.7492); rgb(125pt)=(0.0369,0.74,0.7426); rgb(126pt)=(0.0504,0.7423,0.7359); rgb(127pt)=(0.0638,0.7446,0.7292); rgb(128pt)=(0.077,0.7468,0.7224); rgb(129pt)=(0.0899,0.7489,0.7156); rgb(130pt)=(0.1023,0.751,0.7088); rgb(131pt)=(0.1141,0.7531,0.7019); rgb(132pt)=(0.1252,0.7552,0.695); rgb(133pt)=(0.1354,0.7572,0.6881); rgb(134pt)=(0.1448,0.7593,0.6812); rgb(135pt)=(0.1532,0.7614,0.6741); rgb(136pt)=(0.1609,0.7635,0.6671); rgb(137pt)=(0.1678,0.7656,0.6599); rgb(138pt)=(0.1741,0.7678,0.6527); rgb(139pt)=(0.1799,0.7699,0.6454); rgb(140pt)=(0.1853,0.7721,0.6379); rgb(141pt)=(0.1905,0.7743,0.6303); rgb(142pt)=(0.1954,0.7765,0.6225); rgb(143pt)=(0.2003,0.7787,0.6146); rgb(144pt)=(0.2061,0.7808,0.6065); rgb(145pt)=(0.2118,0.7828,0.5983); rgb(146pt)=(0.2178,0.7849,0.5899); rgb(147pt)=(0.2244,0.7869,0.5813); rgb(148pt)=(0.2318,0.7887,0.5725); rgb(149pt)=(0.2401,0.7905,0.5636); rgb(150pt)=(0.2491,0.7922,0.5546); rgb(151pt)=(0.2589,0.7937,0.5454); rgb(152pt)=(0.2695,0.7951,0.536); rgb(153pt)=(0.2809,0.7964,0.5266); rgb(154pt)=(0.2929,0.7975,0.517); rgb(155pt)=(0.3052,0.7985,0.5074); rgb(156pt)=(0.3176,0.7994,0.4975); rgb(157pt)=(0.3301,0.8002,0.4876); rgb(158pt)=(0.3424,0.8009,0.4774); rgb(159pt)=(0.3548,0.8016,0.4669); rgb(160pt)=(0.3671,0.8021,0.4563); rgb(161pt)=(0.3795,0.8026,0.4454); rgb(162pt)=(0.3921,0.8029,0.4344); rgb(163pt)=(0.405,0.8031,0.4233); rgb(164pt)=(0.4184,0.803,0.4122); rgb(165pt)=(0.4322,0.8028,0.4013); rgb(166pt)=(0.4463,0.8024,0.3904); rgb(167pt)=(0.4608,0.8018,0.3797); rgb(168pt)=(0.4753,0.8011,0.3691); rgb(169pt)=(0.4899,0.8002,0.3586); rgb(170pt)=(0.5044,0.7993,0.348); rgb(171pt)=(0.5187,0.7982,0.3374); rgb(172pt)=(0.5329,0.797,0.3267); rgb(173pt)=(0.547,0.7957,0.3159); rgb(175pt)=(0.5748,0.7929,0.2941); rgb(176pt)=(0.5886,0.7913,0.2833); rgb(177pt)=(0.6024,0.7896,0.2726); rgb(178pt)=(0.6161,0.7878,0.2622); rgb(179pt)=(0.6297,0.7859,0.2521); rgb(180pt)=(0.6433,0.7839,0.2423); rgb(181pt)=(0.6567,0.7818,0.2329); rgb(182pt)=(0.6701,0.7796,0.2239); rgb(183pt)=(0.6833,0.7773,0.2155); rgb(184pt)=(0.6963,0.775,0.2075); rgb(185pt)=(0.7091,0.7727,0.1998); rgb(186pt)=(0.7218,0.7703,0.1924); rgb(187pt)=(0.7344,0.7679,0.1852); rgb(188pt)=(0.7468,0.7654,0.1782); rgb(189pt)=(0.759,0.7629,0.1717); rgb(190pt)=(0.771,0.7604,0.1658); rgb(191pt)=(0.7829,0.7579,0.1608); rgb(192pt)=(0.7945,0.7554,0.157); rgb(193pt)=(0.806,0.7529,0.1546); rgb(194pt)=(0.8172,0.7505,0.1535); rgb(195pt)=(0.8281,0.7481,0.1536); rgb(196pt)=(0.8389,0.7457,0.1546); rgb(197pt)=(0.8495,0.7435,0.1564); rgb(198pt)=(0.86,0.7413,0.1587); rgb(199pt)=(0.8703,0.7392,0.1615); rgb(200pt)=(0.8804,0.7372,0.165); rgb(201pt)=(0.8903,0.7353,0.1695); rgb(202pt)=(0.9,0.7336,0.1749); rgb(203pt)=(0.9093,0.7321,0.1815); rgb(204pt)=(0.9184,0.7308,0.189); rgb(205pt)=(0.9272,0.7298,0.1973); rgb(206pt)=(0.9357,0.729,0.2061); rgb(207pt)=(0.944,0.7285,0.2151); rgb(208pt)=(0.9523,0.7284,0.2237); rgb(209pt)=(0.9606,0.7285,0.2312); rgb(210pt)=(0.9689,0.7292,0.2373); rgb(211pt)=(0.977,0.7304,0.2418); rgb(212pt)=(0.9842,0.733,0.2446); rgb(213pt)=(0.99,0.7365,0.2429); rgb(214pt)=(0.9946,0.7407,0.2394); rgb(215pt)=(0.9966,0.7458,0.2351); rgb(216pt)=(0.9971,0.7513,0.2309); rgb(217pt)=(0.9972,0.7569,0.2267); rgb(218pt)=(0.9971,0.7626,0.2224); rgb(219pt)=(0.9969,0.7683,0.2181); rgb(220pt)=(0.9966,0.774,0.2138); rgb(221pt)=(0.9962,0.7798,0.2095); rgb(222pt)=(0.9957,0.7856,0.2053); rgb(223pt)=(0.9949,0.7915,0.2012); rgb(224pt)=(0.9938,0.7974,0.1974); rgb(225pt)=(0.9923,0.8034,0.1939); rgb(226pt)=(0.9906,0.8095,0.1906); rgb(227pt)=(0.9885,0.8156,0.1875); rgb(228pt)=(0.9861,0.8218,0.1846); rgb(229pt)=(0.9835,0.828,0.1817); rgb(230pt)=(0.9807,0.8342,0.1787); rgb(231pt)=(0.9778,0.8404,0.1757); rgb(232pt)=(0.9748,0.8467,0.1726); rgb(233pt)=(0.972,0.8529,0.1695); rgb(234pt)=(0.9694,0.8591,0.1665); rgb(235pt)=(0.9671,0.8654,0.1636); rgb(236pt)=(0.9651,0.8716,0.1608); rgb(237pt)=(0.9634,0.8778,0.1582); rgb(238pt)=(0.9619,0.884,0.1557); rgb(239pt)=(0.9608,0.8902,0.1532); rgb(240pt)=(0.9601,0.8963,0.1507); rgb(241pt)=(0.9596,0.9023,0.148); rgb(242pt)=(0.9595,0.9084,0.145); rgb(243pt)=(0.9597,0.9143,0.1418); rgb(244pt)=(0.9601,0.9203,0.1382); rgb(245pt)=(0.9608,0.9262,0.1344); rgb(246pt)=(0.9618,0.932,0.1304); rgb(247pt)=(0.9629,0.9379,0.1261); rgb(248pt)=(0.9642,0.9437,0.1216); rgb(249pt)=(0.9657,0.9494,0.1168); rgb(250pt)=(0.9674,0.9552,0.1116); rgb(251pt)=(0.9692,0.9609,0.1061); rgb(252pt)=(0.9711,0.9667,0.1001); rgb(253pt)=(0.973,0.9724,0.0938); rgb(254pt)=(0.9749,0.9782,0.0872); rgb(255pt)=(0.9769,0.9839,0.0805)},
colorbar
]
\addplot [forget plot] graphics [xmin=0.5, xmax=128.5, ymin=0.5, ymax=128.5] {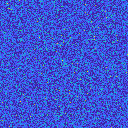};
\end{axis}
\end{tikzpicture}%

%% file: poisson_level.tikz
\begin{tikzpicture}[baseline]
	
	\begin{axis}[%
		width=\fwidth,
		height=\fheight,
		scale only axis,
		xmin=0,
		xmax=1.5,
		ylabel = {Level},
		ymin=0,
		ymax=0.3,
		legend pos = north east,
		y tick label style={
			/pgf/number format/fixed,
			/pgf/number format/precision=2
		}
		]
		\addplot [color=red,solid]
		table[row sep=crcr]{%
			0	0\\
			0.01	0.0123333333333333\\
			0.02	0.018\\
			0.03	0.0254285714285714\\
			0.04	0.049\\
			0.05	0.0578888888888889\\
			0.06	0.0682222222222222\\
			0.07	0.079\\
			0.08	0.0888888888888889\\
			0.09	0.0973333333333333\\
			0.1	0.106222222222222\\
			0.11	0.117777777777778\\
			0.12	0.121777777777778\\
			0.13	0.119444444444444\\
			0.14	0.125444444444444\\
			0.15	0.126777777777778\\
			0.16	0.128\\
			0.17	0.128888888888889\\
			0.18	0.134222222222222\\
			0.19	0.139333333333333\\
			0.2	0.135777777777778\\
			0.21	0.134777777777778\\
			0.22	0.129888888888889\\
			0.23	0.132222222222222\\
			0.24	0.137222222222222\\
			0.25	0.138222222222222\\
			0.26	0.138444444444444\\
			0.27	0.134777777777778\\
			0.28	0.133\\
			0.29	0.133222222222222\\
			0.3	0.135111111111111\\
			0.31	0.139222222222222\\
			0.32	0.135111111111111\\
			0.33	0.132666666666667\\
			0.34	0.128666666666667\\
			0.35	0.128555555555556\\
			0.36	0.127777777777778\\
			0.37	0.127333333333333\\
			0.38	0.127444444444444\\
			0.39	0.126444444444444\\
			0.4	0.123888888888889\\
			0.41	0.122777777777778\\
			0.42	0.124\\
			0.43	0.125111111111111\\
			0.44	0.123777777777778\\
			0.45	0.123666666666667\\
			0.46	0.122111111111111\\
			0.47	0.118555555555556\\
			0.48	0.113666666666667\\
			0.49	0.112444444444444\\
			0.5	0.114777777777778\\
			0.51	0.109666666666667\\
			0.52	0.11\\
			0.53	0.109888888888889\\
			0.54	0.111555555555556\\
			0.55	0.112888888888889\\
			0.56	0.113555555555556\\
			0.57	0.115888888888889\\
			0.58	0.116777777777778\\
			0.59	0.114777777777778\\
			0.6	0.115222222222222\\
			0.61	0.112222222222222\\
			0.62	0.111777777777778\\
			0.63	0.107555555555556\\
			0.64	0.104111111111111\\
			0.65	0.103888888888889\\
			0.66	0.103777777777778\\
			0.67	0.104333333333333\\
			0.68	0.104555555555556\\
			0.69	0.102888888888889\\
			0.7	0.104888888888889\\
			0.71	0.106555555555556\\
			0.72	0.109333333333333\\
			0.73	0.110555555555556\\
			0.74	0.109888888888889\\
			0.75	0.109444444444444\\
			0.76	0.111111111111111\\
			0.77	0.110333333333333\\
			0.78	0.111777777777778\\
			0.79	0.110555555555556\\
			0.8	0.108111111111111\\
			0.81	0.106666666666667\\
			0.82	0.106444444444444\\
			0.83	0.108666666666667\\
			0.84	0.110666666666667\\
			0.85	0.108555555555556\\
			0.86	0.106777777777778\\
			0.87	0.107333333333333\\
			0.88	0.108\\
			0.89	0.108\\
			0.9	0.106666666666667\\
			0.91	0.106777777777778\\
			0.92	0.105666666666667\\
			0.93	0.103666666666667\\
			0.94	0.102777777777778\\
			0.95	0.105222222222222\\
			0.96	0.104666666666667\\
			0.97	0.105111111111111\\
			0.98	0.104222222222222\\
			0.99	0.106444444444444\\
			1	0.108555555555556\\
			1.01	0.107666666666667\\
			1.02	0.109666666666667\\
			1.03	0.110555555555556\\
			1.04	0.109777777777778\\
			1.05	0.110111111111111\\
			1.06	0.111444444444444\\
			1.07	0.114444444444444\\
			1.08	0.114\\
			1.09	0.114111111111111\\
			1.1	0.112888888888889\\
			1.11	0.111111111111111\\
			1.12	0.109222222222222\\
			1.13	0.109444444444444\\
			1.14	0.109333333333333\\
			1.15	0.107\\
			1.16	0.105\\
			1.17	0.104222222222222\\
			1.18	0.103\\
			1.19	0.103333333333333\\
			1.2	0.103777777777778\\
			1.21	0.105666666666667\\
			1.22	0.105666666666667\\
			1.23	0.105222222222222\\
			1.24	0.105111111111111\\
			1.25	0.105222222222222\\
			1.26	0.105444444444444\\
			1.27	0.106111111111111\\
			1.28	0.105444444444444\\
			1.29	0.102666666666667\\
			1.3	0.101333333333333\\
			1.31	0.100111111111111\\
			1.32	0.0995555555555555\\
			1.33	0.100333333333333\\
			1.34	0.0986666666666667\\
			1.35	0.101\\
			1.36	0.0982222222222222\\
			1.37	0.0987777777777778\\
			1.38	0.100333333333333\\
			1.39	0.100444444444444\\
			1.4	0.101888888888889\\
			1.41	0.101777777777778\\
			1.42	0.101888888888889\\
			1.43	0.103111111111111\\
			1.44	0.101555555555556\\
			1.45	0.101777777777778\\
			1.46	0.102777777777778\\
			1.47	0.103428571428571\\
			1.48	0.1036\\
			1.49	0.104\\
			1.5	0.108\\
		};
		\addlegendentry{Oracle}
		\addplot [color=black,solid]
		table[row sep=crcr]{%
			0	0\\
			0.01	0.012\\
			0.02	0.0162\\
			0.03	0.0224285714285714\\
			0.04	0.0367777777777778\\
			0.05	0.0455555555555556\\
			0.06	0.0584444444444444\\
			0.07	0.0675555555555555\\
			0.08	0.0784444444444444\\
			0.09	0.0888888888888889\\
			0.1	0.101222222222222\\
			0.11	0.109111111111111\\
			0.12	0.114222222222222\\
			0.13	0.116666666666667\\
			0.14	0.119666666666667\\
			0.15	0.119\\
			0.16	0.120666666666667\\
			0.17	0.121\\
			0.18	0.123222222222222\\
			0.19	0.125666666666667\\
			0.2	0.124333333333333\\
			0.21	0.123444444444444\\
			0.22	0.126\\
			0.23	0.132333333333333\\
			0.24	0.133111111111111\\
			0.25	0.135555555555556\\
			0.26	0.136333333333333\\
			0.27	0.133444444444444\\
			0.28	0.130333333333333\\
			0.29	0.131\\
			0.3	0.134\\
			0.31	0.132666666666667\\
			0.32	0.127333333333333\\
			0.33	0.126222222222222\\
			0.34	0.121444444444444\\
			0.35	0.118666666666667\\
			0.36	0.118222222222222\\
			0.37	0.118333333333333\\
			0.38	0.119555555555556\\
			0.39	0.116\\
			0.4	0.114888888888889\\
			0.41	0.114444444444444\\
			0.42	0.114444444444444\\
			0.43	0.115666666666667\\
			0.44	0.116222222222222\\
			0.45	0.115\\
			0.46	0.112777777777778\\
			0.47	0.109777777777778\\
			0.48	0.109111111111111\\
			0.49	0.108444444444444\\
			0.5	0.108\\
			0.51	0.104777777777778\\
			0.52	0.105111111111111\\
			0.53	0.105777777777778\\
			0.54	0.108444444444444\\
			0.55	0.108777777777778\\
			0.56	0.109222222222222\\
			0.57	0.109222222222222\\
			0.58	0.109333333333333\\
			0.59	0.110777777777778\\
			0.6	0.114111111111111\\
			0.61	0.115111111111111\\
			0.62	0.113111111111111\\
			0.63	0.113111111111111\\
			0.64	0.113444444444444\\
			0.65	0.113333333333333\\
			0.66	0.112666666666667\\
			0.67	0.111333333333333\\
			0.68	0.109111111111111\\
			0.69	0.107\\
			0.7	0.106555555555556\\
			0.71	0.108222222222222\\
			0.72	0.107111111111111\\
			0.73	0.105555555555556\\
			0.74	0.106111111111111\\
			0.75	0.103333333333333\\
			0.76	0.102666666666667\\
			0.77	0.101777777777778\\
			0.78	0.102222222222222\\
			0.79	0.101111111111111\\
			0.8	0.0988888888888889\\
			0.81	0.0998888888888889\\
			0.82	0.102\\
			0.83	0.101333333333333\\
			0.84	0.106444444444444\\
			0.85	0.108333333333333\\
			0.86	0.113444444444444\\
			0.87	0.111666666666667\\
			0.88	0.111777777777778\\
			0.89	0.111888888888889\\
			0.9	0.112555555555556\\
			0.91	0.112777777777778\\
			0.92	0.112333333333333\\
			0.93	0.112888888888889\\
			0.94	0.113333333333333\\
			0.95	0.111333333333333\\
			0.96	0.109222222222222\\
			0.97	0.108555555555556\\
			0.98	0.11\\
			0.99	0.108666666666667\\
			1	0.106111111111111\\
			1.01	0.104888888888889\\
			1.02	0.100555555555556\\
			1.03	0.101333333333333\\
			1.04	0.097\\
			1.05	0.0996666666666667\\
			1.06	0.1\\
			1.07	0.098\\
			1.08	0.0958888888888889\\
			1.09	0.0975555555555556\\
			1.1	0.0975555555555556\\
			1.11	0.100666666666667\\
			1.12	0.0968888888888889\\
			1.13	0.0992222222222222\\
			1.14	0.0983333333333333\\
			1.15	0.0973333333333333\\
			1.16	0.0972222222222222\\
			1.17	0.0982222222222222\\
			1.18	0.0956666666666667\\
			1.19	0.0956666666666667\\
			1.2	0.095\\
			1.21	0.0961111111111111\\
			1.22	0.0957777777777778\\
			1.23	0.0965555555555555\\
			1.24	0.098\\
			1.25	0.0991111111111111\\
			1.26	0.0974444444444444\\
			1.27	0.0995555555555555\\
			1.28	0.102222222222222\\
			1.29	0.100333333333333\\
			1.3	0.101777777777778\\
			1.31	0.101555555555556\\
			1.32	0.100222222222222\\
			1.33	0.101\\
			1.34	0.100111111111111\\
			1.35	0.0983333333333333\\
			1.36	0.0977777777777778\\
			1.37	0.0946666666666667\\
			1.38	0.0943333333333333\\
			1.39	0.0945555555555556\\
			1.4	0.0954444444444444\\
			1.41	0.0987777777777778\\
			1.42	0.0981111111111111\\
			1.43	0.099\\
			1.44	0.101888888888889\\
			1.45	0.102111111111111\\
			1.46	0.105111111111111\\
			1.47	0.106142857142857\\
			1.48	0.1042\\
			1.49	0.104\\
			1.5	0.116\\
		};
		\addlegendentry{AMS}
		
		\addplot [color=blue,solid]
		table[row sep=crcr]{%
			0	0\\
			0.03	0.06\\
			0.06	0.064\\
			0.09	0.0642857142857143\\
			0.12	0.0688888888888889\\
			0.15	0.0744444444444444\\
			0.18	0.0688888888888889\\
			0.21	0.0711111111111111\\
			0.24	0.0766666666666667\\
			0.27	0.07\\
			0.3	0.0711111111111111\\
			0.33	0.0788888888888889\\
			0.36	0.08\\
			0.39	0.0766666666666667\\
			0.42	0.08\\
			0.45	0.0866666666666667\\
			0.48	0.0844444444444444\\
			0.51	0.0855555555555556\\
			0.54	0.0866666666666667\\
			0.57	0.0922222222222222\\
			0.6	0.0833333333333333\\
			0.63	0.0877777777777778\\
			0.66	0.0933333333333333\\
			0.69	0.0888888888888889\\
			0.72	0.0833333333333333\\
			0.75	0.0966666666666667\\
			0.78	0.1\\
			0.81	0.107777777777778\\
			0.84	0.111111111111111\\
			0.87	0.113333333333333\\
			0.9	0.108888888888889\\
			0.93	0.107777777777778\\
			0.96	0.114444444444444\\
			0.99	0.116666666666667\\
			1.02	0.107777777777778\\
			1.05	0.104444444444444\\
			1.08	0.102222222222222\\
			1.11	0.101111111111111\\
			1.14	0.0955555555555555\\
			1.17	0.0966666666666667\\
			1.2	0.0988888888888889\\
			1.23	0.104444444444444\\
			1.26	0.107777777777778\\
			1.29	0.101111111111111\\
			1.32	0.113333333333333\\
			1.35	0.112222222222222\\
			1.38	0.111111111111111\\
			1.41	0.12\\
			1.44	0.114\\
			1.47	0.133333333333333\\
			1.5	0.1\\
		};
		\addlegendentry{Bootstrap}
		
		\addplot [color=green,solid]
		table[row sep=crcr]{%
			0	0.1\\
			1	0.1\\
		};
	\end{axis}
\end{tikzpicture}%

%% file: poisson_power.tikz
\begin{tikzpicture}[baseline]
	
	\begin{axis}[%
		width=\fwidth,
		height=\fheight,
		scale only axis,
		xmin=1,
		xmax=2.5,
		xlabel={$\lambda$},
		ylabel ={Power},
		ymin=0,
		ymax=1,
		legend pos = south east
		]
		\addplot [color=red,solid]
		table[row sep=crcr]{%
			1	0.114\\
			1.01	0.115\\
			1.02	0.1044\\
			1.03	0.102857142857143\\
			1.04	0.104111111111111\\
			1.05	0.104777777777778\\
			1.06	0.102777777777778\\
			1.07	0.104111111111111\\
			1.08	0.105222222222222\\
			1.09	0.106888888888889\\
			1.1	0.105666666666667\\
			1.11	0.108777777777778\\
			1.12	0.108777777777778\\
			1.13	0.109888888888889\\
			1.14	0.107777777777778\\
			1.15	0.109555555555556\\
			1.16	0.108444444444444\\
			1.17	0.111222222222222\\
			1.18	0.113777777777778\\
			1.19	0.116111111111111\\
			1.2	0.118333333333333\\
			1.21	0.120888888888889\\
			1.22	0.120444444444444\\
			1.23	0.123555555555556\\
			1.24	0.125333333333333\\
			1.25	0.127888888888889\\
			1.26	0.128666666666667\\
			1.27	0.130444444444444\\
			1.28	0.135\\
			1.29	0.137333333333333\\
			1.3	0.139444444444444\\
			1.31	0.146\\
			1.32	0.15\\
			1.33	0.154222222222222\\
			1.34	0.161\\
			1.35	0.169888888888889\\
			1.36	0.178333333333333\\
			1.37	0.184444444444444\\
			1.38	0.193777777777778\\
			1.39	0.205333333333333\\
			1.4	0.214444444444444\\
			1.41	0.228555555555556\\
			1.42	0.243111111111111\\
			1.43	0.257666666666667\\
			1.44	0.267333333333333\\
			1.45	0.283666666666667\\
			1.46	0.299888888888889\\
			1.47	0.313666666666667\\
			1.48	0.327555555555556\\
			1.49	0.344555555555556\\
			1.5	0.363444444444444\\
			1.51	0.379777777777778\\
			1.52	0.394888888888889\\
			1.53	0.417333333333333\\
			1.54	0.435555555555556\\
			1.55	0.453333333333333\\
			1.56	0.474333333333333\\
			1.57	0.494777777777778\\
			1.58	0.514444444444444\\
			1.59	0.532333333333333\\
			1.6	0.552666666666667\\
			1.61	0.574222222222222\\
			1.62	0.593222222222222\\
			1.63	0.611333333333333\\
			1.64	0.632333333333333\\
			1.65	0.652\\
			1.66	0.671777777777778\\
			1.67	0.694222222222222\\
			1.68	0.712666666666667\\
			1.69	0.730111111111111\\
			1.7	0.745444444444444\\
			1.71	0.761444444444444\\
			1.72	0.776888888888889\\
			1.73	0.791888888888889\\
			1.74	0.805777777777778\\
			1.75	0.82\\
			1.76	0.830666666666667\\
			1.77	0.843111111111111\\
			1.78	0.856333333333333\\
			1.79	0.868444444444444\\
			1.8	0.879444444444444\\
			1.81	0.890555555555555\\
			1.82	0.902111111111111\\
			1.83	0.912222222222222\\
			1.84	0.919777777777778\\
			1.85	0.928333333333333\\
			1.86	0.934555555555555\\
			1.87	0.940111111111111\\
			1.88	0.946666666666667\\
			1.89	0.953333333333333\\
			1.9	0.958333333333333\\
			1.91	0.960444444444444\\
			1.92	0.963888888888889\\
			1.93	0.969\\
			1.94	0.972111111111111\\
			1.95	0.975222222222222\\
			1.96	0.978\\
			1.97	0.980444444444444\\
			1.98	0.982111111111111\\
			1.99	0.983555555555556\\
			2	0.986111111111111\\
			2.01	0.988555555555556\\
			2.02	0.989222222222222\\
			2.03	0.990333333333333\\
			2.04	0.991666666666667\\
			2.05	0.993222222222222\\
			2.06	0.994222222222222\\
			2.07	0.995111111111111\\
			2.08	0.996222222222222\\
			2.09	0.996666666666667\\
			2.1	0.997\\
			2.11	0.998\\
			2.12	0.998111111111111\\
			2.13	0.998333333333333\\
			2.14	0.998444444444444\\
			2.15	0.998666666666666\\
			2.16	0.999\\
			2.17	0.999\\
			2.18	0.999\\
			2.19	0.999111111111111\\
			2.2	0.999111111111111\\
		};
		\addlegendentry{Oracle}
		\addplot [color=black,solid]
		table[row sep=crcr]{%
			1	0.087\\
			1.01	0.097\\
			1.02	0.0994\\
			1.03	0.101142857142857\\
			1.04	0.102333333333333\\
			1.05	0.106555555555556\\
			1.06	0.105111111111111\\
			1.07	0.105333333333333\\
			1.08	0.104555555555556\\
			1.09	0.101888888888889\\
			1.1	0.102222222222222\\
			1.11	0.102666666666667\\
			1.12	0.103666666666667\\
			1.13	0.101666666666667\\
			1.14	0.101\\
			1.15	0.104333333333333\\
			1.16	0.105555555555556\\
			1.17	0.108222222222222\\
			1.18	0.111\\
			1.19	0.112\\
			1.2	0.114555555555556\\
			1.21	0.114777777777778\\
			1.22	0.118\\
			1.23	0.118333333333333\\
			1.24	0.118444444444444\\
			1.25	0.123888888888889\\
			1.26	0.126666666666667\\
			1.27	0.129888888888889\\
			1.28	0.132555555555556\\
			1.29	0.132555555555556\\
			1.3	0.136222222222222\\
			1.31	0.139\\
			1.32	0.146555555555556\\
			1.33	0.154222222222222\\
			1.34	0.158\\
			1.35	0.164\\
			1.36	0.172777777777778\\
			1.37	0.179\\
			1.38	0.191\\
			1.39	0.202777777777778\\
			1.4	0.212888888888889\\
			1.41	0.222888888888889\\
			1.42	0.231111111111111\\
			1.43	0.244666666666667\\
			1.44	0.255555555555556\\
			1.45	0.268222222222222\\
			1.46	0.285777777777778\\
			1.47	0.300777777777778\\
			1.48	0.317444444444444\\
			1.49	0.337\\
			1.5	0.351222222222222\\
			1.51	0.371666666666667\\
			1.52	0.387444444444444\\
			1.53	0.408777777777778\\
			1.54	0.427\\
			1.55	0.444777777777778\\
			1.56	0.464111111111111\\
			1.57	0.480555555555556\\
			1.58	0.497222222222222\\
			1.59	0.517222222222222\\
			1.6	0.535777777777778\\
			1.61	0.556888888888889\\
			1.62	0.578555555555556\\
			1.63	0.598666666666667\\
			1.64	0.622444444444444\\
			1.65	0.640333333333333\\
			1.66	0.659888888888889\\
			1.67	0.683333333333333\\
			1.68	0.704777777777778\\
			1.69	0.719444444444444\\
			1.7	0.736333333333333\\
			1.71	0.752444444444444\\
			1.72	0.767\\
			1.73	0.778666666666667\\
			1.74	0.794111111111111\\
			1.75	0.811\\
			1.76	0.823\\
			1.77	0.834333333333333\\
			1.78	0.849555555555556\\
			1.79	0.861111111111111\\
			1.8	0.870777777777778\\
			1.81	0.885\\
			1.82	0.895777777777778\\
			1.83	0.904222222222222\\
			1.84	0.912666666666667\\
			1.85	0.919333333333333\\
			1.86	0.927777777777778\\
			1.87	0.935222222222222\\
			1.88	0.942\\
			1.89	0.948444444444444\\
			1.9	0.953222222222222\\
			1.91	0.957888888888889\\
			1.92	0.963555555555556\\
			1.93	0.965888888888889\\
			1.94	0.97\\
			1.95	0.972666666666667\\
			1.96	0.975333333333333\\
			1.97	0.978\\
			1.98	0.980666666666666\\
			1.99	0.982222222222222\\
			2	0.984777777777778\\
			2.01	0.986777777777778\\
			2.02	0.989555555555556\\
			2.03	0.990333333333333\\
			2.04	0.991666666666667\\
			2.05	0.993333333333333\\
			2.06	0.994444444444444\\
			2.07	0.995222222222222\\
			2.08	0.995666666666666\\
			2.09	0.996111111111111\\
			2.1	0.996333333333333\\
			2.11	0.996555555555556\\
			2.12	0.996888888888889\\
			2.13	0.997333333333333\\
			2.14	0.997444444444444\\
			2.15	0.998\\
			2.16	0.998\\
			2.17	0.998333333333333\\
			2.18	0.998222222222222\\
			2.19	0.998333333333333\\
			2.2	0.998555555555556\\
		};
		\addlegendentry{AMS}

		\addplot [color=blue,solid]
		table[row sep=crcr]{%
			1	0.15\\
			1.03	0.123333333333333\\
			1.06	0.112\\
			1.09	0.115714285714286\\
			1.12	0.122222222222222\\
			1.15	0.115555555555556\\
			1.18	0.121111111111111\\
			1.21	0.124444444444444\\
			1.24	0.132222222222222\\
			1.27	0.134444444444444\\
			1.3	0.147777777777778\\
			1.33	0.161111111111111\\
			1.36	0.18\\
			1.39	0.207777777777778\\
			1.42	0.24\\
			1.45	0.278888888888889\\
			1.48	0.311111111111111\\
			1.51	0.363333333333333\\
			1.54	0.423333333333333\\
			1.57	0.475555555555556\\
			1.6	0.528888888888889\\
			1.63	0.577777777777778\\
			1.66	0.628888888888889\\
			1.69	0.683333333333333\\
			1.72	0.734444444444444\\
			1.75	0.794444444444444\\
			1.78	0.83\\
			1.81	0.86\\
			1.84	0.886666666666667\\
			1.87	0.916666666666667\\
			1.9	0.94\\
			1.93	0.956666666666667\\
			1.96	0.97\\
			1.99	0.975555555555556\\
			2.02	0.976666666666667\\
			2.05	0.982222222222222\\
			2.08	0.99\\
			2.11	0.994285714285714\\
			2.14	0.998\\
			2.17	1\\
			2.2	1\\
		};
		\addlegendentry{Bootstrap}
	\end{axis}
\end{tikzpicture}%

%% file: poisson_mean_for_power_d=1.tikz
\begin{tikzpicture}[baseline]
	
	\begin{axis}[%
		width=\fwidth,
		height=\fheight,
		scale only axis,
xmin=4,
xmax=14,
ymin=1,
ymax=10,
xtick = {5,9,13},
xticklabels = {{$5$},{$9$},{$13$}},
xlabel= {anomaly size},
ylabel = {$\lambda$},
legend pos = north east,
ytick = {1,3,5,7,9}
]
\addplot [color=black]
  table[row sep=crcr]{%
4	6.7578125\\
5	6.09375\\
6	5.46875\\
7	5.41015625\\
8	5.03656616210937\\
9	4.921875\\
10	4.855029296875\\
11	4.8828125\\
12	4.6875\\
13	4.84375\\
14	4.8046875\\
};
\addlegendentry{$n=32$}

\addplot [color=blue]
  table[row sep=crcr]{%
4	6.25\\
5	5.5859375\\
6	5.078125\\
7	4.6484375\\
8	4.4396484375\\
9	4.1796875\\
10	3.99188232421875\\
11	3.988720703125\\
12	3.75\\
13	3.71270484924316\\
14	3.65655517578125\\
};
\addlegendentry{$n=64$}

\addplot [color=red]
  table[row sep=crcr]{%
4	6.17938232421875\\
5	5.5078125\\
6	4.921875\\
7	4.609375\\
8	4.296875\\
9	4.0625\\
10	3.8201171875\\
11	3.73046875\\
12	3.59375\\
13	3.49609375\\
14	3.383837890625\\
};
\addlegendentry{$n=96$}

\addplot [color=green]
  table[row sep=crcr]{%
4	6.328125\\
5	5.60161290322658\\
6	4.931640625\\
7	4.609375\\
8	4.296875\\
9	3.984375\\
10	3.828125\\
11	3.67976379394531\\
12	3.525390625\\
13	3.4375\\
14	3.3203125\\
};
\addlegendentry{$n=128$}

\end{axis}
\end{tikzpicture}%

%% file: poisson_mean_for_power_d=2.tikz
\begin{tikzpicture}[baseline]
	
	\begin{axis}[%
		width=\fwidth,
		height=\fheight,
		scale only axis,
		xmin=4,
		xmax=14,
ymin=1,
ymax=3.5,
xtick = {5,9,13},
xticklabels = {{$5\times 5$},{$9\times 9$},{$13\times 13$}},
xlabel= {anomaly size},
ylabel = {$\lambda$},
legend pos = north east
]
\addplot [color=black]
  table[row sep=crcr]{%
4	2.79296875\\
5	2.36328125\\
6	2.05078125\\
7	1.85546875\\
8	1.73828125\\
9	1.6509765625\\
10	1.5673828125\\
11	1.51884765625\\
12	1.484375\\
13	1.44072265625\\
14	1.40625\\
};
\addlegendentry{$n=32$}

\addplot [color=blue]
  table[row sep=crcr]{%
4	2.92119140625\\
5	2.45244140625\\
6	2.113134765625\\
7	1.9140625\\
8	1.7681640625\\
9	1.664111328125\\
10	1.58096923828125\\
11	1.5234375\\
12	1.4708984375\\
13	1.42809791564941\\
14	1.39280090332031\\
};
\addlegendentry{$n=64$}

\addplot [color=red]
  table[row sep=crcr]{%
4	3.0078125\\
5	2.51369018554688\\
6	2.1484375\\
7	1.9634765625\\
8	1.796875\\
9	1.69504699707031\\
10	1.602734375\\
11	1.55399571955204\\
12	1.4947265625\\
13	1.44790649414062\\
14	1.41602783203125\\
};
\addlegendentry{$n=96$}

\addplot [color=green]
  table[row sep=crcr]{%
4	3.058154296875\\
5	2.57626953125\\
6	2.1875\\
7	1.98952026367187\\
8	1.83277587890625\\
9	1.7291015625\\
10	1.62109375\\
11	1.550439453125\\
12	1.50390625\\
13	1.463037109375\\
14	1.43623352050781\\
};
\addlegendentry{$n=128$}

\end{axis}
\end{tikzpicture}%

%% file: fibro_limited_data.tikz
\begin{tikzpicture}[baseline]

\begin{axis}[%
width=\fwidth,
height=\fheight,
scale only axis,
point meta min=0,
point meta max=8,
axis on top,
xmin=0.5,
xmax=400.5,
y dir=reverse,
ymin=0.5,
ymax=400.5,
xtick = \empty,
ytick = \empty,
colormap={mymap}{[1pt] rgb(0pt)=(1,1,1); rgb(2pt)=(0.870899,0.870899,0.870899); rgb(3pt)=(0.806349,0.806349,0.806349); rgb(4pt)=(0.741799,0.741799,0.741799); rgb(6pt)=(0.698765,0.698765,0.698765); rgb(7pt)=(0.677249,0.677249,0.677249); rgb(10pt)=(0.612698,0.612698,0.612698); rgb(11pt)=(0.591182,0.591182,0.591182); rgb(13pt)=(0.548148,0.548148,0.548148); rgb(16pt)=(0.492819,0.492819,0.492819); rgb(17pt)=(0.474376,0.474376,0.474376); rgb(23pt)=(0.363719,0.363719,0.363719); rgb(24pt)=(0.345276,0.345276,0.345276); rgb(27pt)=(0.289947,0.289947,0.289947); rgb(29pt)=(0.257672,0.257672,0.257672); rgb(30pt)=(0.241534,0.241534,0.241534); rgb(32pt)=(0.209259,0.209259,0.209259); rgb(33pt)=(0.193122,0.193122,0.193122); rgb(36pt)=(0.144709,0.144709,0.144709); rgb(37pt)=(0.128571,0.128571,0.128571); rgb(39pt)=(0.0962963,0.0962963,0.0962963); rgb(40pt)=(0.0801587,0.0801587,0.0801587); rgb(41pt)=(0.0640212,0.0640212,0.0640212); rgb(42pt)=(0.0478836,0.0478836,0.0478836); rgb(43pt)=(0.031746,0.031746,0.031746); rgb(48pt)=(0.0238095,0.0238095,0.0238095); rgb(49pt)=(0.0222222,0.0222222,0.0222222); rgb(55pt)=(0.0126984,0.0126984,0.0126984); rgb(56pt)=(0.0111111,0.0111111,0.0111111); rgb(59pt)=(0.00634921,0.00634921,0.00634921); rgb(60pt)=(0.0047619,0.0047619,0.0047619); rgb(61pt)=(0.0031746,0.0031746,0.0031746); rgb(63pt)=(0,0,0)},
colorbar
]
\addplot [forget plot] graphics [xmin=0.5, xmax=400.5, ymin=0.5, ymax=400.5] {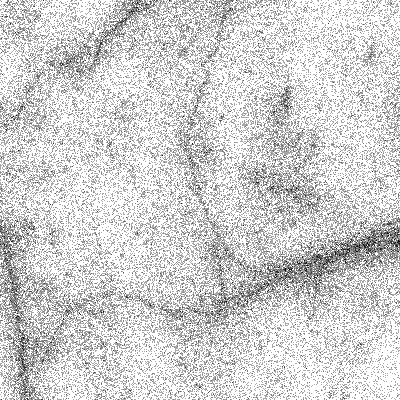};

\coordinate (A) at (330,55);
\coordinate (B) at (354,55);

\draw[color=black,solid,line width=2.0pt] (A) -- (B) node[midway,above] {\tiny 500nm};

\end{axis}

\end{tikzpicture}%

%% file: fibro_result_limited.tikz
\begin{tikzpicture}[baseline]

\begin{axis}[
width=\fwidth,
height=\fheight,
scale only axis,
point meta min=0,
point meta max=0.0625,
axis on top,
xmin=0.5,
xmax=400.5,
y dir=reverse,
ymin=0.5,
ymax=400.5,
xtick = \empty,
ytick = \empty,
colormap={mymap}{[1pt] rgb(0pt)=(1,1,1); rgb(1pt)=(0,0,0.625); rgb(7pt)=(0,0,1); rgb(23pt)=(0,1,1); rgb(39pt)=(1,1,0); rgb(55pt)=(1,0,0); rgb(63pt)=(0.5,0,0)},
colorbar,
colorbar style={
	ytick={0,0.0125,0.025,0.0375,0.05,0.0625},
	yticklabels={{},{32000},{16000},{10668},{8000},{6400}},
	scaled ticks = false
}
]
]
\addplot [forget plot] graphics [xmin=0.5, xmax=400.5, ymin=0.5, ymax=400.5] {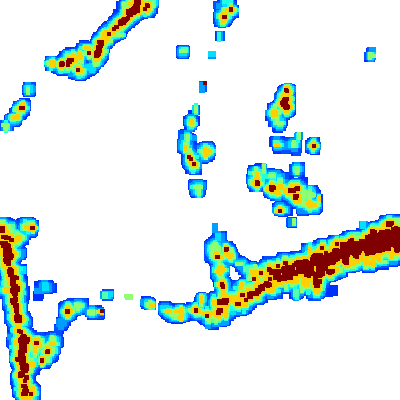};

\coordinate (A) at (330,55);
\coordinate (B) at (354,55);

\draw[color=black,solid,line width=2.0pt] (A) -- (B) node[midway,above] {\tiny 500nm};

\end{axis}
\end{tikzpicture}

%% file: fibro_full_data.tikz
\begin{tikzpicture}[baseline]

\begin{axis}[%
width=\fwidth,
height=\fheight,
scale only axis,
point meta min=0,
point meta max=21,
axis on top,
xmin=0.5,
xmax=400.5,
y dir=reverse,
ymin=0.5,
ymax=400.5,
xtick = \empty,
ytick = \empty,
colormap={mymap}{[1pt] rgb(0pt)=(1,1,1); rgb(2pt)=(0.870899,0.870899,0.870899); rgb(3pt)=(0.806349,0.806349,0.806349); rgb(4pt)=(0.741799,0.741799,0.741799); rgb(6pt)=(0.698765,0.698765,0.698765); rgb(7pt)=(0.677249,0.677249,0.677249); rgb(10pt)=(0.612698,0.612698,0.612698); rgb(11pt)=(0.591182,0.591182,0.591182); rgb(13pt)=(0.548148,0.548148,0.548148); rgb(16pt)=(0.492819,0.492819,0.492819); rgb(17pt)=(0.474376,0.474376,0.474376); rgb(23pt)=(0.363719,0.363719,0.363719); rgb(24pt)=(0.345276,0.345276,0.345276); rgb(27pt)=(0.289947,0.289947,0.289947); rgb(29pt)=(0.257672,0.257672,0.257672); rgb(30pt)=(0.241534,0.241534,0.241534); rgb(32pt)=(0.209259,0.209259,0.209259); rgb(33pt)=(0.193122,0.193122,0.193122); rgb(36pt)=(0.144709,0.144709,0.144709); rgb(37pt)=(0.128571,0.128571,0.128571); rgb(39pt)=(0.0962963,0.0962963,0.0962963); rgb(40pt)=(0.0801587,0.0801587,0.0801587); rgb(41pt)=(0.0640212,0.0640212,0.0640212); rgb(42pt)=(0.0478836,0.0478836,0.0478836); rgb(43pt)=(0.031746,0.031746,0.031746); rgb(48pt)=(0.0238095,0.0238095,0.0238095); rgb(49pt)=(0.0222222,0.0222222,0.0222222); rgb(55pt)=(0.0126984,0.0126984,0.0126984); rgb(56pt)=(0.0111111,0.0111111,0.0111111); rgb(59pt)=(0.00634921,0.00634921,0.00634921); rgb(60pt)=(0.0047619,0.0047619,0.0047619); rgb(61pt)=(0.0031746,0.0031746,0.0031746); rgb(63pt)=(0,0,0)},
colorbar
]
\addplot [forget plot] graphics [xmin=0.5, xmax=400.5, ymin=0.5, ymax=400.5] {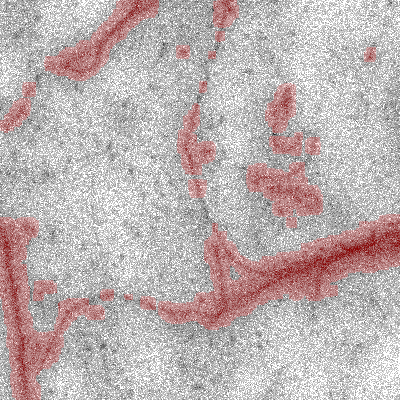};

\coordinate (A) at (330,55);
\coordinate (B) at (354,55);

\draw[color=black,solid,line width=2.0pt] (A) -- (B) node[midway,above] {\tiny 500nm};

\end{axis}

\end{tikzpicture}%

%% file: fibro_full_data2.tikz
\begin{tikzpicture}[baseline]

\begin{axis}[%
width=\fwidth,
height=\fheight,
scale only axis,
point meta min=0,
point meta max=21,
axis on top,
xmin=0.5,
xmax=400.5,
y dir=reverse,
ymin=0.5,
ymax=400.5,
xtick = \empty,
ytick = \empty,
colormap={mymap}{[1pt] rgb(0pt)=(1,1,1); rgb(2pt)=(0.870899,0.870899,0.870899); rgb(3pt)=(0.806349,0.806349,0.806349); rgb(4pt)=(0.741799,0.741799,0.741799); rgb(6pt)=(0.698765,0.698765,0.698765); rgb(7pt)=(0.677249,0.677249,0.677249); rgb(10pt)=(0.612698,0.612698,0.612698); rgb(11pt)=(0.591182,0.591182,0.591182); rgb(13pt)=(0.548148,0.548148,0.548148); rgb(16pt)=(0.492819,0.492819,0.492819); rgb(17pt)=(0.474376,0.474376,0.474376); rgb(23pt)=(0.363719,0.363719,0.363719); rgb(24pt)=(0.345276,0.345276,0.345276); rgb(27pt)=(0.289947,0.289947,0.289947); rgb(29pt)=(0.257672,0.257672,0.257672); rgb(30pt)=(0.241534,0.241534,0.241534); rgb(32pt)=(0.209259,0.209259,0.209259); rgb(33pt)=(0.193122,0.193122,0.193122); rgb(36pt)=(0.144709,0.144709,0.144709); rgb(37pt)=(0.128571,0.128571,0.128571); rgb(39pt)=(0.0962963,0.0962963,0.0962963); rgb(40pt)=(0.0801587,0.0801587,0.0801587); rgb(41pt)=(0.0640212,0.0640212,0.0640212); rgb(42pt)=(0.0478836,0.0478836,0.0478836); rgb(43pt)=(0.031746,0.031746,0.031746); rgb(48pt)=(0.0238095,0.0238095,0.0238095); rgb(49pt)=(0.0222222,0.0222222,0.0222222); rgb(55pt)=(0.0126984,0.0126984,0.0126984); rgb(56pt)=(0.0111111,0.0111111,0.0111111); rgb(59pt)=(0.00634921,0.00634921,0.00634921); rgb(60pt)=(0.0047619,0.0047619,0.0047619); rgb(61pt)=(0.0031746,0.0031746,0.0031746); rgb(63pt)=(0,0,0)},
colorbar
]
\addplot [forget plot] graphics [xmin=0.5, xmax=400.5, ymin=0.5, ymax=400.5] {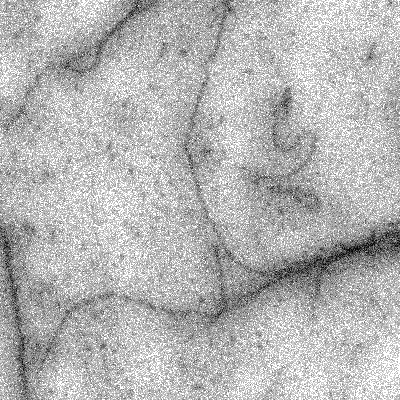};

\coordinate (A) at (330,55);
\coordinate (B) at (354,55);

\draw[color=black,solid,line width=2.0pt] (A) -- (B) node[midway,above] {\tiny 500nm};

\end{axis}

\end{tikzpicture}%

%% file: fibro_result_full.tikz
\begin{tikzpicture}[baseline]

\begin{axis}[
width=\fwidth,
height=\fheight,
scale only axis,
point meta min=0,
point meta max=0.0625,
axis on top,
xmin=0.5,
xmax=400.5,
y dir=reverse,
ymin=0.5,
ymax=400.5,
xtick = \empty,
ytick = \empty,
colormap={mymap}{[1pt] rgb(0pt)=(1,1,1); rgb(1pt)=(0,0,0.625); rgb(7pt)=(0,0,1); rgb(23pt)=(0,1,1); rgb(39pt)=(1,1,0); rgb(55pt)=(1,0,0); rgb(63pt)=(0.5,0,0)},
colorbar,
colorbar style={
	ytick={0,0.0125,0.025,0.0375,0.05,0.0625},
	yticklabels={{},{32000},{16000},{10668},{8000},{6400}},
	scaled ticks = false
}
]
\addplot [forget plot] graphics [xmin=0.5, xmax=400.5, ymin=0.5, ymax=400.5] {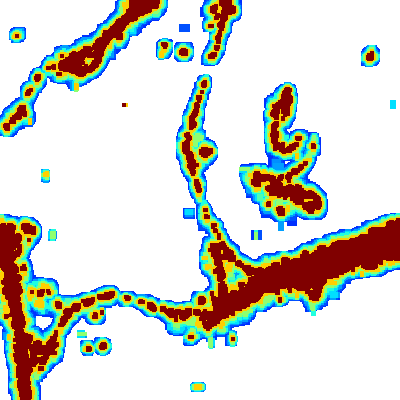};

\coordinate (A) at (330,55);
\coordinate (B) at (354,55);

\draw[color=black,solid,line width=2.0pt] (A) -- (B) node[midway,above] {\tiny 500nm};

\end{axis}
\end{tikzpicture}

%% file: fibro_full_segments.tikz
\begin{tikzpicture}[baseline]

\begin{axis}[
width=\fwidth,
height=\fheight,
scale only axis,
point meta min=0,
point meta max=0.0625,
axis on top,
xmin=0.5,
xmax=400.5,
y dir=reverse,
ymin=0.5,
ymax=400.5,
xtick = \empty,
ytick = \empty,
colormap={mymap}{[1pt] rgb(0pt)=(1,1,1); rgb(63pt)=(1,0,0)},
colorbar,
colorbar style={
	ytick={0,0.0625},
	yticklabels={{inactive},{active}},
	scaled ticks = false
}
]
\addplot [forget plot] graphics [xmin=0.5, xmax=400.5, ymin=0.5, ymax=400.5] {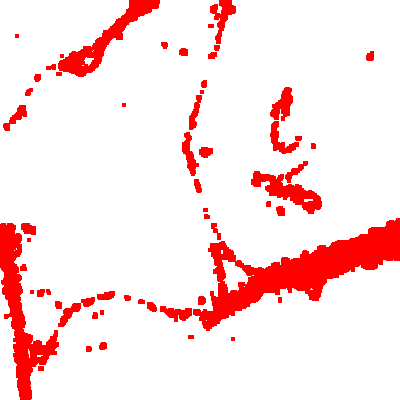};

\coordinate (A) at (330,55);
\coordinate (B) at (354,55);

\draw[color=black,solid,line width=2.0pt] (A) -- (B) node[midway,above] {\tiny 500nm};
\end{axis}
\end{tikzpicture}

%% file: fibro_pixelwise_Hard1.tikz
\begin{tikzpicture}[baseline]

\begin{axis}[%
width=\fwidth,
height=\fheight,
scale only axis,
point meta min=0,
point meta max=0.0625,
axis on top,
xmin=0.5,
xmax=400.5,
y dir=reverse,
ymin=0.5,
ymax=400.5,
xtick = \empty,
ytick = \empty,
colormap={mymap}{[1pt] rgb(0pt)=(1,1,1); rgb(63pt)=(1,0,0)},
colorbar,
colorbar style={
	ytick={0,0.0625},
	yticklabels={{inactive},{active}},
	scaled ticks = false
}
]
\addplot [forget plot] graphics [xmin=0.5, xmax=400.5, ymin=0.5, ymax=400.5] {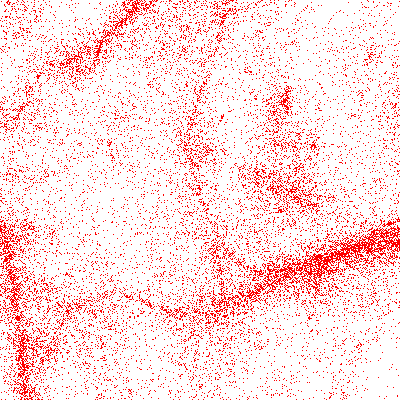};

\coordinate (A) at (330,55);
\coordinate (B) at (354,55);

\draw[color=black,solid,line width=2.0pt] (A) -- (B) node[midway,above] {\tiny 500nm};
\end{axis}
\end{tikzpicture}%

%% file: fibro_pixelwise_Hard2.tikz
\begin{tikzpicture}[baseline]
	
	\begin{axis}[%
		width=\fwidth,
		height=\fheight,
		scale only axis,
		point meta min=0,
		point meta max=0.0625,
		axis on top,
		xmin=0.5,
		xmax=400.5,
		y dir=reverse,
		ymin=0.5,
		ymax=400.5,
		xtick = \empty,
		ytick = \empty,
		colormap={mymap}{[1pt] rgb(0pt)=(1,1,1); rgb(63pt)=(1,0,0)},
		colorbar,
		colorbar style={
			ytick={0,0.0625},
			yticklabels={{inactive},{active}},
			scaled ticks = false
		}
		]
		\addplot [forget plot] graphics [xmin=0.5, xmax=400.5, ymin=0.5, ymax=400.5] {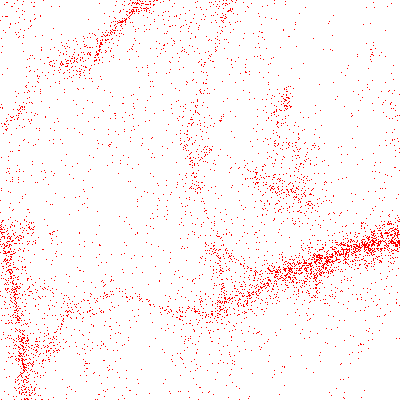};
		
		\coordinate (A) at (330,55);
		\coordinate (B) at (354,55);
		
		\draw[color=black,solid,line width=2.0pt] (A) -- (B) node[midway,above] {\tiny 500nm};
\end{axis}
\end{tikzpicture}%

%% file: fibro_pixelwise_FDR.tikz
\begin{tikzpicture}[baseline]
	
	\begin{axis}[%
		width=\fwidth,
		height=\fheight,
		scale only axis,
		point meta min=0,
		point meta max=0.0625,
		axis on top,
		xmin=0.5,
		xmax=400.5,
		y dir=reverse,
		ymin=0.5,
		ymax=400.5,
		xtick = \empty,
		ytick = \empty,
		colormap={mymap}{[1pt] rgb(0pt)=(1,1,1); rgb(63pt)=(1,0,0)},
		colorbar,
		colorbar style={
			ytick={0,0.0625},
			yticklabels={{inactive},{active}},
			scaled ticks = false
		}
		]
		\addplot [forget plot] graphics [xmin=0.5, xmax=400.5, ymin=0.5, ymax=400.5] {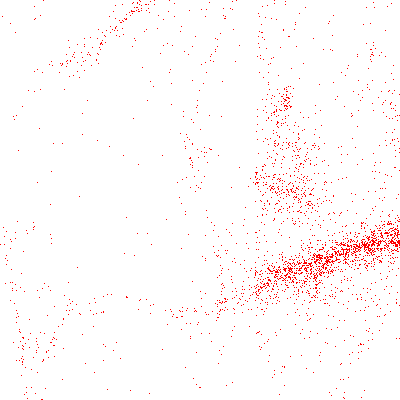};
		
		\coordinate (A) at (330,55);
		\coordinate (B) at (354,55);
		
		\draw[color=black,solid,line width=2.0pt] (A) -- (B) node[midway,above] {\tiny 500nm};
\end{axis}
\end{tikzpicture}%

%% file: density.tikz
\begin{tikzpicture}[baseline]

\begin{axis}[%
width=\fwidth,
height=\fheight,
scale only axis,
xmin=-0.1,
xmax=6.3,
ymin=0,
ymax=1.4,
axis background/.style={fill=white},
axis x line*=bottom,
axis y line*=left
]
]
\addplot [color=black]
  table[row sep=crcr]{%
-0.0232435740047992	9.18394220785659e-06\\
0.00632284054498817	4.5417373512902e-05\\
0.0358892550947756	0.000169221243919604\\
0.0654556696445629	0.00048002569165311\\
0.0950220841943503	0.00108307269349952\\
0.124588498744138	0.00202994244990092\\
0.154154913293925	0.00330861294239715\\
0.183721327843712	0.00489673516563054\\
0.2132877423935	0.00697658255829451\\
0.242854156943287	0.0102822295693039\\
0.272420571493075	0.0161335864179102\\
0.301986986042862	0.0259431635332692\\
0.331553400592649	0.0409187809645566\\
0.361119815142437	0.0623089345635031\\
0.390686229692224	0.0913311184639257\\
0.420252644242011	0.128318913527642\\
0.449819058791799	0.173198247106999\\
0.479385473341586	0.227882831185329\\
0.508951887891373	0.2964423550902\\
0.538518302441161	0.381029401397589\\
0.568084716990948	0.478239994726508\\
0.597651131540736	0.580369146259826\\
0.627217546090523	0.680045705604102\\
0.656783960640311	0.774033065803663\\
0.686350375190098	0.863466638645177\\
0.715916789739885	0.951121811336545\\
0.745483204289673	1.03709332307333\\
0.77504961883946	1.11629383067575\\
0.804616033389247	1.18122145577239\\
0.834182447939035	1.22799034572003\\
0.863748862488822	1.25826845013538\\
0.893315277038609	1.27614971770189\\
0.922881691588397	1.28494709164323\\
0.952448106138184	1.28597884093554\\
0.982014520687972	1.27847219522043\\
1.01158093523776	1.26116593211006\\
1.04114734978755	1.23424146817375\\
1.07071376433733	1.19749050297573\\
1.10028017888712	1.14713961740684\\
1.12984659343691	1.0801681780294\\
1.1594130079867	1.002521675573\\
1.18897942253648	0.926607237987749\\
1.21854583708627	0.860562367217511\\
1.24811225163606	0.804300836525194\\
1.27767866618585	0.75311599031479\\
1.30724508073563	0.701130216473729\\
1.33681149528542	0.644393985461157\\
1.36637790983521	0.584327999355758\\
1.395944324385	0.526989141983386\\
1.42551073893478	0.476879494453697\\
1.45507715348457	0.432898636461159\\
1.48464356803436	0.391843731262126\\
1.51420998258414	0.353242885977493\\
1.54377639713393	0.31837764722697\\
1.57334281168372	0.286858292937098\\
1.60290922623351	0.256869089622146\\
1.63247564078329	0.227807229440154\\
1.66204205533308	0.200593498776764\\
1.69160846988287	0.175909776930105\\
1.72117488443266	0.153954157408628\\
1.75074129898244	0.135039131514296\\
1.78030771353223	0.118605674780458\\
1.80987412808202	0.103319895105879\\
1.83944054263181	0.0891925485140957\\
1.86900695718159	0.0776924286245675\\
1.89857337173138	0.0695346832347885\\
1.92813978628117	0.0637535437989667\\
1.95770620083095	0.0586951688277487\\
1.98727261538074	0.0531410576149391\\
2.01683902993053	0.0466915284253459\\
2.04640544448032	0.0395694376209789\\
2.0759718590301	0.0324562463088711\\
2.10553827357989	0.026352527486937\\
2.13510468812968	0.0220915308610084\\
2.16467110267947	0.0195380604115256\\
2.19423751722925	0.017432373929102\\
2.22380393177904	0.0146379016616546\\
2.25337034632883	0.0114049769631361\\
2.28293676087862	0.00868175085015471\\
2.3125031754284	0.00672520350259801\\
2.34206958997819	0.00524738638990795\\
2.37163600452798	0.00429009866311234\\
2.40120241907777	0.00410052598063315\\
2.43076883362755	0.00448451012889075\\
2.46033524817734	0.00483519809055367\\
2.48990166272713	0.00472050984681099\\
2.51946807727691	0.00429505264568125\\
2.5490344918267	0.00405663151494082\\
2.57860090637649	0.00420080004810565\\
2.60816732092628	0.00440829115243083\\
2.63773373547606	0.00424718737512214\\
2.66730015002585	0.00359517530816439\\
2.69686656457564	0.00269653527649365\\
2.72643297912543	0.00187663664306381\\
2.75599939367521	0.00124484658703287\\
2.785565808225	0.000748541932761825\\
2.81513222277479	0.000373179633288044\\
2.84469863732458	0.000143527765528129\\
2.87426505187436	4.11243235257603e-05\\
2.90383146642415	8.56027717646647e-06\\
};

\addplot [red,dashed]
table[row sep=crcr]{%
	3.46393713865112	8.82668975879867e-06\\
	3.49169073270759	3.97887558206132e-05\\
	3.51944432676405	0.000132599615277214\\
	3.54719792082052	0.000328314884498671\\
	3.57495151487699	0.000614404923155413\\
	3.60270510893345	0.000908391628660173\\
	3.63045870298992	0.00119247870971783\\
	3.65821229704639	0.00167379058684184\\
	3.68596589110285	0.00272669763782137\\
	3.71371948515932	0.00470191039695794\\
	3.74147307921579	0.00794385250284219\\
	3.76922667327225	0.0129910059076202\\
	3.79698026732872	0.0205034120501955\\
	3.82473386138519	0.0309703585295308\\
	3.85248745544165	0.0451748742634266\\
	3.88024104949812	0.0653745073430871\\
	3.90799464355458	0.0952147861091728\\
	3.93574823761105	0.137838580398233\\
	3.96350183166752	0.194389651343485\\
	3.99125542572398	0.263983047730488\\
	4.01900901978045	0.344350374951926\\
	4.04676261383692	0.433449396122298\\
	4.07451620789338	0.530633157491466\\
	4.10226980194985	0.633715802544844\\
	4.13002339600632	0.735540722032079\\
	4.15777699006278	0.828497653010919\\
	4.18553058411925	0.912633493404783\\
	4.21328417817571	0.99441832570645\\
	4.24103777223218	1.07754111251404\\
	4.26879136628865	1.15874480160034\\
	4.29654496034511	1.23176860436596\\
	4.32429855440158	1.29138079483829\\
	4.35205214845805	1.33329870053584\\
	4.37980574251451	1.35335245790998\\
	4.40755933657098	1.34969536086972\\
	4.43531293062745	1.32699334535214\\
	4.46306652468391	1.29556783690265\\
	4.49082011874038	1.2637533367949\\
	4.51857371279684	1.23324440225888\\
	4.54632730685331	1.20167998830726\\
	4.57408090090978	1.1660607644725\\
	4.60183449496624	1.12438219012092\\
	4.62958808902271	1.07670709183835\\
	4.65734168307918	1.02367682643849\\
	4.68509527713564	0.964988758031864\\
	4.71284887119211	0.901621209347156\\
	4.74060246524858	0.837061956561412\\
	4.76835605930504	0.773631101304075\\
	4.79610965336151	0.709798668372078\\
	4.82386324741797	0.643039256397916\\
	4.85161684147444	0.574184215735055\\
	4.87937043553091	0.508412026064579\\
	4.90712402958737	0.45236502961563\\
	4.93487762364384	0.409457984332174\\
	4.96263121770031	0.377031090580556\\
	4.99038481175677	0.348588833477291\\
	5.01813840581324	0.318891235438428\\
	5.04589199986971	0.28674701853016\\
	5.07364559392617	0.254189151538465\\
	5.10139918798264	0.224370579210445\\
	5.1291527820391	0.199343186679914\\
	5.15690637609557	0.178615076574896\\
	5.18465997015204	0.159759438876364\\
	5.2124135642085	0.140817325130944\\
	5.24016715826497	0.121948265138221\\
	5.26792075232144	0.104867380828013\\
	5.2956743463779	0.0908310657174514\\
	5.32342794043437	0.0794695418828584\\
	5.35118153449084	0.0696377698598736\\
	5.3789351285473	0.0608664471697418\\
	5.40668872260377	0.0534474647098733\\
	5.43444231666023	0.0473682545697597\\
	5.4621959107167	0.0419262617921737\\
	5.48994950477317	0.036440253929575\\
	5.51770309882963	0.030957898070549\\
	5.5454566928861	0.0261392776307864\\
	5.57321028694257	0.0226098862126675\\
	5.60096388099903	0.020408903863235\\
	5.6287174750555	0.018986471139052\\
	5.65647106911197	0.0177104027866711\\
	5.68422466316843	0.0162024670722898\\
	5.7119782572249	0.0142690057964849\\
	5.73973185128136	0.01191271657537\\
	5.76748544533783	0.00941482736985623\\
	5.7952390393943	0.00714945077945509\\
	5.82299263345076	0.00534846456782052\\
	5.85074622750723	0.004138680762706\\
	5.8784998215637	0.0036099126834758\\
	5.90625341562016	0.00374484442959631\\
	5.93400700967663	0.00443321993373416\\
	5.9617606037331	0.00548264850240409\\
	5.98951419778956	0.00649367204258748\\
	6.01726779184603	0.00688164036121425\\
	6.0450213859025	0.0062333150215052\\
	6.07277497995896	0.00466798919618372\\
	6.10052857401543	0.00280972344654702\\
	6.12828216807189	0.00132644006616496\\
	6.15603576212836	0.000482175020065208\\
	6.18378935618483	0.000132926922414565\\
	6.21154295024129	2.71217099041485e-05\\
};

\end{axis}
\end{tikzpicture}%

%% file: cdf.tikz
\begin{tikzpicture}[baseline]

\begin{axis}[%
width=\fwidth,
height=\fheight,
scale only axis,
xmin=-0.1,
xmax=6.3,
ymin=0,
ymax=1,
axis background/.style={fill=white},
axis x line*=bottom,
axis y line*=left
]
\addplot [color=black]
  table[row sep=crcr]{%
-0.0232435740047992	1.42852240211963e-07\\
0.00632284054498817	8.30180699666801e-07\\
0.0358892550947756	3.68714964272098e-06\\
0.0654556696445629	1.26845174604602e-05\\
0.0950220841943503	3.49891166273394e-05\\
0.124588498744138	8.01653323446776e-05\\
0.154154913293925	0.000158350688250394\\
0.183721327843712	0.000278881300455\\
0.2132877423935	0.000452673982838855\\
0.242854156943287	0.000703488046337113\\
0.272420571493075	0.00108608826187248\\
0.301986986042862	0.00169698265077612\\
0.331553400592649	0.00267154745964664\\
0.361119815142437	0.00418045525342097\\
0.390686229692224	0.00643246676532385\\
0.420252644242011	0.00966097461509578\\
0.449819058791799	0.014098648692111\\
0.479385473341586	0.019999824738004\\
0.508951887891373	0.0277129909648063\\
0.538518302441161	0.0376911087954855\\
0.568084716990948	0.0503722781444315\\
0.597651131540736	0.0660213026092808\\
0.627217546090523	0.0846677423337772\\
0.656783960640311	0.10617931121988\\
0.686350375190098	0.130395749385039\\
0.715916789739885	0.157224627974086\\
0.745483204289673	0.18662643234319\\
0.77504961883946	0.218488289361935\\
0.804616033389247	0.252497335689668\\
0.834182447939035	0.288159727584625\\
0.863748862488822	0.324952034427467\\
0.893315277038609	0.36244655934705\\
0.922881691588397	0.400329416545207\\
0.952448106138184	0.438357104921475\\
0.982014520687972	0.476292695381571\\
1.01158093523776	0.513863426172753\\
1.04114734978755	0.550778137598085\\
1.07071376433733	0.586755948630098\\
1.10028017888712	0.621458939339505\\
1.12984659343691	0.654426028379719\\
1.1594130079867	0.685228094924366\\
1.18897942253648	0.713730277032189\\
1.21854583708627	0.740124687110721\\
1.24811225163606	0.764718949101789\\
1.27767866618585	0.787738808184286\\
1.30724508073563	0.809247213743678\\
1.33681149528542	0.829152700691077\\
1.36637790983521	0.847320152413829\\
1.395944324385	0.863735506082336\\
1.42551073893478	0.878558124415228\\
1.45507715348457	0.89199791891101\\
1.48464356803436	0.904186110400881\\
1.51420998258414	0.915193997393323\\
1.54377639713393	0.92511364011284\\
1.57334281168372	0.934055605062832\\
1.60290922623351	0.942092179996741\\
1.63247564078329	0.949254644252439\\
1.66204205533308	0.955582489570653\\
1.69160846988287	0.961142403901119\\
1.72117488443266	0.966012113544276\\
1.75074129898244	0.970277283023502\\
1.78030771353223	0.974022888192135\\
1.80987412808202	0.977302122401354\\
1.83944054263181	0.980143878843579\\
1.86900695718159	0.982603034175394\\
1.89857337173138	0.984772056739868\\
1.92813978628117	0.986738852487296\\
1.95770620083095	0.988549238883743\\
1.98727261538074	0.990204727116541\\
2.01683902993053	0.991682862916336\\
2.04640544448032	0.992959166211783\\
2.0759718590301	0.994022852065782\\
2.10553827357989	0.994888552169291\\
2.13510468812968	0.99559979039132\\
2.16467110267947	0.996212275356129\\
2.19423751722925	0.996759643922015\\
2.22380393177904	0.997235856975491\\
2.25337034632883	0.997620729590543\\
2.28293676087862	0.997915678431439\\
2.3125031754284	0.998141939271222\\
2.34206958997819	0.998317955186786\\
2.37163600452798	0.998457297061984\\
2.40120241907777	0.998579425722348\\
2.43076883362755	0.998705626550329\\
2.46033524817734	0.998844245080922\\
2.48990166272713	0.99898674423339\\
2.51946807727691	0.999120193850677\\
2.5490344918267	0.999242702149171\\
2.57860090637649	0.99936404991837\\
2.60816732092628	0.999491767044152\\
2.63773373547606	0.999620979180813\\
2.66730015002585	0.99973795063012\\
2.69686656457564	0.999831122158364\\
2.72643297912543	0.999898281772333\\
2.75599939367521	0.999944009693896\\
2.785565808225	0.99997321338501\\
2.81513222277479	0.999989466175526\\
2.84469863732458	0.999996746270203\\
2.87426505187436	0.999999232506224\\
2.90383146642415	0.999999865010296\\
};
\label{cal}

\addplot [color=red, dashed]
table[row sep=crcr]{%
	3.46393713865112	1.35066752269032e-07\\
	3.49169073270759	7.19857417150328e-07\\
	3.51944432676405	2.9129745208689e-06\\
	3.54719792082052	9.05048606151863e-06\\
	3.57495151487699	2.20085360148588e-05\\
	3.60270510893345	4.32152174991653e-05\\
	3.63045870298992	7.22744187220475e-05\\
	3.65821229704639	0.00011122322151279\\
	3.68596589110285	0.000170560656222068\\
	3.71371948515932	0.000271189531432318\\
	3.74147307921579	0.000443324602857681\\
	3.76922667327225	0.000729020463187482\\
	3.79698026732872	0.00118763828350431\\
	3.82473386138519	0.00189485264032578\\
	3.85248745544165	0.00294147409346572\\
	3.88024104949812	0.00445818568133112\\
	3.90799464355458	0.0066605617669788\\
	3.93574823761105	0.00986342402434782\\
	3.96350183166752	0.0144424462889537\\
	3.99125542572398	0.0207760585346448\\
	4.01900901978045	0.0291966815027575\\
	4.04676261383692	0.0399719384352518\\
	4.07451620789338	0.0533333673915544\\
	4.10226980194985	0.0694844760434304\\
	4.13002339600632	0.0885001247001483\\
	4.15777699006278	0.110229998226253\\
	4.18553058411925	0.134406503577247\\
	4.21328417817571	0.16086996800829\\
	4.24103777223218	0.18962173002729\\
	4.26879136628865	0.220667338126003\\
	4.29654496034511	0.253867271391579\\
	4.32429855440158	0.288918204974281\\
	4.35205214845805	0.32538792014886\\
	4.37980574251451	0.362726559714945\\
	4.40755933657098	0.400290887411395\\
	4.43531293062745	0.437469767711365\\
	4.46306652468391	0.473872337974075\\
	4.49082011874038	0.509385706406052\\
	4.51857371279684	0.54403692759428\\
	4.54632730685331	0.577833788084331\\
	4.57408090090978	0.610704821581539\\
	4.60183449496624	0.642505485713758\\
	4.62958808902271	0.67306414354242\\
	4.65734168307918	0.702224905261854\\
	4.68509527713564	0.729835986786034\\
	4.71284887119211	0.755747469012018\\
	4.74060246524858	0.779874754966004\\
	4.76835605930504	0.802225070414833\\
	4.79610965336151	0.822816219839891\\
	4.82386324741797	0.841598375503788\\
	4.85161684147444	0.858491101687247\\
	4.87937043553091	0.873500038595613\\
	4.90712402958737	0.886804848974197\\
	4.93487762364384	0.898735262518792\\
	4.96263121770031	0.909632784037238\\
	4.99038481175677	0.919701335760904\\
	5.01813840581324	0.928970620785147\\
	5.04589199986971	0.937379874337195\\
	5.07364559392617	0.944884244304274\\
	5.10139918798264	0.951515937814024\\
	5.1291527820391	0.957384448864468\\
	5.15690637609557	0.962622151308312\\
	5.18465997015204	0.967316776022939\\
	5.2124135642085	0.971489164584372\\
	5.24016715826497	0.975133925078689\\
	5.26792075232144	0.978275505372172\\
	5.2956743463779	0.980984094867178\\
	5.32342794043437	0.98334272200079\\
	5.35118153449084	0.98540950246764\\
	5.3789351285473	0.987217949746837\\
	5.40668872260377	0.988800952011088\\
	5.43444231666023	0.990197603363249\\
	5.4621959107167	0.991436370502366\\
	5.48994950477317	0.992524326593478\\
	5.51770309882963	0.99345906422033\\
	5.5454566928861	0.994249044881134\\
	5.57321028694257	0.994922245235143\\
	5.60096388099903	0.995516660341568\\
	5.6287174750555	0.996062390431728\\
	5.65647106911197	0.996571882180285\\
	5.68422466316843	0.997043333603247\\
	5.7119782572249	0.997467295716361\\
	5.73973185128136	0.997831406037402\\
	5.76748544533783	0.998127243283705\\
	5.7952390393943	0.998356236903481\\
	5.82299263345076	0.998528459456016\\
	5.85074622750723	0.998658642778638\\
	5.8784998215637	0.998764586003983\\
	5.90625341562016	0.998865242327358\\
	5.93400700967663	0.998977629294542\\
	5.9617606037331	0.99911474642035\\
	5.98951419778956	0.999281685628023\\
	6.01726779184603	0.999469390946334\\
	6.0450213859025	0.999653911530877\\
	6.07277497995896	0.999806728432919\\
	6.10052857401543	0.999910295729925\\
	6.12828216807189	0.999966312011762\\
	6.15603576212836	0.9999899852326\\
	6.18378935618483	0.999997685701395\\
	6.21154295024129	0.999999594520399\\
};
\label{uncal}

\end{axis}
\end{tikzpicture}%

%% file: Gaussian_means_sim.tikz
\begin{tikzpicture}[baseline]

\begin{axis}[%
width=\fwidth,
height=\fheight,
scale only axis,
point meta min=0,
point meta max=0.5,
axis on top,
xmin=0.5,
xmax=128.5,
y dir=reverse,
ymin=0.5,
ymax=128.5,
xtick = \empty,
ytick = \empty,
axis background/.style={fill=white},
legend style={legend cell align=left, align=left, draw=white!15!black},
colormap={mymap}{[1pt] rgb(0pt)=(0.2422,0.1504,0.6603); rgb(1pt)=(0.2444,0.1534,0.6728); rgb(2pt)=(0.2464,0.1569,0.6847); rgb(3pt)=(0.2484,0.1607,0.6961); rgb(4pt)=(0.2503,0.1648,0.7071); rgb(5pt)=(0.2522,0.1689,0.7179); rgb(6pt)=(0.254,0.1732,0.7286); rgb(7pt)=(0.2558,0.1773,0.7393); rgb(8pt)=(0.2576,0.1814,0.7501); rgb(9pt)=(0.2594,0.1854,0.761); rgb(11pt)=(0.2628,0.1932,0.7828); rgb(12pt)=(0.2645,0.1972,0.7937); rgb(13pt)=(0.2661,0.2011,0.8043); rgb(14pt)=(0.2676,0.2052,0.8148); rgb(15pt)=(0.2691,0.2094,0.8249); rgb(16pt)=(0.2704,0.2138,0.8346); rgb(17pt)=(0.2717,0.2184,0.8439); rgb(18pt)=(0.2729,0.2231,0.8528); rgb(19pt)=(0.274,0.228,0.8612); rgb(20pt)=(0.2749,0.233,0.8692); rgb(21pt)=(0.2758,0.2382,0.8767); rgb(22pt)=(0.2766,0.2435,0.884); rgb(23pt)=(0.2774,0.2489,0.8908); rgb(24pt)=(0.2781,0.2543,0.8973); rgb(25pt)=(0.2788,0.2598,0.9035); rgb(26pt)=(0.2794,0.2653,0.9094); rgb(27pt)=(0.2798,0.2708,0.915); rgb(28pt)=(0.2802,0.2764,0.9204); rgb(29pt)=(0.2806,0.2819,0.9255); rgb(30pt)=(0.2809,0.2875,0.9305); rgb(31pt)=(0.2811,0.293,0.9352); rgb(32pt)=(0.2813,0.2985,0.9397); rgb(33pt)=(0.2814,0.304,0.9441); rgb(34pt)=(0.2814,0.3095,0.9483); rgb(35pt)=(0.2813,0.315,0.9524); rgb(36pt)=(0.2811,0.3204,0.9563); rgb(37pt)=(0.2809,0.3259,0.96); rgb(38pt)=(0.2807,0.3313,0.9636); rgb(39pt)=(0.2803,0.3367,0.967); rgb(40pt)=(0.2798,0.3421,0.9702); rgb(41pt)=(0.2791,0.3475,0.9733); rgb(42pt)=(0.2784,0.3529,0.9763); rgb(43pt)=(0.2776,0.3583,0.9791); rgb(44pt)=(0.2766,0.3638,0.9817); rgb(45pt)=(0.2754,0.3693,0.984); rgb(46pt)=(0.2741,0.3748,0.9862); rgb(47pt)=(0.2726,0.3804,0.9881); rgb(48pt)=(0.271,0.386,0.9898); rgb(49pt)=(0.2691,0.3916,0.9912); rgb(50pt)=(0.267,0.3973,0.9924); rgb(51pt)=(0.2647,0.403,0.9935); rgb(52pt)=(0.2621,0.4088,0.9946); rgb(53pt)=(0.2591,0.4145,0.9955); rgb(54pt)=(0.2556,0.4203,0.9965); rgb(55pt)=(0.2517,0.4261,0.9974); rgb(56pt)=(0.2473,0.4319,0.9983); rgb(57pt)=(0.2424,0.4378,0.9991); rgb(58pt)=(0.2369,0.4437,0.9996); rgb(59pt)=(0.2311,0.4497,0.9995); rgb(60pt)=(0.225,0.4559,0.9985); rgb(61pt)=(0.2189,0.462,0.9968); rgb(62pt)=(0.2128,0.4682,0.9948); rgb(63pt)=(0.2066,0.4743,0.9926); rgb(64pt)=(0.2006,0.4803,0.9906); rgb(65pt)=(0.195,0.4861,0.9887); rgb(66pt)=(0.1903,0.4919,0.9867); rgb(67pt)=(0.1869,0.4975,0.9844); rgb(68pt)=(0.1847,0.503,0.9819); rgb(69pt)=(0.1831,0.5084,0.9793); rgb(70pt)=(0.1818,0.5138,0.9766); rgb(71pt)=(0.1806,0.5191,0.9738); rgb(72pt)=(0.1795,0.5244,0.9709); rgb(73pt)=(0.1785,0.5296,0.9677); rgb(74pt)=(0.1778,0.5349,0.9641); rgb(75pt)=(0.1773,0.5401,0.9602); rgb(76pt)=(0.1768,0.5452,0.956); rgb(77pt)=(0.1764,0.5504,0.9516); rgb(78pt)=(0.1755,0.5554,0.9473); rgb(79pt)=(0.174,0.5605,0.9432); rgb(80pt)=(0.1716,0.5655,0.9393); rgb(81pt)=(0.1686,0.5705,0.9357); rgb(82pt)=(0.1649,0.5755,0.9323); rgb(83pt)=(0.161,0.5805,0.9289); rgb(84pt)=(0.1573,0.5854,0.9254); rgb(85pt)=(0.154,0.5902,0.9218); rgb(86pt)=(0.1513,0.595,0.9182); rgb(87pt)=(0.1492,0.5997,0.9147); rgb(88pt)=(0.1475,0.6043,0.9113); rgb(89pt)=(0.1461,0.6089,0.908); rgb(90pt)=(0.1446,0.6135,0.905); rgb(91pt)=(0.1429,0.618,0.9022); rgb(92pt)=(0.1408,0.6226,0.8998); rgb(93pt)=(0.1383,0.6272,0.8975); rgb(94pt)=(0.1354,0.6317,0.8953); rgb(95pt)=(0.1321,0.6363,0.8932); rgb(96pt)=(0.1288,0.6408,0.891); rgb(97pt)=(0.1253,0.6453,0.8887); rgb(98pt)=(0.1219,0.6497,0.8862); rgb(99pt)=(0.1185,0.6541,0.8834); rgb(100pt)=(0.1152,0.6584,0.8804); rgb(101pt)=(0.1119,0.6627,0.877); rgb(102pt)=(0.1085,0.6669,0.8734); rgb(103pt)=(0.1048,0.671,0.8695); rgb(104pt)=(0.1009,0.675,0.8653); rgb(105pt)=(0.0964,0.6789,0.8609); rgb(106pt)=(0.0914,0.6828,0.8562); rgb(107pt)=(0.0855,0.6865,0.8513); rgb(108pt)=(0.0789,0.6902,0.8462); rgb(109pt)=(0.0713,0.6938,0.8409); rgb(110pt)=(0.0628,0.6972,0.8355); rgb(111pt)=(0.0535,0.7006,0.8299); rgb(112pt)=(0.0433,0.7039,0.8242); rgb(113pt)=(0.0328,0.7071,0.8183); rgb(114pt)=(0.0234,0.7103,0.8124); rgb(115pt)=(0.0155,0.7133,0.8064); rgb(116pt)=(0.0091,0.7163,0.8003); rgb(117pt)=(0.0046,0.7192,0.7941); rgb(118pt)=(0.0019,0.722,0.7878); rgb(119pt)=(0.0009,0.7248,0.7815); rgb(120pt)=(0.0018,0.7275,0.7752); rgb(121pt)=(0.0046,0.7301,0.7688); rgb(122pt)=(0.0094,0.7327,0.7623); rgb(123pt)=(0.0162,0.7352,0.7558); rgb(124pt)=(0.0253,0.7376,0.7492); rgb(125pt)=(0.0369,0.74,0.7426); rgb(126pt)=(0.0504,0.7423,0.7359); rgb(127pt)=(0.0638,0.7446,0.7292); rgb(128pt)=(0.077,0.7468,0.7224); rgb(129pt)=(0.0899,0.7489,0.7156); rgb(130pt)=(0.1023,0.751,0.7088); rgb(131pt)=(0.1141,0.7531,0.7019); rgb(132pt)=(0.1252,0.7552,0.695); rgb(133pt)=(0.1354,0.7572,0.6881); rgb(134pt)=(0.1448,0.7593,0.6812); rgb(135pt)=(0.1532,0.7614,0.6741); rgb(136pt)=(0.1609,0.7635,0.6671); rgb(137pt)=(0.1678,0.7656,0.6599); rgb(138pt)=(0.1741,0.7678,0.6527); rgb(139pt)=(0.1799,0.7699,0.6454); rgb(140pt)=(0.1853,0.7721,0.6379); rgb(141pt)=(0.1905,0.7743,0.6303); rgb(142pt)=(0.1954,0.7765,0.6225); rgb(143pt)=(0.2003,0.7787,0.6146); rgb(144pt)=(0.2061,0.7808,0.6065); rgb(145pt)=(0.2118,0.7828,0.5983); rgb(146pt)=(0.2178,0.7849,0.5899); rgb(147pt)=(0.2244,0.7869,0.5813); rgb(148pt)=(0.2318,0.7887,0.5725); rgb(149pt)=(0.2401,0.7905,0.5636); rgb(150pt)=(0.2491,0.7922,0.5546); rgb(151pt)=(0.2589,0.7937,0.5454); rgb(152pt)=(0.2695,0.7951,0.536); rgb(153pt)=(0.2809,0.7964,0.5266); rgb(154pt)=(0.2929,0.7975,0.517); rgb(155pt)=(0.3052,0.7985,0.5074); rgb(156pt)=(0.3176,0.7994,0.4975); rgb(157pt)=(0.3301,0.8002,0.4876); rgb(158pt)=(0.3424,0.8009,0.4774); rgb(159pt)=(0.3548,0.8016,0.4669); rgb(160pt)=(0.3671,0.8021,0.4563); rgb(161pt)=(0.3795,0.8026,0.4454); rgb(162pt)=(0.3921,0.8029,0.4344); rgb(163pt)=(0.405,0.8031,0.4233); rgb(164pt)=(0.4184,0.803,0.4122); rgb(165pt)=(0.4322,0.8028,0.4013); rgb(166pt)=(0.4463,0.8024,0.3904); rgb(167pt)=(0.4608,0.8018,0.3797); rgb(168pt)=(0.4753,0.8011,0.3691); rgb(169pt)=(0.4899,0.8002,0.3586); rgb(170pt)=(0.5044,0.7993,0.348); rgb(171pt)=(0.5187,0.7982,0.3374); rgb(172pt)=(0.5329,0.797,0.3267); rgb(173pt)=(0.547,0.7957,0.3159); rgb(175pt)=(0.5748,0.7929,0.2941); rgb(176pt)=(0.5886,0.7913,0.2833); rgb(177pt)=(0.6024,0.7896,0.2726); rgb(178pt)=(0.6161,0.7878,0.2622); rgb(179pt)=(0.6297,0.7859,0.2521); rgb(180pt)=(0.6433,0.7839,0.2423); rgb(181pt)=(0.6567,0.7818,0.2329); rgb(182pt)=(0.6701,0.7796,0.2239); rgb(183pt)=(0.6833,0.7773,0.2155); rgb(184pt)=(0.6963,0.775,0.2075); rgb(185pt)=(0.7091,0.7727,0.1998); rgb(186pt)=(0.7218,0.7703,0.1924); rgb(187pt)=(0.7344,0.7679,0.1852); rgb(188pt)=(0.7468,0.7654,0.1782); rgb(189pt)=(0.759,0.7629,0.1717); rgb(190pt)=(0.771,0.7604,0.1658); rgb(191pt)=(0.7829,0.7579,0.1608); rgb(192pt)=(0.7945,0.7554,0.157); rgb(193pt)=(0.806,0.7529,0.1546); rgb(194pt)=(0.8172,0.7505,0.1535); rgb(195pt)=(0.8281,0.7481,0.1536); rgb(196pt)=(0.8389,0.7457,0.1546); rgb(197pt)=(0.8495,0.7435,0.1564); rgb(198pt)=(0.86,0.7413,0.1587); rgb(199pt)=(0.8703,0.7392,0.1615); rgb(200pt)=(0.8804,0.7372,0.165); rgb(201pt)=(0.8903,0.7353,0.1695); rgb(202pt)=(0.9,0.7336,0.1749); rgb(203pt)=(0.9093,0.7321,0.1815); rgb(204pt)=(0.9184,0.7308,0.189); rgb(205pt)=(0.9272,0.7298,0.1973); rgb(206pt)=(0.9357,0.729,0.2061); rgb(207pt)=(0.944,0.7285,0.2151); rgb(208pt)=(0.9523,0.7284,0.2237); rgb(209pt)=(0.9606,0.7285,0.2312); rgb(210pt)=(0.9689,0.7292,0.2373); rgb(211pt)=(0.977,0.7304,0.2418); rgb(212pt)=(0.9842,0.733,0.2446); rgb(213pt)=(0.99,0.7365,0.2429); rgb(214pt)=(0.9946,0.7407,0.2394); rgb(215pt)=(0.9966,0.7458,0.2351); rgb(216pt)=(0.9971,0.7513,0.2309); rgb(217pt)=(0.9972,0.7569,0.2267); rgb(218pt)=(0.9971,0.7626,0.2224); rgb(219pt)=(0.9969,0.7683,0.2181); rgb(220pt)=(0.9966,0.774,0.2138); rgb(221pt)=(0.9962,0.7798,0.2095); rgb(222pt)=(0.9957,0.7856,0.2053); rgb(223pt)=(0.9949,0.7915,0.2012); rgb(224pt)=(0.9938,0.7974,0.1974); rgb(225pt)=(0.9923,0.8034,0.1939); rgb(226pt)=(0.9906,0.8095,0.1906); rgb(227pt)=(0.9885,0.8156,0.1875); rgb(228pt)=(0.9861,0.8218,0.1846); rgb(229pt)=(0.9835,0.828,0.1817); rgb(230pt)=(0.9807,0.8342,0.1787); rgb(231pt)=(0.9778,0.8404,0.1757); rgb(232pt)=(0.9748,0.8467,0.1726); rgb(233pt)=(0.972,0.8529,0.1695); rgb(234pt)=(0.9694,0.8591,0.1665); rgb(235pt)=(0.9671,0.8654,0.1636); rgb(236pt)=(0.9651,0.8716,0.1608); rgb(237pt)=(0.9634,0.8778,0.1582); rgb(238pt)=(0.9619,0.884,0.1557); rgb(239pt)=(0.9608,0.8902,0.1532); rgb(240pt)=(0.9601,0.8963,0.1507); rgb(241pt)=(0.9596,0.9023,0.148); rgb(242pt)=(0.9595,0.9084,0.145); rgb(243pt)=(0.9597,0.9143,0.1418); rgb(244pt)=(0.9601,0.9203,0.1382); rgb(245pt)=(0.9608,0.9262,0.1344); rgb(246pt)=(0.9618,0.932,0.1304); rgb(247pt)=(0.9629,0.9379,0.1261); rgb(248pt)=(0.9642,0.9437,0.1216); rgb(249pt)=(0.9657,0.9494,0.1168); rgb(250pt)=(0.9674,0.9552,0.1116); rgb(251pt)=(0.9692,0.9609,0.1061); rgb(252pt)=(0.9711,0.9667,0.1001); rgb(253pt)=(0.973,0.9724,0.0938); rgb(254pt)=(0.9749,0.9782,0.0872); rgb(255pt)=(0.9769,0.9839,0.0805)},
colorbar
]
\addplot [forget plot] graphics [xmin=0.5, xmax=128.5, ymin=0.5, ymax=128.5] {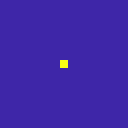};
\end{axis}

\end{tikzpicture}

%% file: Gaussian_data_sim.tikz
\begin{tikzpicture}[baseline]

\begin{axis}[%
width=\fwidth,
height=\fheight,
scale only axis,
point meta min=-4.31931630567483,
point meta max=4.49514175607893,
axis on top,
xmin=0.5,
xmax=128.5,
y dir=reverse,
ymin=0.5,
ymax=128.5,
xtick = \empty,
ytick = \empty,
axis background/.style={fill=white},
legend style={legend cell align=left, align=left, draw=white!15!black},
colormap={mymap}{[1pt] rgb(0pt)=(0.2422,0.1504,0.6603); rgb(1pt)=(0.2444,0.1534,0.6728); rgb(2pt)=(0.2464,0.1569,0.6847); rgb(3pt)=(0.2484,0.1607,0.6961); rgb(4pt)=(0.2503,0.1648,0.7071); rgb(5pt)=(0.2522,0.1689,0.7179); rgb(6pt)=(0.254,0.1732,0.7286); rgb(7pt)=(0.2558,0.1773,0.7393); rgb(8pt)=(0.2576,0.1814,0.7501); rgb(9pt)=(0.2594,0.1854,0.761); rgb(11pt)=(0.2628,0.1932,0.7828); rgb(12pt)=(0.2645,0.1972,0.7937); rgb(13pt)=(0.2661,0.2011,0.8043); rgb(14pt)=(0.2676,0.2052,0.8148); rgb(15pt)=(0.2691,0.2094,0.8249); rgb(16pt)=(0.2704,0.2138,0.8346); rgb(17pt)=(0.2717,0.2184,0.8439); rgb(18pt)=(0.2729,0.2231,0.8528); rgb(19pt)=(0.274,0.228,0.8612); rgb(20pt)=(0.2749,0.233,0.8692); rgb(21pt)=(0.2758,0.2382,0.8767); rgb(22pt)=(0.2766,0.2435,0.884); rgb(23pt)=(0.2774,0.2489,0.8908); rgb(24pt)=(0.2781,0.2543,0.8973); rgb(25pt)=(0.2788,0.2598,0.9035); rgb(26pt)=(0.2794,0.2653,0.9094); rgb(27pt)=(0.2798,0.2708,0.915); rgb(28pt)=(0.2802,0.2764,0.9204); rgb(29pt)=(0.2806,0.2819,0.9255); rgb(30pt)=(0.2809,0.2875,0.9305); rgb(31pt)=(0.2811,0.293,0.9352); rgb(32pt)=(0.2813,0.2985,0.9397); rgb(33pt)=(0.2814,0.304,0.9441); rgb(34pt)=(0.2814,0.3095,0.9483); rgb(35pt)=(0.2813,0.315,0.9524); rgb(36pt)=(0.2811,0.3204,0.9563); rgb(37pt)=(0.2809,0.3259,0.96); rgb(38pt)=(0.2807,0.3313,0.9636); rgb(39pt)=(0.2803,0.3367,0.967); rgb(40pt)=(0.2798,0.3421,0.9702); rgb(41pt)=(0.2791,0.3475,0.9733); rgb(42pt)=(0.2784,0.3529,0.9763); rgb(43pt)=(0.2776,0.3583,0.9791); rgb(44pt)=(0.2766,0.3638,0.9817); rgb(45pt)=(0.2754,0.3693,0.984); rgb(46pt)=(0.2741,0.3748,0.9862); rgb(47pt)=(0.2726,0.3804,0.9881); rgb(48pt)=(0.271,0.386,0.9898); rgb(49pt)=(0.2691,0.3916,0.9912); rgb(50pt)=(0.267,0.3973,0.9924); rgb(51pt)=(0.2647,0.403,0.9935); rgb(52pt)=(0.2621,0.4088,0.9946); rgb(53pt)=(0.2591,0.4145,0.9955); rgb(54pt)=(0.2556,0.4203,0.9965); rgb(55pt)=(0.2517,0.4261,0.9974); rgb(56pt)=(0.2473,0.4319,0.9983); rgb(57pt)=(0.2424,0.4378,0.9991); rgb(58pt)=(0.2369,0.4437,0.9996); rgb(59pt)=(0.2311,0.4497,0.9995); rgb(60pt)=(0.225,0.4559,0.9985); rgb(61pt)=(0.2189,0.462,0.9968); rgb(62pt)=(0.2128,0.4682,0.9948); rgb(63pt)=(0.2066,0.4743,0.9926); rgb(64pt)=(0.2006,0.4803,0.9906); rgb(65pt)=(0.195,0.4861,0.9887); rgb(66pt)=(0.1903,0.4919,0.9867); rgb(67pt)=(0.1869,0.4975,0.9844); rgb(68pt)=(0.1847,0.503,0.9819); rgb(69pt)=(0.1831,0.5084,0.9793); rgb(70pt)=(0.1818,0.5138,0.9766); rgb(71pt)=(0.1806,0.5191,0.9738); rgb(72pt)=(0.1795,0.5244,0.9709); rgb(73pt)=(0.1785,0.5296,0.9677); rgb(74pt)=(0.1778,0.5349,0.9641); rgb(75pt)=(0.1773,0.5401,0.9602); rgb(76pt)=(0.1768,0.5452,0.956); rgb(77pt)=(0.1764,0.5504,0.9516); rgb(78pt)=(0.1755,0.5554,0.9473); rgb(79pt)=(0.174,0.5605,0.9432); rgb(80pt)=(0.1716,0.5655,0.9393); rgb(81pt)=(0.1686,0.5705,0.9357); rgb(82pt)=(0.1649,0.5755,0.9323); rgb(83pt)=(0.161,0.5805,0.9289); rgb(84pt)=(0.1573,0.5854,0.9254); rgb(85pt)=(0.154,0.5902,0.9218); rgb(86pt)=(0.1513,0.595,0.9182); rgb(87pt)=(0.1492,0.5997,0.9147); rgb(88pt)=(0.1475,0.6043,0.9113); rgb(89pt)=(0.1461,0.6089,0.908); rgb(90pt)=(0.1446,0.6135,0.905); rgb(91pt)=(0.1429,0.618,0.9022); rgb(92pt)=(0.1408,0.6226,0.8998); rgb(93pt)=(0.1383,0.6272,0.8975); rgb(94pt)=(0.1354,0.6317,0.8953); rgb(95pt)=(0.1321,0.6363,0.8932); rgb(96pt)=(0.1288,0.6408,0.891); rgb(97pt)=(0.1253,0.6453,0.8887); rgb(98pt)=(0.1219,0.6497,0.8862); rgb(99pt)=(0.1185,0.6541,0.8834); rgb(100pt)=(0.1152,0.6584,0.8804); rgb(101pt)=(0.1119,0.6627,0.877); rgb(102pt)=(0.1085,0.6669,0.8734); rgb(103pt)=(0.1048,0.671,0.8695); rgb(104pt)=(0.1009,0.675,0.8653); rgb(105pt)=(0.0964,0.6789,0.8609); rgb(106pt)=(0.0914,0.6828,0.8562); rgb(107pt)=(0.0855,0.6865,0.8513); rgb(108pt)=(0.0789,0.6902,0.8462); rgb(109pt)=(0.0713,0.6938,0.8409); rgb(110pt)=(0.0628,0.6972,0.8355); rgb(111pt)=(0.0535,0.7006,0.8299); rgb(112pt)=(0.0433,0.7039,0.8242); rgb(113pt)=(0.0328,0.7071,0.8183); rgb(114pt)=(0.0234,0.7103,0.8124); rgb(115pt)=(0.0155,0.7133,0.8064); rgb(116pt)=(0.0091,0.7163,0.8003); rgb(117pt)=(0.0046,0.7192,0.7941); rgb(118pt)=(0.0019,0.722,0.7878); rgb(119pt)=(0.0009,0.7248,0.7815); rgb(120pt)=(0.0018,0.7275,0.7752); rgb(121pt)=(0.0046,0.7301,0.7688); rgb(122pt)=(0.0094,0.7327,0.7623); rgb(123pt)=(0.0162,0.7352,0.7558); rgb(124pt)=(0.0253,0.7376,0.7492); rgb(125pt)=(0.0369,0.74,0.7426); rgb(126pt)=(0.0504,0.7423,0.7359); rgb(127pt)=(0.0638,0.7446,0.7292); rgb(128pt)=(0.077,0.7468,0.7224); rgb(129pt)=(0.0899,0.7489,0.7156); rgb(130pt)=(0.1023,0.751,0.7088); rgb(131pt)=(0.1141,0.7531,0.7019); rgb(132pt)=(0.1252,0.7552,0.695); rgb(133pt)=(0.1354,0.7572,0.6881); rgb(134pt)=(0.1448,0.7593,0.6812); rgb(135pt)=(0.1532,0.7614,0.6741); rgb(136pt)=(0.1609,0.7635,0.6671); rgb(137pt)=(0.1678,0.7656,0.6599); rgb(138pt)=(0.1741,0.7678,0.6527); rgb(139pt)=(0.1799,0.7699,0.6454); rgb(140pt)=(0.1853,0.7721,0.6379); rgb(141pt)=(0.1905,0.7743,0.6303); rgb(142pt)=(0.1954,0.7765,0.6225); rgb(143pt)=(0.2003,0.7787,0.6146); rgb(144pt)=(0.2061,0.7808,0.6065); rgb(145pt)=(0.2118,0.7828,0.5983); rgb(146pt)=(0.2178,0.7849,0.5899); rgb(147pt)=(0.2244,0.7869,0.5813); rgb(148pt)=(0.2318,0.7887,0.5725); rgb(149pt)=(0.2401,0.7905,0.5636); rgb(150pt)=(0.2491,0.7922,0.5546); rgb(151pt)=(0.2589,0.7937,0.5454); rgb(152pt)=(0.2695,0.7951,0.536); rgb(153pt)=(0.2809,0.7964,0.5266); rgb(154pt)=(0.2929,0.7975,0.517); rgb(155pt)=(0.3052,0.7985,0.5074); rgb(156pt)=(0.3176,0.7994,0.4975); rgb(157pt)=(0.3301,0.8002,0.4876); rgb(158pt)=(0.3424,0.8009,0.4774); rgb(159pt)=(0.3548,0.8016,0.4669); rgb(160pt)=(0.3671,0.8021,0.4563); rgb(161pt)=(0.3795,0.8026,0.4454); rgb(162pt)=(0.3921,0.8029,0.4344); rgb(163pt)=(0.405,0.8031,0.4233); rgb(164pt)=(0.4184,0.803,0.4122); rgb(165pt)=(0.4322,0.8028,0.4013); rgb(166pt)=(0.4463,0.8024,0.3904); rgb(167pt)=(0.4608,0.8018,0.3797); rgb(168pt)=(0.4753,0.8011,0.3691); rgb(169pt)=(0.4899,0.8002,0.3586); rgb(170pt)=(0.5044,0.7993,0.348); rgb(171pt)=(0.5187,0.7982,0.3374); rgb(172pt)=(0.5329,0.797,0.3267); rgb(173pt)=(0.547,0.7957,0.3159); rgb(175pt)=(0.5748,0.7929,0.2941); rgb(176pt)=(0.5886,0.7913,0.2833); rgb(177pt)=(0.6024,0.7896,0.2726); rgb(178pt)=(0.6161,0.7878,0.2622); rgb(179pt)=(0.6297,0.7859,0.2521); rgb(180pt)=(0.6433,0.7839,0.2423); rgb(181pt)=(0.6567,0.7818,0.2329); rgb(182pt)=(0.6701,0.7796,0.2239); rgb(183pt)=(0.6833,0.7773,0.2155); rgb(184pt)=(0.6963,0.775,0.2075); rgb(185pt)=(0.7091,0.7727,0.1998); rgb(186pt)=(0.7218,0.7703,0.1924); rgb(187pt)=(0.7344,0.7679,0.1852); rgb(188pt)=(0.7468,0.7654,0.1782); rgb(189pt)=(0.759,0.7629,0.1717); rgb(190pt)=(0.771,0.7604,0.1658); rgb(191pt)=(0.7829,0.7579,0.1608); rgb(192pt)=(0.7945,0.7554,0.157); rgb(193pt)=(0.806,0.7529,0.1546); rgb(194pt)=(0.8172,0.7505,0.1535); rgb(195pt)=(0.8281,0.7481,0.1536); rgb(196pt)=(0.8389,0.7457,0.1546); rgb(197pt)=(0.8495,0.7435,0.1564); rgb(198pt)=(0.86,0.7413,0.1587); rgb(199pt)=(0.8703,0.7392,0.1615); rgb(200pt)=(0.8804,0.7372,0.165); rgb(201pt)=(0.8903,0.7353,0.1695); rgb(202pt)=(0.9,0.7336,0.1749); rgb(203pt)=(0.9093,0.7321,0.1815); rgb(204pt)=(0.9184,0.7308,0.189); rgb(205pt)=(0.9272,0.7298,0.1973); rgb(206pt)=(0.9357,0.729,0.2061); rgb(207pt)=(0.944,0.7285,0.2151); rgb(208pt)=(0.9523,0.7284,0.2237); rgb(209pt)=(0.9606,0.7285,0.2312); rgb(210pt)=(0.9689,0.7292,0.2373); rgb(211pt)=(0.977,0.7304,0.2418); rgb(212pt)=(0.9842,0.733,0.2446); rgb(213pt)=(0.99,0.7365,0.2429); rgb(214pt)=(0.9946,0.7407,0.2394); rgb(215pt)=(0.9966,0.7458,0.2351); rgb(216pt)=(0.9971,0.7513,0.2309); rgb(217pt)=(0.9972,0.7569,0.2267); rgb(218pt)=(0.9971,0.7626,0.2224); rgb(219pt)=(0.9969,0.7683,0.2181); rgb(220pt)=(0.9966,0.774,0.2138); rgb(221pt)=(0.9962,0.7798,0.2095); rgb(222pt)=(0.9957,0.7856,0.2053); rgb(223pt)=(0.9949,0.7915,0.2012); rgb(224pt)=(0.9938,0.7974,0.1974); rgb(225pt)=(0.9923,0.8034,0.1939); rgb(226pt)=(0.9906,0.8095,0.1906); rgb(227pt)=(0.9885,0.8156,0.1875); rgb(228pt)=(0.9861,0.8218,0.1846); rgb(229pt)=(0.9835,0.828,0.1817); rgb(230pt)=(0.9807,0.8342,0.1787); rgb(231pt)=(0.9778,0.8404,0.1757); rgb(232pt)=(0.9748,0.8467,0.1726); rgb(233pt)=(0.972,0.8529,0.1695); rgb(234pt)=(0.9694,0.8591,0.1665); rgb(235pt)=(0.9671,0.8654,0.1636); rgb(236pt)=(0.9651,0.8716,0.1608); rgb(237pt)=(0.9634,0.8778,0.1582); rgb(238pt)=(0.9619,0.884,0.1557); rgb(239pt)=(0.9608,0.8902,0.1532); rgb(240pt)=(0.9601,0.8963,0.1507); rgb(241pt)=(0.9596,0.9023,0.148); rgb(242pt)=(0.9595,0.9084,0.145); rgb(243pt)=(0.9597,0.9143,0.1418); rgb(244pt)=(0.9601,0.9203,0.1382); rgb(245pt)=(0.9608,0.9262,0.1344); rgb(246pt)=(0.9618,0.932,0.1304); rgb(247pt)=(0.9629,0.9379,0.1261); rgb(248pt)=(0.9642,0.9437,0.1216); rgb(249pt)=(0.9657,0.9494,0.1168); rgb(250pt)=(0.9674,0.9552,0.1116); rgb(251pt)=(0.9692,0.9609,0.1061); rgb(252pt)=(0.9711,0.9667,0.1001); rgb(253pt)=(0.973,0.9724,0.0938); rgb(254pt)=(0.9749,0.9782,0.0872); rgb(255pt)=(0.9769,0.9839,0.0805)},
colorbar
]
\addplot [forget plot] graphics [xmin=0.5, xmax=128.5, ymin=0.5, ymax=128.5] {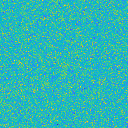};
\end{axis}
\end{tikzpicture}%

%% file: gauss_level.tikz
\begin{tikzpicture}[baseline]

\begin{axis}[%
width=\fwidth,
height=\fheight,
scale only axis,
separate axis lines,
xmin=0.1,
xmax=1,
ylabel = {Level},
ymin=0,
ymax=0.3,
legend pos = north east,
y tick label style={
	/pgf/number format/fixed,
	/pgf/number format/precision=2
}
]
\addplot [color=red,solid]
  table[row sep=crcr]{%
0.1	0.097\\
0.11	0.0993333333333333\\
0.12	0.0976\\
0.13	0.0951428571428571\\
0.14	0.0974444444444444\\
0.15	0.0991111111111111\\
0.16	0.0994444444444444\\
0.17	0.101\\
0.18	0.101444444444444\\
0.19	0.101\\
0.2	0.0984444444444444\\
0.21	0.101333333333333\\
0.22	0.101\\
0.23	0.0997777777777778\\
0.24	0.0981111111111111\\
0.25	0.0975555555555555\\
0.26	0.0971111111111111\\
0.27	0.0973333333333333\\
0.28	0.0966666666666667\\
0.29	0.0998888888888889\\
0.3	0.100555555555556\\
0.31	0.100222222222222\\
0.32	0.100444444444444\\
0.33	0.101555555555556\\
0.34	0.103222222222222\\
0.35	0.102\\
0.36	0.102222222222222\\
0.37	0.105666666666667\\
0.38	0.105555555555556\\
0.39	0.103777777777778\\
0.4	0.105444444444444\\
0.41	0.104666666666667\\
0.42	0.103666666666667\\
0.43	0.101666666666667\\
0.44	0.100111111111111\\
0.45	0.0996666666666667\\
0.46	0.0988888888888889\\
0.47	0.0988888888888889\\
0.48	0.0976666666666667\\
0.49	0.0962222222222222\\
0.5	0.0982222222222222\\
0.51	0.0996666666666667\\
0.52	0.100777777777778\\
0.53	0.101666666666667\\
0.54	0.101555555555556\\
0.55	0.0982222222222222\\
0.56	0.0985555555555555\\
0.57	0.101333333333333\\
0.58	0.101333333333333\\
0.59	0.0981111111111111\\
0.6	0.0978888888888889\\
0.61	0.0965555555555556\\
0.62	0.0981111111111111\\
0.63	0.0996666666666667\\
0.64	0.100888888888889\\
0.65	0.100666666666667\\
0.66	0.0993333333333333\\
0.67	0.0991111111111111\\
0.68	0.101\\
0.69	0.101333333333333\\
0.7	0.102111111111111\\
0.71	0.102666666666667\\
0.72	0.100777777777778\\
0.73	0.0986666666666667\\
0.74	0.0986666666666667\\
0.75	0.0986666666666667\\
0.76	0.0987777777777778\\
0.77	0.0993333333333333\\
0.78	0.098\\
0.79	0.0975555555555555\\
0.8	0.097\\
0.81	0.0985555555555555\\
0.82	0.101\\
0.83	0.101222222222222\\
0.84	0.101555555555556\\
0.85	0.102222222222222\\
0.86	0.102888888888889\\
0.87	0.101777777777778\\
0.88	0.100444444444444\\
0.89	0.0997777777777778\\
0.9	0.0996666666666667\\
0.91	0.100888888888889\\
0.92	0.101444444444444\\
0.93	0.100888888888889\\
0.94	0.0993333333333333\\
0.95	0.0964444444444444\\
0.96	0.0965555555555556\\
0.97	0.0975714285714286\\
0.98	0.0942\\
0.99	0.088\\
1	0.091\\
};
\addlegendentry{Oracle}
\addplot [color=black,solid]
  table[row sep=crcr]{%
0.1	0.098\\
0.11	0.100666666666667\\
0.12	0.103\\
0.13	0.0992857142857143\\
0.14	0.101555555555556\\
0.15	0.102\\
0.16	0.102666666666667\\
0.17	0.103222222222222\\
0.18	0.103333333333333\\
0.19	0.103777777777778\\
0.2	0.104333333333333\\
0.21	0.105777777777778\\
0.22	0.103888888888889\\
0.23	0.103333333333333\\
0.24	0.103444444444444\\
0.25	0.104666666666667\\
0.26	0.104\\
0.27	0.103333333333333\\
0.28	0.102777777777778\\
0.29	0.102333333333333\\
0.3	0.103222222222222\\
0.31	0.102666666666667\\
0.32	0.102555555555556\\
0.33	0.102444444444444\\
0.34	0.0998888888888889\\
0.35	0.098\\
0.36	0.0988888888888889\\
0.37	0.0968888888888889\\
0.38	0.0983333333333333\\
0.39	0.0987777777777778\\
0.4	0.0991111111111111\\
0.41	0.100555555555556\\
0.42	0.1\\
0.43	0.101222222222222\\
0.44	0.103444444444444\\
0.45	0.102888888888889\\
0.46	0.104888888888889\\
0.47	0.101666666666667\\
0.48	0.102777777777778\\
0.49	0.101444444444444\\
0.5	0.100888888888889\\
0.51	0.102111111111111\\
0.52	0.103222222222222\\
0.53	0.102111111111111\\
0.54	0.101777777777778\\
0.55	0.100888888888889\\
0.56	0.104111111111111\\
0.57	0.102888888888889\\
0.58	0.103333333333333\\
0.59	0.101777777777778\\
0.6	0.100222222222222\\
0.61	0.0995555555555556\\
0.62	0.100888888888889\\
0.63	0.100555555555556\\
0.64	0.1\\
0.65	0.098\\
0.66	0.0974444444444444\\
0.67	0.0972222222222222\\
0.68	0.0962222222222222\\
0.69	0.0953333333333333\\
0.7	0.0973333333333333\\
0.71	0.0975555555555555\\
0.72	0.0975555555555555\\
0.73	0.0964444444444444\\
0.74	0.0964444444444444\\
0.75	0.0977777777777778\\
0.76	0.099\\
0.77	0.101333333333333\\
0.78	0.101111111111111\\
0.79	0.0988888888888889\\
0.8	0.098\\
0.81	0.0982222222222222\\
0.82	0.0998888888888889\\
0.83	0.1\\
0.84	0.100111111111111\\
0.85	0.101666666666667\\
0.86	0.101222222222222\\
0.87	0.102\\
0.88	0.0996666666666667\\
0.89	0.099\\
0.9	0.0985555555555555\\
0.91	0.0983333333333333\\
0.92	0.0978888888888889\\
0.93	0.0965555555555556\\
0.94	0.0957777777777778\\
0.95	0.0947777777777778\\
0.96	0.096\\
0.97	0.0974285714285714\\
0.98	0.098\\
0.99	0.099\\
1	0.102\\
};
\addlegendentry{AMS}
\addplot [color=green,solid]
  table[row sep=crcr]{%
0.1	0.1\\
1	0.1\\
};
\addlegendentry{$\alpha$}
\end{axis}
\end{tikzpicture}%

%% file: gauss_power.tikz
\begin{tikzpicture}[baseline]

\begin{axis}[%
width=\fwidth,
height=\fheight,
scale only axis,
xmin=0,
xmax=1.5,
xlabel={$\mu$},
ylabel={Power},
ymin=0,
ymax=1,
legend pos = south east
]
\addplot [color=red,solid]
  table[row sep=crcr]{%
0	0.112\\
0.05	0.105\\
0.1	0.104\\
0.15	0.113714285714286\\
0.2	0.134333333333333\\
0.25	0.156333333333333\\
0.3	0.195111111111111\\
0.35	0.243333333333333\\
0.4	0.308888888888889\\
0.45	0.385\\
0.5	0.468444444444444\\
0.55	0.556333333333333\\
0.6	0.641333333333333\\
0.65	0.728222222222222\\
0.7	0.804111111111111\\
0.75	0.866666666666667\\
0.8	0.916666666666667\\
0.85	0.951222222222222\\
0.9	0.974444444444444\\
0.95	0.986666666666667\\
1	0.994666666666667\\
1.05	0.997888888888889\\
1.1	0.999111111111111\\
1.15	0.999888888888889\\
1.2	1\\
1.25	1\\
1.3	1\\
1.35	1\\
1.4	1\\
1.45	1\\
1.5	1\\
};
\addlegendentry{Oracle}
\addplot [color=black,solid]
  table[row sep=crcr]{%
0	0.09\\
0.05	0.0923333333333333\\
0.1	0.0964\\
0.15	0.108285714285714\\
0.2	0.129444444444444\\
0.25	0.154111111111111\\
0.3	0.195\\
0.35	0.246\\
0.4	0.310333333333333\\
0.45	0.386444444444444\\
0.5	0.470444444444444\\
0.55	0.558555555555555\\
0.6	0.645111111111111\\
0.65	0.731\\
0.7	0.807333333333333\\
0.75	0.868444444444444\\
0.8	0.916888888888889\\
0.85	0.952222222222222\\
0.9	0.975888888888889\\
0.95	0.989333333333333\\
1	0.995333333333333\\
1.05	0.998666666666667\\
1.1	0.999888888888889\\
1.15	1\\
1.2	1\\
1.25	1\\
1.3	1\\
1.35	1\\
1.4	1\\
1.45	1\\
1.5	1\\
};
\addlegendentry{AMS}
\end{axis}
\end{tikzpicture}%

%% file: gauss_mean_for_power_d=1.tikz
\begin{tikzpicture}[baseline]
	
\begin{axis}[%
width=\fwidth,
height=\fheight,
scale only axis,
xmin=4,
xmax=14,
ymin=0,
ymax=5.5,
xtick = {5,9,13},
xticklabels = {{$5$},{$9$},{$13$}},
ylabel = {$\mu$},
xlabel = {anomaly size},
legend pos = north east,
ytick = {0,1,2,3,4,5}
]
\addplot [color=black]
  table[row sep=crcr]{%
4	4.258056640625\\
5	3.8375\\
6	3.05104166666667\\
7	2.75\\
8	2.375\\
9	2.19670138888889\\
10	1.975\\
11	1.86363636363636\\
12	1.70833333333333\\
13	1.63461538461538\\
14	1.54732142857143\\
};
\addlegendentry{$n=32$}

\addplot [color=blue]
  table[row sep=crcr]{%
4	3.555859375\\
5	3.3\\
6	2.75\\
7	2.52762276785714\\
8	2.28092651367188\\
9	2.13819444444444\\
10	1.97001953125\\
11	1.81818181818182\\
12	1.773046875\\
13	1.65384615384615\\
14	1.6125\\
};
\addlegendentry{$n=64$}

\addplot [color=red]
  table[row sep=crcr]{%
4	3.46875\\
5	3.09375\\
6	2.75\\
7	2.46428571428571\\
8	2.25\\
9	2.11111111111111\\
10	1.95\\
11	1.85767045454545\\
12	1.75\\
13	1.69230769230769\\
14	1.64285714285714\\
};
\addlegendentry{$n=96$}

\addplot [color=green]
  table[row sep=crcr]{%
4	3.40625\\
5	3.09375\\
6	2.67122395833333\\
7	2.46517857142857\\
8	2.2703125\\
9	2.11111111111111\\
10	1.975\\
11	1.86363636363636\\
12	1.77083333333333\\
13	1.69230769230769\\
14	1.64285714285714\\
};
\addlegendentry{$n=128$}

\end{axis}
\end{tikzpicture}%

%% file: gauss_mean_for_power_d=2.tikz
\begin{tikzpicture}[baseline]
	
	\begin{axis}[%
		width=\fwidth,
		height=\fheight,
		scale only axis,
		xmin=4,
		xmax=14,
		ymin=0,
		ymax=2,
xtick = {5,9,13},
		xticklabels = {{$5\times 5$},{$9\times 9$},{$13\times 13$}},
		ylabel = {$\mu$},
		xlabel = {anomaly size},
		legend pos = north east
		]
		\addplot [color=black]
  table[row sep=crcr]{%
4	1.38828125\\
5	1.10625\\
6	0.823958333333333\\
7	0.702678571428571\\
8	0.59375\\
9	0.521457248263889\\
10	0.4300537109375\\
11	0.401704545454545\\
12	0.354166666666667\\
13	0.334474651630108\\
14	0.313504464285714\\
};
\addlegendentry{$n=32$}

\addplot [color=blue]
  table[row sep=crcr]{%
4	1.50283203125\\
5	1.225\\
6	0.918977864583333\\
7	0.785714285714286\\
8	0.65625\\
9	0.583333333333333\\
10	0.49658203125\\
11	0.465305397727273\\
12	0.4109375\\
13	0.384615384615385\\
14	0.3488037109375\\
};
\addlegendentry{$n=64$}

\addplot [color=red]
  table[row sep=crcr]{%
4	1.5625\\
5	1.225\\
6	0.981591796875\\
7	0.821428571428571\\
8	0.7064453125\\
9	0.615277777777778\\
10	0.540731811523438\\
11	0.492613636363636\\
12	0.4375\\
13	0.403846153846154\\
14	0.371003069196429\\
};
\addlegendentry{$n=96$}

\addplot [color=green]
  table[row sep=crcr]{%
4	1.6228759765625\\
5	1.2939453125\\
6	1.023046875\\
7	0.852566964285714\\
8	0.71875\\
9	0.638888888888889\\
10	0.55\\
11	0.507386363636364\\
12	0.456022135416667\\
13	0.417427884615385\\
14	0.38828125\\
};
\addlegendentry{$n=128$}

\end{axis}
\end{tikzpicture}%

%% file: beads_limited_data.tikz
\begin{tikzpicture}[baseline]

\begin{axis}[%
width=\fwidth,
height=\fheight,
scale only axis,
point meta min=0,
point meta max=6,
axis on top,
xmin=0.5,
xmax=400.5,
y dir=reverse,
ymin=0.5,
ymax=400.5,
xtick = \empty,
ytick = \empty,
colormap={mymap}{[1pt] rgb(0pt)=(1,1,1); rgb(2pt)=(0.870899,0.870899,0.870899); rgb(3pt)=(0.806349,0.806349,0.806349); rgb(4pt)=(0.741799,0.741799,0.741799); rgb(6pt)=(0.698765,0.698765,0.698765); rgb(7pt)=(0.677249,0.677249,0.677249); rgb(10pt)=(0.612698,0.612698,0.612698); rgb(11pt)=(0.591182,0.591182,0.591182); rgb(13pt)=(0.548148,0.548148,0.548148); rgb(16pt)=(0.492819,0.492819,0.492819); rgb(17pt)=(0.474376,0.474376,0.474376); rgb(23pt)=(0.363719,0.363719,0.363719); rgb(24pt)=(0.345276,0.345276,0.345276); rgb(27pt)=(0.289947,0.289947,0.289947); rgb(29pt)=(0.257672,0.257672,0.257672); rgb(30pt)=(0.241534,0.241534,0.241534); rgb(32pt)=(0.209259,0.209259,0.209259); rgb(33pt)=(0.193122,0.193122,0.193122); rgb(36pt)=(0.144709,0.144709,0.144709); rgb(37pt)=(0.128571,0.128571,0.128571); rgb(39pt)=(0.0962963,0.0962963,0.0962963); rgb(40pt)=(0.0801587,0.0801587,0.0801587); rgb(41pt)=(0.0640212,0.0640212,0.0640212); rgb(42pt)=(0.0478836,0.0478836,0.0478836); rgb(43pt)=(0.031746,0.031746,0.031746); rgb(48pt)=(0.0238095,0.0238095,0.0238095); rgb(49pt)=(0.0222222,0.0222222,0.0222222); rgb(55pt)=(0.0126984,0.0126984,0.0126984); rgb(56pt)=(0.0111111,0.0111111,0.0111111); rgb(59pt)=(0.00634921,0.00634921,0.00634921); rgb(60pt)=(0.0047619,0.0047619,0.0047619); rgb(61pt)=(0.0031746,0.0031746,0.0031746); rgb(63pt)=(0,0,0)},
colorbar
]
\addplot [forget plot] graphics [xmin=0.5, xmax=400.5, ymin=0.5, ymax=400.5] {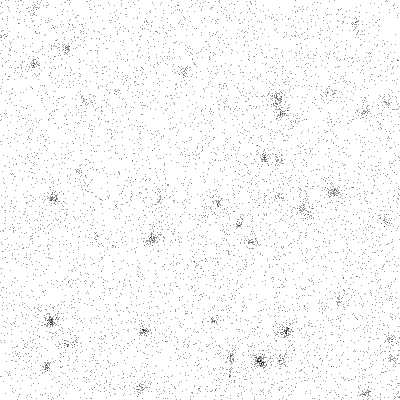};

\coordinate (A) at (350,55);
\coordinate (B) at (374,55);

\draw[color=black,solid,line width=2.0pt] (A) -- (B) node[midway,above] {\tiny 500nm};

\end{axis}

\end{tikzpicture}%

%% file: beads_result_limited.tikz
\begin{tikzpicture}[baseline]

\begin{axis}[
width=\fwidth,
height=\fheight,
scale only axis,
point meta min=0,
point meta max=0.0625,
axis on top,
xmin=0.5,
xmax=400.5,
y dir=reverse,
ymin=0.5,
ymax=400.5,
xtick = \empty,
ytick = \empty,
colormap={mymap}{[1pt] rgb(0pt)=(1,1,1); rgb(1pt)=(0,0,0.625); rgb(7pt)=(0,0,1); rgb(23pt)=(0,1,1); rgb(39pt)=(1,1,0); rgb(55pt)=(1,0,0); rgb(63pt)=(0.5,0,0)},
colorbar,
colorbar style={
	ytick={0,0.0125,0.025,0.0375,0.05,0.0625},
	yticklabels={{},{32000},{16000},{10668},{8000},{6400}},
	scaled ticks = false
}
]
\addplot [forget plot] graphics [xmin=0.5, xmax=400.5, ymin=0.5, ymax=400.5] {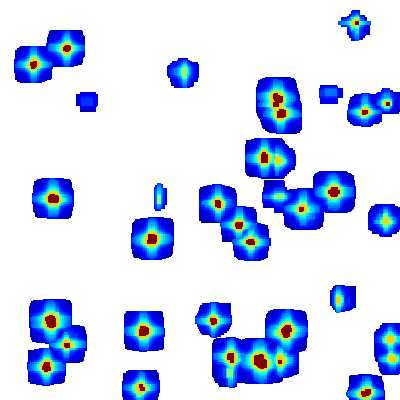};

\coordinate (A) at (350,55);
\coordinate (B) at (374,55);

\draw[color=black,solid,line width=2.0pt] (A) -- (B) node[midway,above] {\tiny 500nm};

\end{axis}

\end{tikzpicture}

%% file: beads_full_data.tikz
\begin{tikzpicture}[baseline]

\begin{axis}[%
width=\fwidth,
height=\fheight,
scale only axis,
point meta min=0,
point meta max=64,
axis on top,
xmin=0.5,
xmax=400.5,
y dir=reverse,
ymin=0.5,
ymax=400.5,
xtick = \empty,
ytick = \empty,
colormap={mymap}{[1pt] rgb(0pt)=(1,1,1); rgb(2pt)=(0.870899,0.870899,0.870899); rgb(3pt)=(0.806349,0.806349,0.806349); rgb(4pt)=(0.741799,0.741799,0.741799); rgb(6pt)=(0.698765,0.698765,0.698765); rgb(7pt)=(0.677249,0.677249,0.677249); rgb(10pt)=(0.612698,0.612698,0.612698); rgb(11pt)=(0.591182,0.591182,0.591182); rgb(13pt)=(0.548148,0.548148,0.548148); rgb(16pt)=(0.492819,0.492819,0.492819); rgb(17pt)=(0.474376,0.474376,0.474376); rgb(23pt)=(0.363719,0.363719,0.363719); rgb(24pt)=(0.345276,0.345276,0.345276); rgb(27pt)=(0.289947,0.289947,0.289947); rgb(29pt)=(0.257672,0.257672,0.257672); rgb(30pt)=(0.241534,0.241534,0.241534); rgb(32pt)=(0.209259,0.209259,0.209259); rgb(33pt)=(0.193122,0.193122,0.193122); rgb(36pt)=(0.144709,0.144709,0.144709); rgb(37pt)=(0.128571,0.128571,0.128571); rgb(39pt)=(0.0962963,0.0962963,0.0962963); rgb(40pt)=(0.0801587,0.0801587,0.0801587); rgb(41pt)=(0.0640212,0.0640212,0.0640212); rgb(42pt)=(0.0478836,0.0478836,0.0478836); rgb(43pt)=(0.031746,0.031746,0.031746); rgb(48pt)=(0.0238095,0.0238095,0.0238095); rgb(49pt)=(0.0222222,0.0222222,0.0222222); rgb(55pt)=(0.0126984,0.0126984,0.0126984); rgb(56pt)=(0.0111111,0.0111111,0.0111111); rgb(59pt)=(0.00634921,0.00634921,0.00634921); rgb(60pt)=(0.0047619,0.0047619,0.0047619); rgb(61pt)=(0.0031746,0.0031746,0.0031746); rgb(63pt)=(0,0,0)},
colorbar
]
\addplot [forget plot] graphics [xmin=0.5, xmax=400.5, ymin=0.5, ymax=400.5] {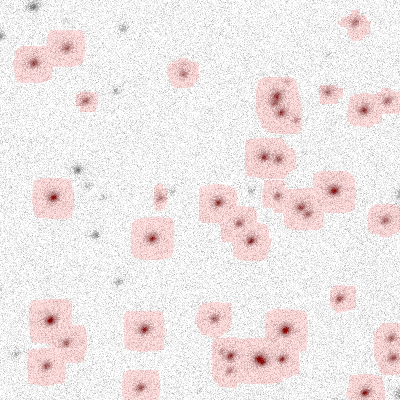};

\coordinate (A) at (350,55);
\coordinate (B) at (374,55);

\draw[color=black,solid,line width=2.0pt] (A) -- (B) node[midway,above] {\tiny 500nm};

\end{axis}

\end{tikzpicture}%

%% file: beads_full_data2.tikz
\begin{tikzpicture}[baseline]

\begin{axis}[%
width=\fwidth,
height=\fheight,
scale only axis,
point meta min=0,
point meta max=64,
axis on top,
xmin=0.5,
xmax=400.5,
y dir=reverse,
ymin=0.5,
ymax=400.5,
xtick = \empty,
ytick = \empty,
colormap={mymap}{[1pt] rgb(0pt)=(1,1,1); rgb(2pt)=(0.870899,0.870899,0.870899); rgb(3pt)=(0.806349,0.806349,0.806349); rgb(4pt)=(0.741799,0.741799,0.741799); rgb(6pt)=(0.698765,0.698765,0.698765); rgb(7pt)=(0.677249,0.677249,0.677249); rgb(10pt)=(0.612698,0.612698,0.612698); rgb(11pt)=(0.591182,0.591182,0.591182); rgb(13pt)=(0.548148,0.548148,0.548148); rgb(16pt)=(0.492819,0.492819,0.492819); rgb(17pt)=(0.474376,0.474376,0.474376); rgb(23pt)=(0.363719,0.363719,0.363719); rgb(24pt)=(0.345276,0.345276,0.345276); rgb(27pt)=(0.289947,0.289947,0.289947); rgb(29pt)=(0.257672,0.257672,0.257672); rgb(30pt)=(0.241534,0.241534,0.241534); rgb(32pt)=(0.209259,0.209259,0.209259); rgb(33pt)=(0.193122,0.193122,0.193122); rgb(36pt)=(0.144709,0.144709,0.144709); rgb(37pt)=(0.128571,0.128571,0.128571); rgb(39pt)=(0.0962963,0.0962963,0.0962963); rgb(40pt)=(0.0801587,0.0801587,0.0801587); rgb(41pt)=(0.0640212,0.0640212,0.0640212); rgb(42pt)=(0.0478836,0.0478836,0.0478836); rgb(43pt)=(0.031746,0.031746,0.031746); rgb(48pt)=(0.0238095,0.0238095,0.0238095); rgb(49pt)=(0.0222222,0.0222222,0.0222222); rgb(55pt)=(0.0126984,0.0126984,0.0126984); rgb(56pt)=(0.0111111,0.0111111,0.0111111); rgb(59pt)=(0.00634921,0.00634921,0.00634921); rgb(60pt)=(0.0047619,0.0047619,0.0047619); rgb(61pt)=(0.0031746,0.0031746,0.0031746); rgb(63pt)=(0,0,0)},
colorbar
]
\addplot [forget plot] graphics [xmin=0.5, xmax=400.5, ymin=0.5, ymax=400.5] {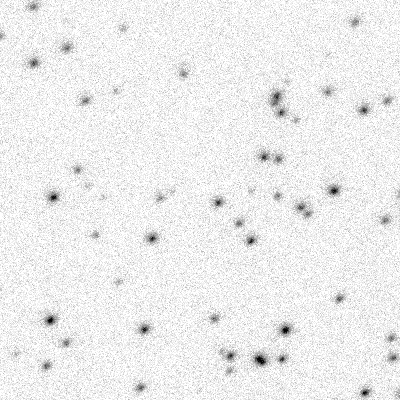};

\coordinate (A) at (350,55);
\coordinate (B) at (374,55);

\draw[color=black,solid,line width=2.0pt] (A) -- (B) node[midway,above] {\tiny 500nm};

\end{axis}

\end{tikzpicture}%

%% file: beads_result_full.tikz
\begin{tikzpicture}[baseline]

\begin{axis}[
width=\fwidth,
height=\fheight,
scale only axis,
point meta min=0,
point meta max=0.0625,
axis on top,
xmin=0.5,
xmax=400.5,
y dir=reverse,
ymin=0.5,
ymax=400.5,
xtick = \empty,
ytick = \empty,
colormap={mymap}{[1pt] rgb(0pt)=(1,1,1); rgb(1pt)=(0,0,0.625); rgb(7pt)=(0,0,1); rgb(23pt)=(0,1,1); rgb(39pt)=(1,1,0); rgb(55pt)=(1,0,0); rgb(63pt)=(0.5,0,0)},
colorbar,
colorbar style={
	ytick={0,0.0125,0.025,0.0375,0.05,0.0625},
	yticklabels={{},{32000},{16000},{10668},{8000},{6400}},
	scaled ticks = false
}
]
\addplot [forget plot] graphics [xmin=0.5, xmax=400.5, ymin=0.5, ymax=400.5] {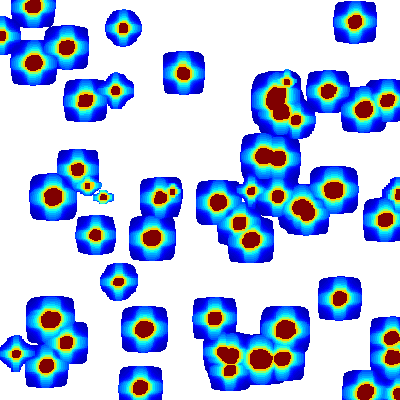};

\coordinate (A) at (350,55);
\coordinate (B) at (374,55);

\draw[color=black,solid,line width=2.0pt] (A) -- (B) node[midway,above] {\tiny 500nm};

\end{axis}

\end{tikzpicture}

%% file: beads_full_segments.tikz
\begin{tikzpicture}[baseline]
	
\begin{axis}[
width=\fwidth,
height=\fheight,
scale only axis,
point meta min=0,
point meta max=0.0625,
axis on top,
xmin=0.5,
xmax=400.5,
y dir=reverse,
ymin=0.5,
ymax=400.5,
xtick = \empty,
ytick = \empty,
colormap={mymap}{[1pt] rgb(0pt)=(1,1,1); rgb(63pt)=(1,0,0)},
colorbar,
colorbar style={
	ytick={0,0.0625},
	yticklabels={{inactive},{active}},
	scaled ticks = false
}
]
\addplot [forget plot] graphics [xmin=0.5, xmax=400.5, ymin=0.5, ymax=400.5] {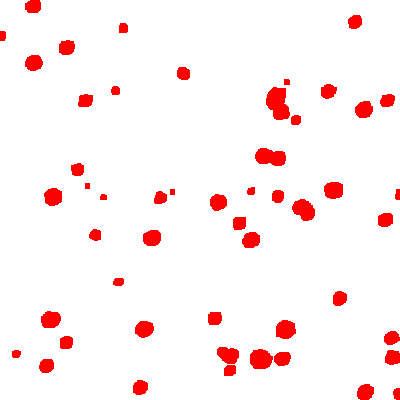};

\coordinate (A) at (330,55);
\coordinate (B) at (354,55);

\draw[color=black,solid,line width=2.0pt] (A) -- (B) node[midway,above] {\tiny 500nm};
\end{axis}
\end{tikzpicture}%

%% file: mixed_beads_limited_data.tikz
\begin{tikzpicture}[baseline]

\begin{axis}[%
width=\fwidth,
height=\fheight,
scale only axis,
point meta min=0,
point meta max=1.74036268949424,
axis on top,
xmin=0.5,
xmax=400.5,
y dir=reverse,
ymin=0.5,
ymax=400.5,
xtick = \empty,
ytick = \empty,
colormap={mymap}{[1pt] rgb(0pt)=(1,1,1); rgb(2pt)=(0.870899,0.870899,0.870899); rgb(3pt)=(0.806349,0.806349,0.806349); rgb(4pt)=(0.741799,0.741799,0.741799); rgb(6pt)=(0.698765,0.698765,0.698765); rgb(7pt)=(0.677249,0.677249,0.677249); rgb(10pt)=(0.612698,0.612698,0.612698); rgb(11pt)=(0.591182,0.591182,0.591182); rgb(13pt)=(0.548148,0.548148,0.548148); rgb(16pt)=(0.492819,0.492819,0.492819); rgb(17pt)=(0.474376,0.474376,0.474376); rgb(23pt)=(0.363719,0.363719,0.363719); rgb(24pt)=(0.345276,0.345276,0.345276); rgb(27pt)=(0.289947,0.289947,0.289947); rgb(29pt)=(0.257672,0.257672,0.257672); rgb(30pt)=(0.241534,0.241534,0.241534); rgb(32pt)=(0.209259,0.209259,0.209259); rgb(33pt)=(0.193122,0.193122,0.193122); rgb(36pt)=(0.144709,0.144709,0.144709); rgb(37pt)=(0.128571,0.128571,0.128571); rgb(39pt)=(0.0962963,0.0962963,0.0962963); rgb(40pt)=(0.0801587,0.0801587,0.0801587); rgb(41pt)=(0.0640212,0.0640212,0.0640212); rgb(42pt)=(0.0478836,0.0478836,0.0478836); rgb(43pt)=(0.031746,0.031746,0.031746); rgb(48pt)=(0.0238095,0.0238095,0.0238095); rgb(49pt)=(0.0222222,0.0222222,0.0222222); rgb(55pt)=(0.0126984,0.0126984,0.0126984); rgb(56pt)=(0.0111111,0.0111111,0.0111111); rgb(59pt)=(0.00634921,0.00634921,0.00634921); rgb(60pt)=(0.0047619,0.0047619,0.0047619); rgb(61pt)=(0.0031746,0.0031746,0.0031746); rgb(63pt)=(0,0,0)},
colorbar,
colorbar style={
	ytick={0,1},
	yticklabels={{0},{10}},
	scaled ticks = false
}
]
\addplot [forget plot] graphics [xmin=0.5, xmax=400.5, ymin=0.5, ymax=400.5] {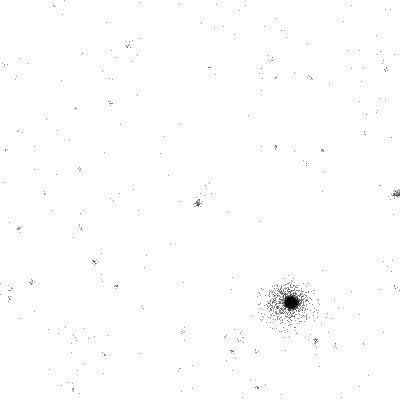};

\coordinate (A) at (350,55);
\coordinate (B) at (374,55);

\draw[color=black,solid,line width=2.0pt] (A) -- (B) node[midway,above] {\tiny 500nm};

\end{axis}

\end{tikzpicture}%

%% file: mixed_beads_result_limited.tikz
\begin{tikzpicture}[baseline]

\begin{axis}[
width=\fwidth,
height=\fheight,
scale only axis,
point meta min=0,
point meta max=0.0625,
axis on top,
xmin=0.5,
xmax=400.5,
y dir=reverse,
ymin=0.5,
ymax=400.5,
xtick = \empty,
ytick = \empty,
colormap={mymap}{[1pt] rgb(0pt)=(1,1,1); rgb(1pt)=(0,0,0.625); rgb(7pt)=(0,0,1); rgb(23pt)=(0,1,1); rgb(39pt)=(1,1,0); rgb(55pt)=(1,0,0); rgb(63pt)=(0.5,0,0)},
colorbar,
colorbar style={
	ytick={0,0.0125,0.025,0.0375,0.05,0.0625},
	yticklabels={{},{32000},{16000},{10668},{8000},{6400}},
	scaled ticks = false
}
]
\addplot [forget plot] graphics [xmin=0.5, xmax=400.5, ymin=0.5, ymax=400.5] {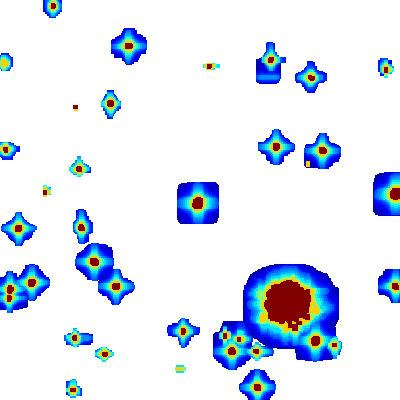};

\coordinate (A) at (350,55);
\coordinate (B) at (374,55);

\draw[color=black,solid,line width=2.0pt] (A) -- (B) node[midway,above] {\tiny 500nm};

\end{axis}

\end{tikzpicture}

%% file: mixed_beads_full_data.tikz
\begin{tikzpicture}[baseline]

\begin{axis}[%
width=\fwidth,
height=\fheight,
scale only axis,
point meta min=0,
point meta max=2.51054501020661,
axis on top,
xmin=0.5,
xmax=400.5,
y dir=reverse,
ymin=0.5,
ymax=400.5,
xtick = \empty,
ytick = \empty,
colormap={mymap}{[1pt] rgb(0pt)=(1,1,1); rgb(2pt)=(0.870899,0.870899,0.870899); rgb(3pt)=(0.806349,0.806349,0.806349); rgb(4pt)=(0.741799,0.741799,0.741799); rgb(6pt)=(0.698765,0.698765,0.698765); rgb(7pt)=(0.677249,0.677249,0.677249); rgb(10pt)=(0.612698,0.612698,0.612698); rgb(11pt)=(0.591182,0.591182,0.591182); rgb(13pt)=(0.548148,0.548148,0.548148); rgb(16pt)=(0.492819,0.492819,0.492819); rgb(17pt)=(0.474376,0.474376,0.474376); rgb(23pt)=(0.363719,0.363719,0.363719); rgb(24pt)=(0.345276,0.345276,0.345276); rgb(27pt)=(0.289947,0.289947,0.289947); rgb(29pt)=(0.257672,0.257672,0.257672); rgb(30pt)=(0.241534,0.241534,0.241534); rgb(32pt)=(0.209259,0.209259,0.209259); rgb(33pt)=(0.193122,0.193122,0.193122); rgb(36pt)=(0.144709,0.144709,0.144709); rgb(37pt)=(0.128571,0.128571,0.128571); rgb(39pt)=(0.0962963,0.0962963,0.0962963); rgb(40pt)=(0.0801587,0.0801587,0.0801587); rgb(41pt)=(0.0640212,0.0640212,0.0640212); rgb(42pt)=(0.0478836,0.0478836,0.0478836); rgb(43pt)=(0.031746,0.031746,0.031746); rgb(48pt)=(0.0238095,0.0238095,0.0238095); rgb(49pt)=(0.0222222,0.0222222,0.0222222); rgb(55pt)=(0.0126984,0.0126984,0.0126984); rgb(56pt)=(0.0111111,0.0111111,0.0111111); rgb(59pt)=(0.00634921,0.00634921,0.00634921); rgb(60pt)=(0.0047619,0.0047619,0.0047619); rgb(61pt)=(0.0031746,0.0031746,0.0031746); rgb(63pt)=(0,0,0)},
colorbar,
colorbar style={
	ytick={0,1,2},
	yticklabels={{0},{10},{100}},
	scaled ticks = false
}
]
\addplot [forget plot] graphics [xmin=0.5, xmax=400.5, ymin=0.5, ymax=400.5] {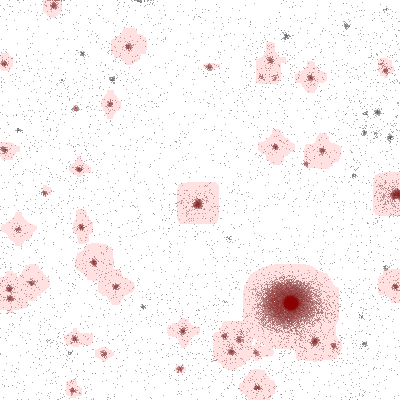};

\coordinate (A) at (350,55);
\coordinate (B) at (374,55);

\draw[color=black,solid,line width=2.0pt] (A) -- (B) node[midway,above] {\tiny 500nm};

\end{axis}

\end{tikzpicture}%

%% file: mixed_beads_full_data2.tikz
\begin{tikzpicture}[baseline]

\begin{axis}[%
width=\fwidth,
height=\fheight,
scale only axis,
point meta min=0,
point meta max=2.51054501020661,
axis on top,
xmin=0.5,
xmax=400.5,
y dir=reverse,
ymin=0.5,
ymax=400.5,
xtick = \empty,
ytick = \empty,
colormap={mymap}{[1pt] rgb(0pt)=(1,1,1); rgb(2pt)=(0.870899,0.870899,0.870899); rgb(3pt)=(0.806349,0.806349,0.806349); rgb(4pt)=(0.741799,0.741799,0.741799); rgb(6pt)=(0.698765,0.698765,0.698765); rgb(7pt)=(0.677249,0.677249,0.677249); rgb(10pt)=(0.612698,0.612698,0.612698); rgb(11pt)=(0.591182,0.591182,0.591182); rgb(13pt)=(0.548148,0.548148,0.548148); rgb(16pt)=(0.492819,0.492819,0.492819); rgb(17pt)=(0.474376,0.474376,0.474376); rgb(23pt)=(0.363719,0.363719,0.363719); rgb(24pt)=(0.345276,0.345276,0.345276); rgb(27pt)=(0.289947,0.289947,0.289947); rgb(29pt)=(0.257672,0.257672,0.257672); rgb(30pt)=(0.241534,0.241534,0.241534); rgb(32pt)=(0.209259,0.209259,0.209259); rgb(33pt)=(0.193122,0.193122,0.193122); rgb(36pt)=(0.144709,0.144709,0.144709); rgb(37pt)=(0.128571,0.128571,0.128571); rgb(39pt)=(0.0962963,0.0962963,0.0962963); rgb(40pt)=(0.0801587,0.0801587,0.0801587); rgb(41pt)=(0.0640212,0.0640212,0.0640212); rgb(42pt)=(0.0478836,0.0478836,0.0478836); rgb(43pt)=(0.031746,0.031746,0.031746); rgb(48pt)=(0.0238095,0.0238095,0.0238095); rgb(49pt)=(0.0222222,0.0222222,0.0222222); rgb(55pt)=(0.0126984,0.0126984,0.0126984); rgb(56pt)=(0.0111111,0.0111111,0.0111111); rgb(59pt)=(0.00634921,0.00634921,0.00634921); rgb(60pt)=(0.0047619,0.0047619,0.0047619); rgb(61pt)=(0.0031746,0.0031746,0.0031746); rgb(63pt)=(0,0,0)},
colorbar,
colorbar style={
	ytick={0,1,2},
	yticklabels={{0},{10},{100}},
	scaled ticks = false
}
]
\addplot [forget plot] graphics [xmin=0.5, xmax=400.5, ymin=0.5, ymax=400.5] {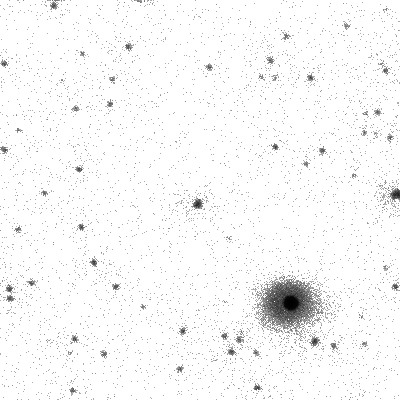};

\coordinate (A) at (350,55);
\coordinate (B) at (374,55);

\draw[color=black,solid,line width=2.0pt] (A) -- (B) node[midway,above] {\tiny 500nm};

\end{axis}

\end{tikzpicture}%

%% file: mixed_beads_result_full.tikz
\begin{tikzpicture}[baseline]

\begin{axis}[
width=\fwidth,
height=\fheight,
scale only axis,
point meta min=0,
point meta max=0.0625,
axis on top,
xmin=0.5,
xmax=400.5,
y dir=reverse,
ymin=0.5,
ymax=400.5,
xtick = \empty,
ytick = \empty,
colormap={mymap}{[1pt] rgb(0pt)=(1,1,1); rgb(1pt)=(0,0,0.625); rgb(7pt)=(0,0,1); rgb(23pt)=(0,1,1); rgb(39pt)=(1,1,0); rgb(55pt)=(1,0,0); rgb(63pt)=(0.5,0,0)},
colorbar,
colorbar style={
	ytick={0,0.0125,0.025,0.0375,0.05,0.0625},
	yticklabels={{},{32000},{16000},{10668},{8000},{6400}},
	scaled ticks = false
}
]
\addplot [forget plot] graphics [xmin=0.5, xmax=400.5, ymin=0.5, ymax=400.5] {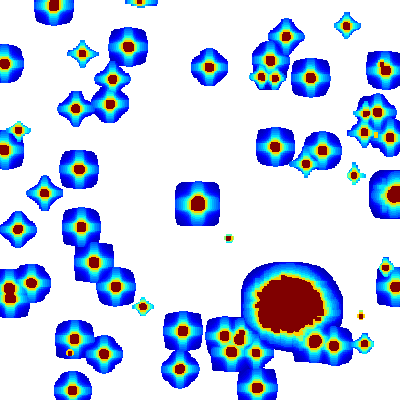};

\coordinate (A) at (350,55);
\coordinate (B) at (374,55);

\draw[color=black,solid,line width=2.0pt] (A) -- (B) node[midway,above] {\tiny 500nm};

\end{axis}

\end{tikzpicture}

%% file: mixed_beads_full_segments.tikz
\begin{tikzpicture}[baseline]
	
	\begin{axis}[
		width=\fwidth,
		height=\fheight,
		scale only axis,
		point meta min=0,
		point meta max=0.0625,
		axis on top,
		xmin=0.5,
		xmax=400.5,
		y dir=reverse,
		ymin=0.5,
		ymax=400.5,
		xtick = \empty,
		ytick = \empty,
		colormap={mymap}{[1pt] rgb(0pt)=(1,1,1); rgb(63pt)=(1,0,0)},
		colorbar,
		colorbar style={
			ytick={0,0.0625},
			yticklabels={{inactive},{active}},
			scaled ticks = false
		}
		]
\addplot [forget plot] graphics [xmin=0.5, xmax=400.5, ymin=0.5, ymax=400.5] {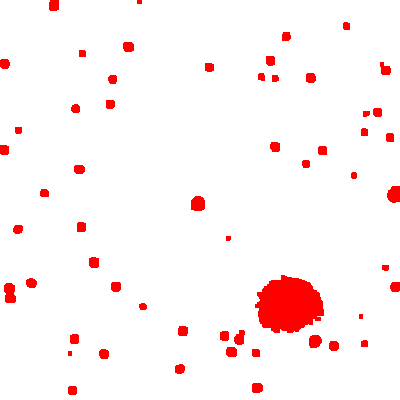};

\coordinate (A) at (330,55);
\coordinate (B) at (354,55);

\draw[color=black,solid,line width=2.0pt] (A) -- (B) node[midway,above] {\tiny 500nm};
\end{axis}
\end{tikzpicture}%

%% file: fibro_alpha_05.tikz
\begin{tikzpicture}[baseline]

\begin{axis}[
width=\fwidth,
height=\fheight,
scale only axis,
point meta min=0,
point meta max=0.0625,
axis on top,
xmin=0.5,
xmax=400.5,
y dir=reverse,
ymin=0.5,
ymax=400.5,
xtick = \empty,
ytick = \empty,
colormap={mymap}{[1pt] rgb(0pt)=(1,1,1); rgb(1pt)=(0,0,0.625); rgb(7pt)=(0,0,1); rgb(23pt)=(0,1,1); rgb(39pt)=(1,1,0); rgb(55pt)=(1,0,0); rgb(63pt)=(0.5,0,0)},
colorbar,
colorbar style={
	ytick={0,0.0125,0.025,0.0375,0.05,0.0625},
	yticklabels={{},{32000},{16000},{10668},{8000},{6400}},
	scaled ticks = false
}
]
]
\addplot [forget plot] graphics [xmin=0.5, xmax=400.5, ymin=0.5, ymax=400.5] {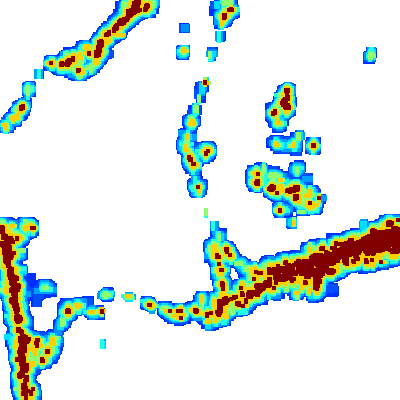};

\coordinate (A) at (330,55);
\coordinate (B) at (354,55);

\draw[color=black,solid,line width=2.0pt] (A) -- (B) node[midway,above] {\tiny 500nm};

\end{axis}
\end{tikzpicture}

%% file: fibro_alpha_07.tikz
\begin{tikzpicture}[baseline]

\begin{axis}[
width=\fwidth,
height=\fheight,
scale only axis,
point meta min=0,
point meta max=0.0625,
axis on top,
xmin=0.5,
xmax=400.5,
y dir=reverse,
ymin=0.5,
ymax=400.5,
xtick = \empty,
ytick = \empty,
colormap={mymap}{[1pt] rgb(0pt)=(1,1,1); rgb(1pt)=(0,0,0.625); rgb(7pt)=(0,0,1); rgb(23pt)=(0,1,1); rgb(39pt)=(1,1,0); rgb(55pt)=(1,0,0); rgb(63pt)=(0.5,0,0)},
colorbar,
colorbar style={
	ytick={0,0.0125,0.025,0.0375,0.05,0.0625},
	yticklabels={{},{32000},{16000},{10668},{8000},{6400}},
	scaled ticks = false
}
]
]
\addplot [forget plot] graphics [xmin=0.5, xmax=400.5, ymin=0.5, ymax=400.5] {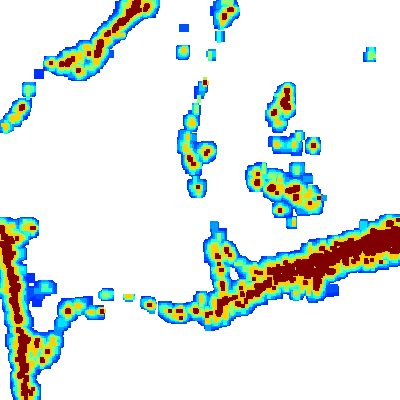};

\coordinate (A) at (330,55);
\coordinate (B) at (354,55);

\draw[color=black,solid,line width=2.0pt] (A) -- (B) node[midway,above] {\tiny 500nm};

\end{axis}
\end{tikzpicture}